\RequirePackage{lineno}
\documentclass[12pt]{JINST}

\usepackage{graphics}

\usepackage{epsfig}
\usepackage{rotate}
\usepackage{cite} 
\usepackage{amssymb}
\usepackage{amsbsy}

\preprint{DESY 10-032 \\ March 2010}

\title{Construction and Commissioning of the CALICE Analog Hadron Calorimeter Prototype}

\author{\centering

\LARGE\bf The CALICE Collaboration
}

\author{\centering
C.\,Adloff, 
Y.\,Karyotakis
\\ \it
Laboratoire d'Annecy-le-Vieux de Physique des Particules, Universit\'{e} de Savoie, CNRS/IN2P3,
9 Chemin du Bellevue BP 110, F-74941 Annecy-le-Vieux Cedex, France
}

\author{\centering
J.\,Repond
\\ \it
Argonne National Laboratory,
9700 S.\ Cass Avenue,
Argonne, IL 60439-4815,
USA}

\author{\centering
A.\,Brandt, H.\,Brown, K.\,De, C.\,Medina, J.\,Smith, J.\,Li, M.\,Sosebee, A.\,White, 
J.\,Yu
\\ \it
Department of Physics, SH108, University of Texas, Arlington, TX 76019, USA
}

\author{\centering
T.\,Buanes, G.\,Eigen
\\ \it
University of Bergen, Inst. of Physics, Allegaten 55, N-5007 Bergen, Norway
}

\author{\centering
Y.\,Mikami, O.\,Miller, N.\,K.\,Watson,  J.\,A.\,Wilson
\\ \it
University of Birmingham,
School of Physics and Astronomy,
Edgbaston, Birmingham B15 2TT, UK
}

\author{\centering 
T.\,Goto, 
G.\,Mavromanolakis\footnote{Now at CERN, Geneva}, M.\,A.\,Thomson, D.\,R.\,Ward, W.\,Yan\footnote{Now at University of Science and Technology of China, Hefei}
\\ \it
University of Cambridge, Cavendish Laboratory, J J Thomson Avenue, CB3 0HE, UK
}

\author{\centering 
D.\,Benchekroun, A.\,Hoummada, Y.\,Khoulaki
\\ \it
Universit\'{e} Hassan II A\"{\i}n Chock, Facult\'{e} des sciences. B.P. 5366 Maarif, Casablanca, Morocco
}

\author{\centering 
M.\,Oreglia
\\ \it
University of Chicago, Dept. of Physics, 5720 So. Ellis Ave., KPTC 201 Chicago, 
IL 60637-1434, USA
}

\author{\centering
M.\,Benyamna, C.\,C\^{a}rloganu, P.\,Gay
\\ \it
Laboratoire de Physique Corpusculaire de Clermont-Ferrand (LPC),
24 avenue des Landais,
63177 Aubi\`ere CEDEX, France
}

\author{\centering
J.\,Ha
\\ \it
Korea Atomic Energy Research Institute,
Taejon 305-600,
South Korea
}

\author{\centering
G.\,C.\,Blazey, D.\,Chakraborty, A.\,Dyshkant, K.\,Francis, D.\,Hedin, G.\,Lima, V.\,Zutshi
\\ \it
NICADD, Northern  Illinois University,
Department of Physics,
DeKalb, IL 60115,
USA
}

\author{\centering 
V.\,A.\,Babkin, S.\,N.\,Bazylev, Yu.\,I.\,Fedotov, V.\,M.\,Slepnev, I.\,A.\,Tiapkin, S.\,V.\,Volgin
\\ \it
Joint Institute for Nuclear Research, Joliot-Curie 6,
141980, Dubna,
Moscow Region, Russia
}

\author{\centering 
J.\,-Y.\,Hostachy, L.\,Morin
\\ \it
Laboratoire de Physique Subatomique et de Cosmologie - Universit\'{e} Joseph Fourier Grenoble 1 -
CNRS/IN2P3 - Institut Polytechnique de Grenoble,
53, rue des Martyrs,
38026 Grenoble CEDEX, France
}

\author{\centering 
N.\,D'Ascenzo, U.\,Cornett, D.\,David, R.\,Fabbri, G.\,Falley, N.\,Feege, K.\,Gadow, E.\,Garutti, P.\,G\"{o}ttlicher, T.\,Jung, S.\,Karstensen, V.\,Korbel, A.\,-I.\,Lucaci-Timoce, B.\,Lutz, N.\,Meyer, V.\,Morgunov, M.\,Reinecke, S.\,Sch\"{a}tzel, S.\,Schmidt, F.\,Sefkow, P.\,Smirnov, A.\,Vargas-Trevino, N.\,Wattimena, O.\,Wendt
\\ \it
DESY, Notkestrasse 85,
D-22603 Hamburg, Germany
}

\author{\centering  
M.\,Groll, R.\,-D.\,Heuer, S.\,Richter\footnote{now at University of Heidelberg}, J.\,Samson
\\ \it
Univ. Hamburg,
Physics Department,
Institut f\"ur Experimentalphysik,
Luruper Chaussee 149,
22761 Hamburg, Germany
}

\author{\centering 
A.\,Kaplan, H.\,-Ch.\,Schultz-Coulon, W.\,Shen, A.\,Tadday
\\ \it
 University of Heidelberg, Fakultat fur Physik und Astronomie,
Albert Uberle Str. 3-5 2.OG Ost,
D-69120 Heidelberg, Germany
}

\author{\centering 
B.\,Bilki, E.\,Norbeck, Y.\,Onel
\\ \it
University of Iowa, Dept. of Physics and Astronomy,
203 Van Allen Hall, Iowa City, IA 52242-1479, USA
}

\author{\centering 
E.\,J.\,Kim
\\ \it
Chonbuk National University, Jeonju, 561-756, South Korea
}

\author{\centering 
G.\,Kim, D-W.\,Kim, K.\,Lee, S.\,C.\,Lee
\\ \it
Kangnung National University, HEP/PD, Kangnung, South Korea
}

\author{\centering 
K.\,Kawagoe, Y.\,Tamura
\\ \it
 Department of Physics, Kobe University, Kobe, 657-8501, Japan
}

\author{\centering 
J.\,A.\,Ballin, 
P.\,D.\,Dauncey, A.\,-M.\,Magnan, 
H.\,Yilmaz, O.\,Zorba
\\ \it
Imperial College, Blackett Laboratory,
Department of Physics,
Prince Consort Road,
London SW7 2AZ, UK 
}

\author{\centering 
V.\,Bartsch\footnote{Now at University of Sussex, Physics and Astronomy Department, Brighton, Sussex, BN1 9QH, UK},  
M.\,Postranecky, M.\,Warren, M.\,Wing
\\ \it
Department of Physics and Astronomy, University College London,
Gower Street,
London WC1E 6BT, UK
}

\author{\centering 
M.\,Faucci Giannelli, 
M.\,G.\,Green, 
F.\,Salvatore\footnote{Now at University of Sussex, Physics and Astronomy Department, Brighton, Sussex, BN1 9QH, UK}
\\ \it
Royal Holloway University of London,
Dept. of Physics,
Egham, Surrey TW20 0EX, UK
}

\author{\centering 
R.\,Kieffer, 
I.\,Laktineh
\\ \it
Universit\'{e} de Lyon, F-69622, Lyon, France ;
Universit\'{e} de Lyon 1, Villeurbanne ;
CNRS/IN2P3, Institut de Physique Nucl\'{e}aire de Lyon
}

\author{\centering 
M.\,C Fouz 
\\ \it
CIEMAT, Centro de Investigaciones Energeticas, Medioambientales y Tecnologicas, Madrid.\, Spain 
}

\author{\centering 
D.\,S.\,Bailey, 
R.\,J.\,Barlow, 
R.\,J.\,Thompson 
\\ \it
The University of Manchester, School of Physics and Astronomy,
Schuster Lab,
Manchester M13 9PL,
UK
}

\author{\centering 
M.\,Batouritski, O.\,Dvornikov, Yu.\,Shulhevich, N.\,Shumeiko, A.\,Solin, P.\,Starovoitov, V.\,Tchekhovski, A.\,Terletski
\\ \it
National Centre of Particle and High Energy Physics of the
Belarusian State University, M.Bogdanovich str. 153, 220040 Minsk, Belarus
}

\author{\centering 

B.\,Bobchenko, M.\,Chadeeva, M.\,Danilov, O.\,Markin, R.\,Mizuk, V.\,Morgunov, E.\,Novikov, V.\,Rusinov, E.\,Tarkovsky
\\ \it
Institute of Theoretical and Experimental Physics, B. Cheremushkinskaya ul. 25,
RU-117218 Moscow, Russia
}

\author{\centering 
V.\,Andreev, N.\,Kirikova,  A.\,Komar, V.\,Kozlov, P.\,Smirnov, Y.\,Soloviev, A.\,Terkulov 
\\ \it
P.N. Lebedev Physical Institute,
Russian Academy of Sciences,
117924 GSP-1 Moscow, B-333, Russia
}

\author{\centering 
P.\,Buzhan, B.\,Dolgoshein, A.\,Ilyin, V.\,Kantserov, V.\,Kaplin, A.\,Karakash, E.\,Popova, S.\,Smirnov 
\\ \it
Moscow Physical Engineering Inst., MEPhI,
Dept. of Physics,
31, Kashirskoye shosse,
115409 Moscow, Russia
}

\author{\centering 
N.\,Baranova, E.\,Boos, L.\,Gladilin, D.\,Karmanov, M.\,Korolev, M.\,Merkin, A.\,Savin, A.\,Voronin
\\ \it
M.V.Lomonosov Moscow State University, D.V.Skobeltsyn Institute of Nuclear
Physics (SINP MSU),
1/2 Leninskiye Gory, Moscow, 119991, Russia
}

\author{\centering 
A.\,Topkar
\\ \it
Bhabha Atomic Research Center,
Mumbai 400085, India
}

\author{\centering 
A.\,Frey\footnote{now at University G\"{o}ttingen}, C.\,Kiesling, S.\,Lu, K.\,Prothmann, K.\,
Seidel, F.\,Simon, C.\, Soldner, L.\, Weuste
\\ \it
Max Planck Inst. f\"ur Physik,
F\"ohringer Ring 6,
D-80805 Munich, Germany
}

\author{\centering 
B.\,Bouquet,    S.\,Callier, P.\,Cornebise, F.\,Dulucq, 
J.\,Fleury, H.\,Li,  G.\,Martin-Chassard, 
F.\,Richard, Ch.\,de la Taille, R.\,Poeschl, L.\,Raux, M.\,Ruan,  N.\,Seguin-Moreau, F.\,Wicek
\\ \it
Laboratoire de L'acc\'elerateur Lin\'eaire,
Centre d'Orsay, Universit\'e de Paris-Sud XI,
BP 34, B\^atiment 200,
F-91898 Orsay CEDEX, France
}

\author{\centering 
M.\,Anduze, V.\,Boudry, J-C.\,Brient, 
G.\,Gaycken, R.\,Cornat, D.\,Jeans, 
P.\,Mora de Freitas, G.\,Musat, M.\,Reinhard, A.\,Roug\'{e},  
J-Ch.\,Vanel, H.\,Videau
\\ \it
\'Ecole Polytechnique,
Laboratoire Leprince-Ringuet (LLR),
Route de Saclay,
F-91128 Palaiseau,
CEDEX France
}

\author{\centering 
K-H.\,Park
\\ \it
Pohang Accelerator Laboratory, Pohang 790-784, South Korea
}

\author{\centering 
J.\,Zacek 
\\ \it
Charles University, Institute of Particle \& Nuclear Physics,
V Holesovickach 2,
CZ-18000 Prague 8, Czech Republic  
}

\author{\centering 
J.\,Cvach, P.\,Gallus, M.\,Havranek, M.\,Janata, 
J.\,Kvasnicka, M.\,Marcisovsky, I.\,Polak, J.\,Popule, L.\,Tomasek, M.\,Tomasek, P.\,Ruzicka, P.\,Sicho, J.\, Smolik, V.\,Vrba, J.\,Zalesak 
\\ \it
Institute of Physics, Academy of Sciences of the Czech Republic, Na Slovance 2,
CZ-18221 Prague 8, Czech Republic
}

\author{\centering 
Yu.\,Arestov, 
V.\,Ammosov, B.\,Chuiko, V.\,Gapienko,
Y.\,Gilitski,V.\,Koreshev, A.\,Semak, Yu.\,Sviridov, V.\,Zaets
\\ \it
Institute of High Energy Physics,
Moscow Region,
RU-142284 Protvino,
Russia
}

\author{\centering 
B.\,Belhorma, M.\, Belmir
\\ \it
Centre National de l'Energie, des Sciences et des Techniques Nucl\'{e}aires, 
B.P. 1382, R.P. 10001, Rabat, Morocco
}

\author{\centering 
 A.\,Baird, R.\,N.\,Halsall
\\ \it
Rutherford Appleton Laboratory, Chilton, Didcot,
Oxon OX110QX, UK 
}

\author{\centering 
S.\,W.\,Nam, I.\,H.\,Park, J.\,Yang 
\\ \it
Ewha Womans University, Dept. of Physics,
Seoul 120,
South Korea
}

\author{\centering 
Jong-Seo Chai, Jong-Tae Kim, Geun-Bum Kim
\\ \it
Sungkyunkwan University,
300 Cheoncheon-dong, Jangan-gu, Suwon, Gyeonggi-do  440-746, South Korea
}

\author{\centering 

Y.\,Kim
\\ \it
Korea Institute of Radiological and
Medical Sciences,
215-4 Gangeung-dong,
Nowon-gu, Seoul 139-706,
SOUTH KOREA
}

\author{\centering 
J.\,Kang, Y.\,-J.\,Kwon  
\\ \it
Yonsei  University, Dept. of Physics,
134 Sinchon-dong,
Sudaemoon-gu, Seoul 120-749,
South Korea
}

\author{\centering 
Ilgoo Kim, Taeyun Lee, Jaehong Park, Jinho Sung
\\ \it
School of Electric Engineering and Computing Science, Seoul National University,
Seoul 151-742, South Korea
}

\author{\centering              
S.\, Itoh,  K.\,Kotera, M.\, Nishiyama,T.\,Takeshita
\\ \it
Shinshu Univ.,
Dept. of Physics,
3-1-1 Asaki,
Matsumoto-shi, Nagano 390-861,
Japan
}

\author{\centering 
S.\,Weber, C.\,Zeitnitz
\\ \it
Bergische Universit\"{a}t Wuppertal
Fachbereich 8 Physik,
Gausstrasse 20,
D-42097 Wuppertal, GERMANY
}

\abstract{
An analog hadron calorimeter (AHCAL) prototype of 5.3 nuclear interaction lengths thickness has been constructed by members of the CALICE Collaboration. The AHCAL prototype consists of a 38-layer sandwich structure of steel plates and highly-segmented scintillator tiles that are read out by wavelength-shifting fibers coupled to SiPMs. The signal is amplified and shaped with a custom-designed ASIC. A calibration/monitoring system based on LED light was developed to monitor the SiPM gain and to measure the full SiPM response curve in order to correct for non-linearity. Ultimately, the physics goals are the study of hadron shower shapes and testing the concept of particle flow. The technical goal consists of measuring the performance and reliability of 7608 SiPMs. The AHCAL was commissioned in test beams at DESY and CERN. The entire prototype was completed in 2007 and recorded hadron showers, electron showers and muons at different energies and incident angles in test beams at CERN and Fermilab.
}

\begin{document}
\section{Introduction}
\label{intro}

The physics of the International Linear Collider (ILC) demands jet energy resolution of $\sigma_E/E =30\%/\sqrt{E}$ in order to efficiently separate $Z^0$, $W^\pm$ and Higgs bosons via reconstruction of their dijet invariant mass~\cite{goal}. Boson separation depends on the success of the particle flow concept~\cite{pf}. The basic idea consists of reconstructing jets by separating the contributions of charged particles, photon showers and neutral hadron showers and measuring each of these components using the best suited detector subsystems. A typical jet consists of $60\%$ charged particles, $30\%$ photons and $10\%$ neutral hadrons. For example, charged hadrons and muons are reconstructed in the tracking system. Since their momenta are reconstructed with excellent resolution even for high-momentum particles, their contribution to the jet-energy resolution is negligible. Photons are reconstructed in the electromagnetic calorimeter with an energy resolution of typically $15\% /\sqrt{E}$, while neutral hadrons are reconstructed in the hadron calorimeter with a typical energy resolution of $>50\%/ \sqrt{E}$. 

In order to  reconstruct accurately neutral hadron showers, energy deposits in the calorimeter from charged hadrons, photons, electrons and muons must be reconstructed and their energy removed from consideration. Due to overlapping showers the assignment of a cell in the calorimeter to a particular shower is ambiguous. This effect is accounted for by a `confusion' term in the jet-energy resolution, which becomes the dominant term as particle separation decreases. Thus, the success of this technique relies on the accuracy with which individual cells in the hadron calorimeter are assigned to the correct shower. For successful associations both the electromagnetic and hadronic calorimeters must have high granularity in the longitudinal and transverse directions.

We present herein a description of the design, construction, and commissioning of the $\rm 1~m^3$ CALICE\footnote{CALICE stands for Calorimeter for the Linear Collider Experiment~\cite{calice}.} analog hadron calorimeter (AHCAL) prototype that consists of a steel-scintillator sandwich structure. The scintillator planes are segmented into square tiles that are individually read out by multipixel Geiger-mode-operated avalanche photodiodes, here called SiPMs \cite{sipma, sipmb, sipmc}. The main technical goal is to test the performance and reliability of the SiPM readout on a large scale, since this detector uses thousands of SiPMs (7608) for photon readout in a test beam. Ultimately, our physics goals are the study of hadron shower shapes, reproduction of observed shower shapes in simulations, and first studies of particle flow (PFLOW) algorithms. 

First tests of the AHCAL prototype were accomplished with test beams at CERN in 2006 and 2007.
With a Si-W electromagnetic calorimeter (ECAL) prototype \cite{ecal} in front and a tail catcher \cite{tcmt} behind the AHCAL, we have measured energy deposits of muons, electrons and pions. In addition, some data were taken with the AHCAL alone. Electron energies were varied between $\rm 10~GeV$ and $\rm 60~GeV$, and pion energies between $\rm 2~GeV$ and $\rm 120~GeV$. We also varied the incident beam angle from normal incidence ($90^\circ$) to $60^\circ$ in steps of $10^\circ$. Further studies with a focus on lower energy data points continued at Fermilab in 2008 and 2009. Here, we also tested an electromagnetic calorimeter with W- scintillator-strips.

This article is organized as follows. In Section Two we discuss goals and design considerations. In Section Three we present the detector layout and in Section Four we give details on the readout system. In Sections Five and Six we respectively give a short summary of the detector electronics and online software and data processing. In Sections Seven and Eight we discuss our calibration/monitoring system and our calibration method. We present commissioning and initial performance in Section Nine before concluding in Section Ten.


\section{Goals and Design Considerations} 
\label{goals}

The design of the AHCAL prototype described here is inspired by the calorimeter layout of the Large Detector concept (LDC)~\cite{ldc-dod}, which has evolved to the International Large Detector Concept (ILD)\cite{ild-loi}. The AHCAL is constructed as a sampling calorimeter using a material of low magnetic permeability ($\mu <1.01$) as absorbers and scintillator plates, subdivided into tiles, as the active medium. The millimeter-size SiPM devices that are mounted on each tile allow operation in high magnetic fields~\cite{groll}. The scintillation light is collected from the tile via wavelength-shifting (WLS) fibers  embedded in a groove. This concept is different from existing tile calorimeters that have long fiber readout. Ultimately, this allows us to integrate both  the photosensors and the front-end electronics into the detector volume. The high granularity imposed by particle flow methods is realizable with scintillators in a very compact way at reasonable cost. The baseline absorber material is stainless steel for reasons of cost and mechanical rigidity. However, since we do not anticipate operating the AHCAL prototype in magnetic fields in test beams, we have used absorber plates manufactured from standard steel (see Section~\ref{absorber}).

We have two main goals for the physics prototype. First, we want to test the novel SiPM readout technology on a large scale, identify critical operational issues, develop quality control procedures and establish reliable calibration concepts that include test bench data. We operate here nearly two orders of magnitude more SiPMs than in our first test calorimeter~\cite{minical}. Second, we want to accumulate very large data samples of hadronic showers in several test beams. These samples are needed to investigate hadronic  shower shapes and to test simulation models, since it is not possible to extract this information from existing calorimeter data. The test beam data samples are also very useful for studying and tuning particle flow reconstruction algorithms with real events. 

Transverse dimensions of $\rm 1~m \times 1~m$  and a depth of $1\rm ~m$ represent an adequate choice considering performance and cost. This guarantees that the core of hadron showers with energies up to several tens of GeV (the range being most relevant for the ILC) is laterally contained. The depth is comparable to that of a realistic ILC AHCAL which has to fit
inside the diameter of the solenoid. The longitudinal segmentation should be of the order of one $X_0$ and the transverse dimension of the order of a Moli\`ere radius to resolve the electromagnetic substructure in the shower. Detailed simulation studies of overlapping hadronic showers in ILC events have shown that a transverse tile dimension of $\rm 3~cm\times 3~cm$ provides optimal two-particle separation~\cite{raspereza} and the anticipated jet energy resolution~\cite{pf}. The total thickness of  the AHCAL prototype is 5.3 nuclear interaction lengths $(\lambda_n)$ or 4.3 pion interaction lengths ($\lambda_\pi$).

The design is largely based on established technologies and contains a high degree of redundancy to ensure stable and reliable operation of the AHCAL prototype over several years while testing the novel SiPM readout. Since this layout is not a section of a full detector for the ILC, it is not scalable and many of its external components still need to be integrated into the ILC detector volume. 
   
In test beam operation, the hadron calorimeter prototype is augmented with a tail catcher and muon tracker (TCMT) system~\cite{tcmt} to record leakage out of the rear of the AHCAL prototype
which becomes important particularly for high-energy hadrons. While for a 10~GeV pion the energy leakage is $ 3 \%$, it reaches $8 \%$  for a 80~GeV pion \cite{zz}. The TCMT is a sandwich structure of steel absorber plates and scintillator strips read out by SiPMs connected to the same electronics as the AHCAL prototype. The first eight layers of the TCMT ($1.1 \lambda_n$) have the same longitudinal sampling as the AHCAL prototype, while the next eight layers have a coarse sampling corresponding to $4\lambda_n$. In addition, the endplate of the AHCAL ($0.12 \lambda_n$) contributes. The ECAL prototype \cite{ecal} in front of the AHCAL prototype has a depth of 0.9~$\lambda_n$. The overall calorimeter depth is $0.88\lambda_n + 5.26\lambda_n + 0.1\lambda_n+5.76\lambda_n = 12.0\lambda_n$.

Since particles in a colliding-beam detector are produced over a large range of polar angles and charged-particle tracks are curved in the strong magnetic field of the ILC detector, the typical angle of incidence at the calorimeter front face will differ from $90^\circ$. Thus, we have built a mechanical support structure that accommodates incidence angles up to $55^\circ$ without turning the support structure itself. The detector layout is modular such that active layers are exchangeable in order to test different readout technologies within the same absorber structure.


\section{Detector Layout}
\label{layout}

\begin{figure}
\begin{center}
\includegraphics[width=140mm]{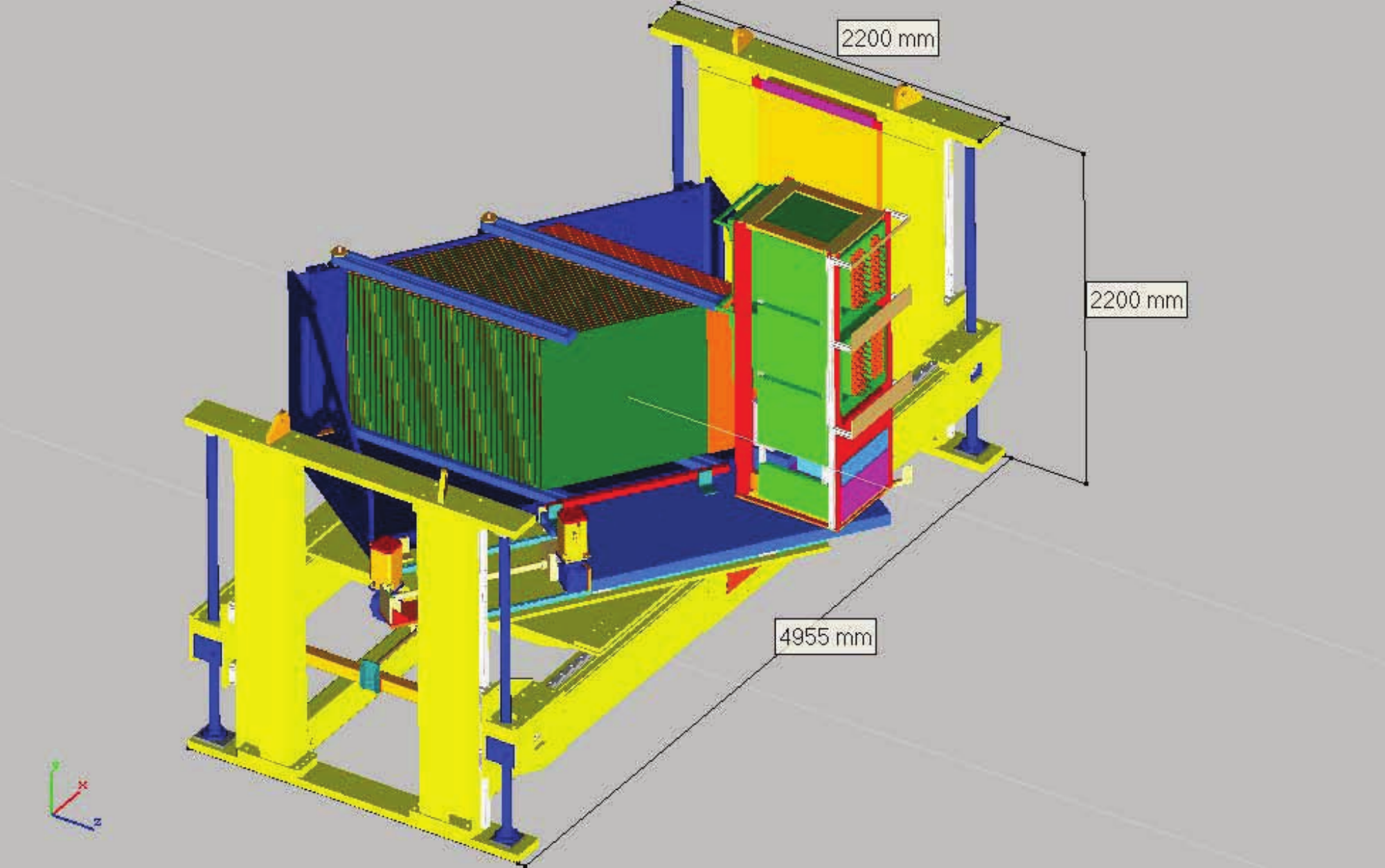}
\caption{Schematic layout of the AHCAL prototype placed on a moving stage. The steel plates mounted on rods are shown rotated with respect to the beam that enters from the right hand side. The rack in the front houses the supply voltages, trigger electronics, and data acquisition system.}
\label{fig:ahcal}
\end{center}
\end{figure}

\begin{figure}
\includegraphics[width=70mm]{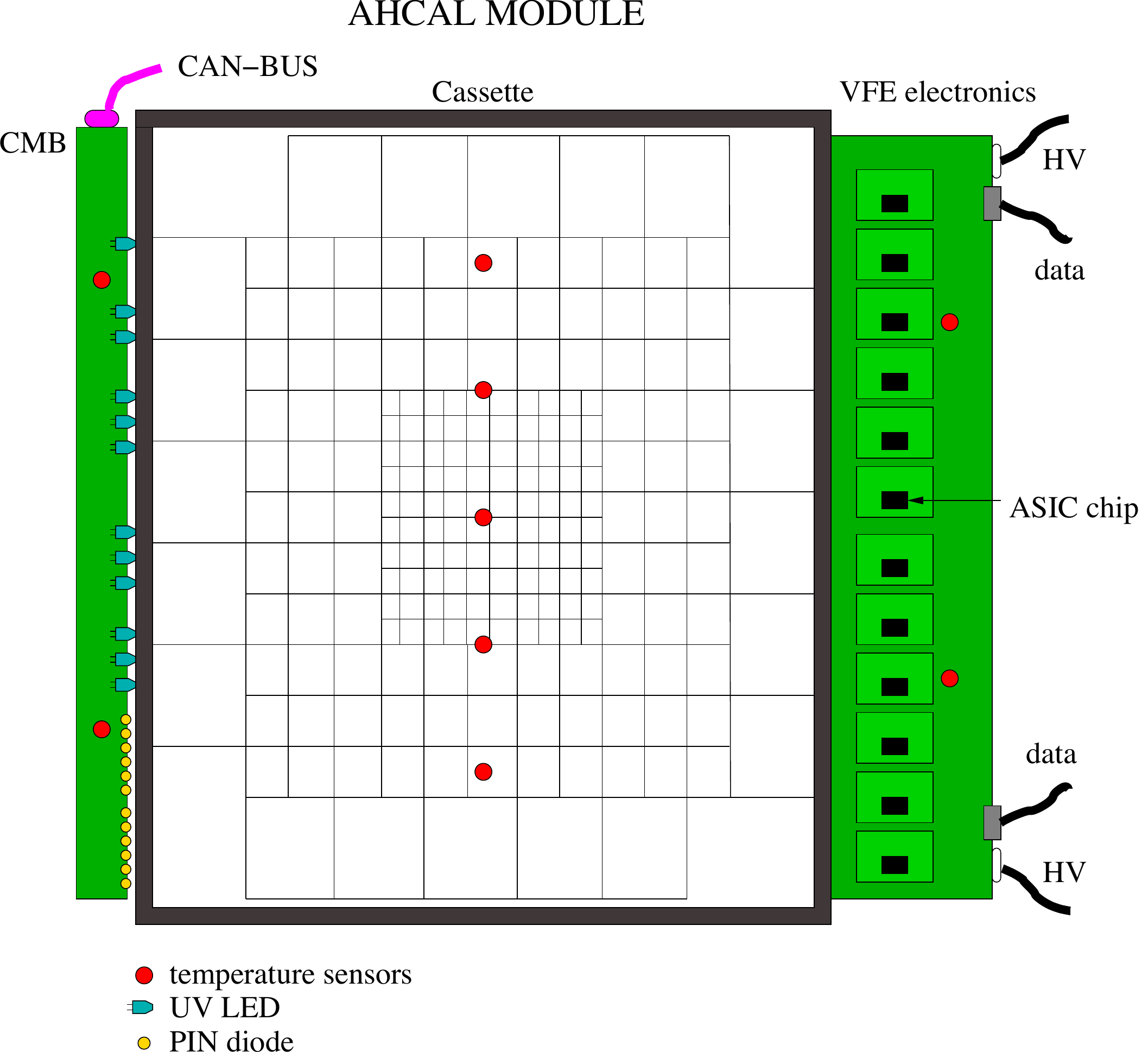}
\includegraphics[width=75mm]{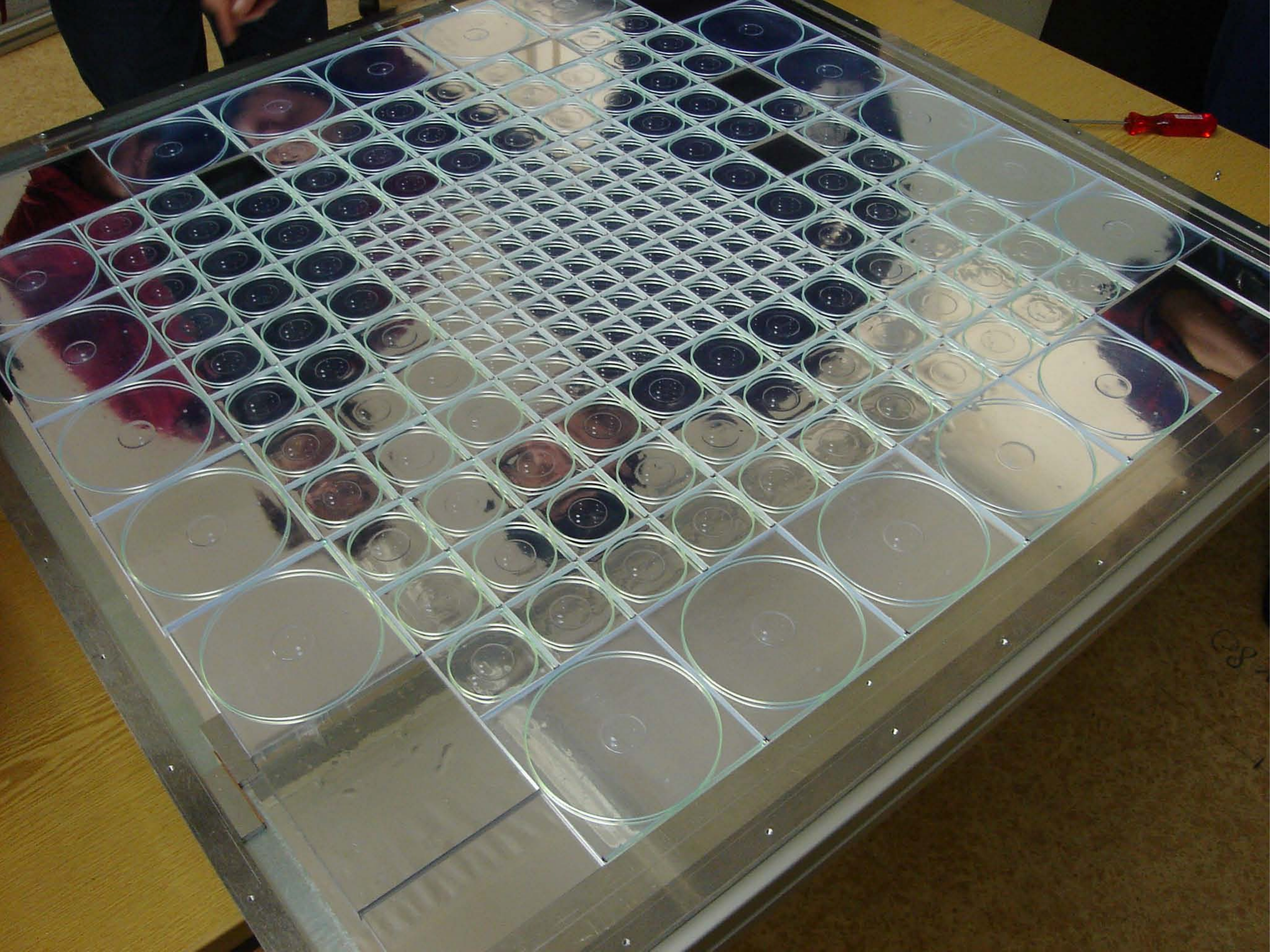}
\caption{Schematic tile layout of a scintillator module for layers 1--30 (left) and a photograph of tiles in a module (right). The red dots indicate the position of the thermosensors.}
\label{fig:module}
\vskip 0.5cm
\end{figure}

\begin{figure}
\begin{center}
\includegraphics[width=100mm]{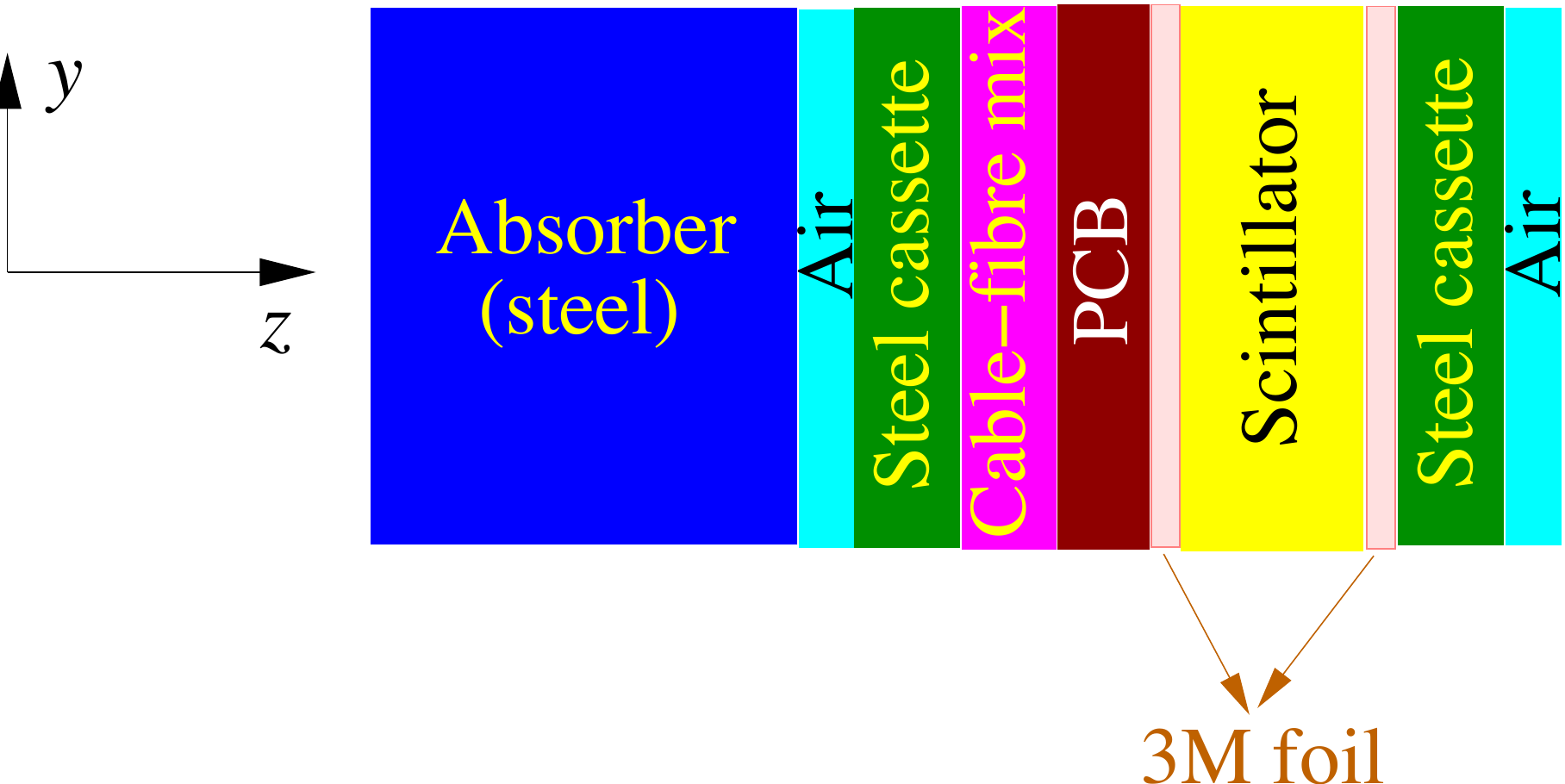}
\caption{Schematic cross section of a cassette (not to scale).}
\label{fig:layer-xsect}
\vskip 0.5cm
\end{center}
\end{figure}

The AHCAL prototype consists of a sandwich structure of 38 absorber plates, 38 active layers containing 7608 scintillator cells and an endplate. A schematic layout is displayed in Figure~\ref{fig:ahcal}. All scintillator tiles in a layer are housed inside a rigid cassette, {\it i.e.} a closed box with steel sheet top and bottom covers.  A cassette with the calibration/monitoring board on one side and the readout electronics on the other side is called a module. Figure~\ref{fig:module} displays the schematic layout of a module and a photograph of the scintillator tiles in a cassette for layers 1--30. The segmentation is $\rm 3~cm \times 3~cm$ in the core and coarser elsewhere (see Section~\ref{system}). Figure~\ref{fig:layer-xsect} shows a schematic cross section of a cassette and Table~\ref{tab:layer}summarizes dimensions, $X_0$, $\lambda_\pi$ and $\lambda_n$ of the individual components.
A VME-based system for digitization and data acquisition is placed in a separate crate. The AHCAL prototype is placed on a movable stage that allows us to move the prototype up/down and left/right as well as to rotate it. In a rotated configuration,  the individual layers have to be realigned to ensure that the beam still traverses through the center of each layer.


\subsection{Absorber Plates}
\label{absorber}

The steel plates are $\rm 1~m \times 1~m$ wide and on average $\rm 17.4~mm$ thick.  We use standard S235 steel that is a composite of  iron, carbon, manganese,  phosphorus and sulphur. Table~\ref{tab:layer} summarizes the characteristic parameters  of each layer\footnote{The steel plates are covered with a thin layer of Zn to prevent rusting. The thickness is typically $\rm 100~ \mu m $ on each side which occasionally may increase to $\rm 250 \mu m$. In the calculation of the properties we have assumed steel for the entire thickness. Including $\rm 100~\mu m$ thick Zn coating on each side decreases $\lambda_\pi$ by 0.0047 pion interaction lengths and increases $X_0$ by 0.0037 radiation lengths.} while
Table~\ref{tab:ahcaltot} shows the number of $\lambda_\pi$, $\lambda_n$ and $X_0$ for the entire AHCAL. The magnetic properties are irrelevant, because we do not plan any measurements inside a magnetic field. The absorber plates are mounted on support bars with four bolts. The gap width between plates is adjustable. For a perpendicular beam direction, the gaps are 1.4~cm wide including a $\rm 2~mm$ tolerance to account for the aplanarity of the steel plates and to allow a smooth insertion of the cassettes. The cover sheets, each 2mm thick, together with the absorber plates yield on average a total absorber thickness of 21.4~mm per layer.

\begin{table}
  \begin{center}
    
\begin{tabular}{|l|c|c|c|c|c|c|c|c|c|}
      \hline
      material       & $\rho$ & $\lambda_\pi$ &$ \lambda_\pi/\rho$& $\lambda_n$ & $\lambda_n/\rho$ & $X_0$ &  $X_0/\rho$ & $R_M$ & f \\             
  & [g/cm$^3]$ &[cm]        & [g/cm$^2]$    &[cm]        & [g/cm$^2]$   & [cm]    &  [g/cm$^2]$  & [cm]  & [\%]    \\
      \hline
      Fe             & 7.87   & 20.4  & 160.8   & 16.8  & 132.1  & 1.76 &  13.8  & 1.72  &  98.34 \\
      Mn            & 7.44   & 21.5  & 160.2   & 17.7  & 131.4  & 1.97 &  14.6  &  1.85 &     1.4   \\
      C              & 2.27   & 52.0    &  117.8  &  37.9 &  85.8   &  18.9  &  42.7 &  4.89 &    0.17 \\
      S              & 2.0     &  70.9 &  141.7  & 56.2 & 112.4 &  9.75 & 19.5 & 5.77 & 0.045 \\
      P              &  2.2    &  64.0& 140.7 &  50.6  & 111.4  & 9.64 & 21.2  & 5.39 & 0.045 \\
            \hline
      steel          & 7.86 & 20.5 & 160.8  & 16.8     & 132.1  & 1.76   &  13.9 & 1.72 & 100  \\  
      \hline
      tile         & 1.06 & 107.2 &  113.7&  77.1   & 81.7   & 41.3  &  43.8 & 9.41 & 100\\
      \hline \hline
      Si           & 2.33 & 59.1 & 137.7 & 46.5  & 108.4 & 9.37  & 40.2 & 4.94 & 18.1 \\
      O           &           & 106.8 & 121.9 & 79.0 & 90.2    &30.01    & 34.2&  9.52 & 40.6 \\
      C           & 2.27  & 52.0   &  117.8 &  37.9 &  85.8  &  18.9  &  42.7 &  4.89 & 27.8 \\
      H           &           & 1134  &  80.3   & 734.6 & 52.0    & 890.4  & 63.0 & 67.92 & 6.8 \\
      Br          &   3.1   & 56.6 & 175.5 & 47.5 & 147.2 & 3.68   & 11.4 & 4.52 & 6.7 \\
      \hline
      FR4      & 1.7  & 71.4& 121.4 & 52.6        & 89.45      & 17.5  &  29.8 & 6.06 & 100  \\
      \hline \hline
      3M foil  & 1.06 & 107.2 &  113.7&  77.1        & 81.7       & 41.3 &  43.8 & 9.41 & 100 \\
      \hline \hline 
      PVC     &1.3 & 98.9& 128.5 & 74.6 & 97.0 & 19.6 & 25.5 & 8.34 & 87.2 \\
      polystyr.  & 1.06 & 107.2 &  113.7&  77.1   & 81.7   & 41.3  &  43.8 & 9.41 & 11.9 \\
      \hline 
      Cable & 1.35 & 93.7 & 126.5 & 70.2 & 94.8 & 19.9 & 26.9 & 7.95 & 100 \\
      \hline 
       air         & & 101k & 122 & 74.8k & 90.1 &  30.4k & 36.6 & 7.3k & 0.9 \\
       \hline
 \end{tabular} 
 \vskip0.0cm
  \end{center}
  \caption{Composition and properties of the absorber (cassette) plates and materials used in a cassette of one AHCAL layer  \cite{pdg}, where $\rho$, $R_M$ and $f$ respectively denote the density, Moli\`ere radius and fraction of components in composite materials (steel, PCB boards and cables) while other quantities are defined in the text.}
  \vskip 0.5 cm
    \label{tab:layer}
 \end{table}

\begin{table}
  \begin{center}
    
\begin{tabular}{|l|c|c|c|c|c|}
      \hline
      material  &  $ \# \lambda_\pi$ & $ \# \lambda_n$  & $\# X_0$ & t [cm] & $\rm t_{layer}~ [cm]$ \\        
      \hline
      Steel plate &   3.237 & 3.941& 37.555  &   66.19 & 1.74 \\  
      Cassette plates & 0.743  & 0.905 & 8.624 & 15.2  & $2 \times 0.2$ \\
      Scintillator tile       &  0.177 & 0.247 & 0.460 & 19.0 & 0.5 \\
      FR4    &  0.053 & 0.072 & 0.217 & 3.8  & 0.1\\
      3M foil  &0.008& 0.011  & 0.021 &  0.9 &0.023 \\
      Air gaps    & & & & 9.5 & $2 \times 0.125$  \\
      Cable mix &  0.061 & 0.081 & 0.286 & 5.7 & 0.15 \\
      \hline
      AHCAL & 4.28 & 5.26 & 47.16& 120.26  & 3.163\\
      \hline
 \end{tabular} 
 \vskip0.0cm
  \end{center}
  \caption{Number of pion interaction lengths, nuclear interaction lengths, radiation lengths, and thickness in the 38 layers of the AHCAL. The last column shows the average thickness of an individual layer. The exact thickness per layer varies from 3.093~cm to 3.183~cm due to variable sizes of the steel plates.}
    \label{tab:ahcaltot}
 \end{table}

\subsection{Active Layers}
\label{active}

Active layers one through 30 contain 216 scintillator tiles, while the last eight layers contain 141 scintillator tiles. Table~\ref{tab:totchannels} summarizes the details of the layout of the AHCAL layers. Housing all scintillator tiles of a layer in cassettes allows us to  independently test modules in electron beams, where typically four of them were stacked together without additional absorber plates. These tests are important for obtaining a first set of calibration constants. The modular design allows us to exchange individual modules easily in case of problems.

\begin{table}
\vskip 0.4 cm
  \begin{center}
    \begin{tabular}{|l|c|c|c|c|c|c|}
      \hline
      granularity & \#~layers &\#~tiles  &\#~tiles  & \#~tiles  & tiles/ & total  \\
             & &\ 3~cm$\times$3~cm & 6~cm$\times$6~cm & 12~cm$\times$12~cm & layer &  \#~tiles \\
      \hline
      fine        & 30   & 100 & 96 & 20     & 216             & 6480             \\
      coarse      &  8   & &121 & 20     & 141             & 1128             \\
      \hline
                  & 38        &  &  &  &              & 7608             \\
      \hline
            \end{tabular}
            \vskip 0.0cm  
  \end{center}
\caption{AHCAL total amount of readout channels.}
   \label{tab:totchannels}
\end{table}

\subsection{The Scintillator SiPM System }
\label{system}

\begin{figure}
\begin{center}
\includegraphics[width=45mm]{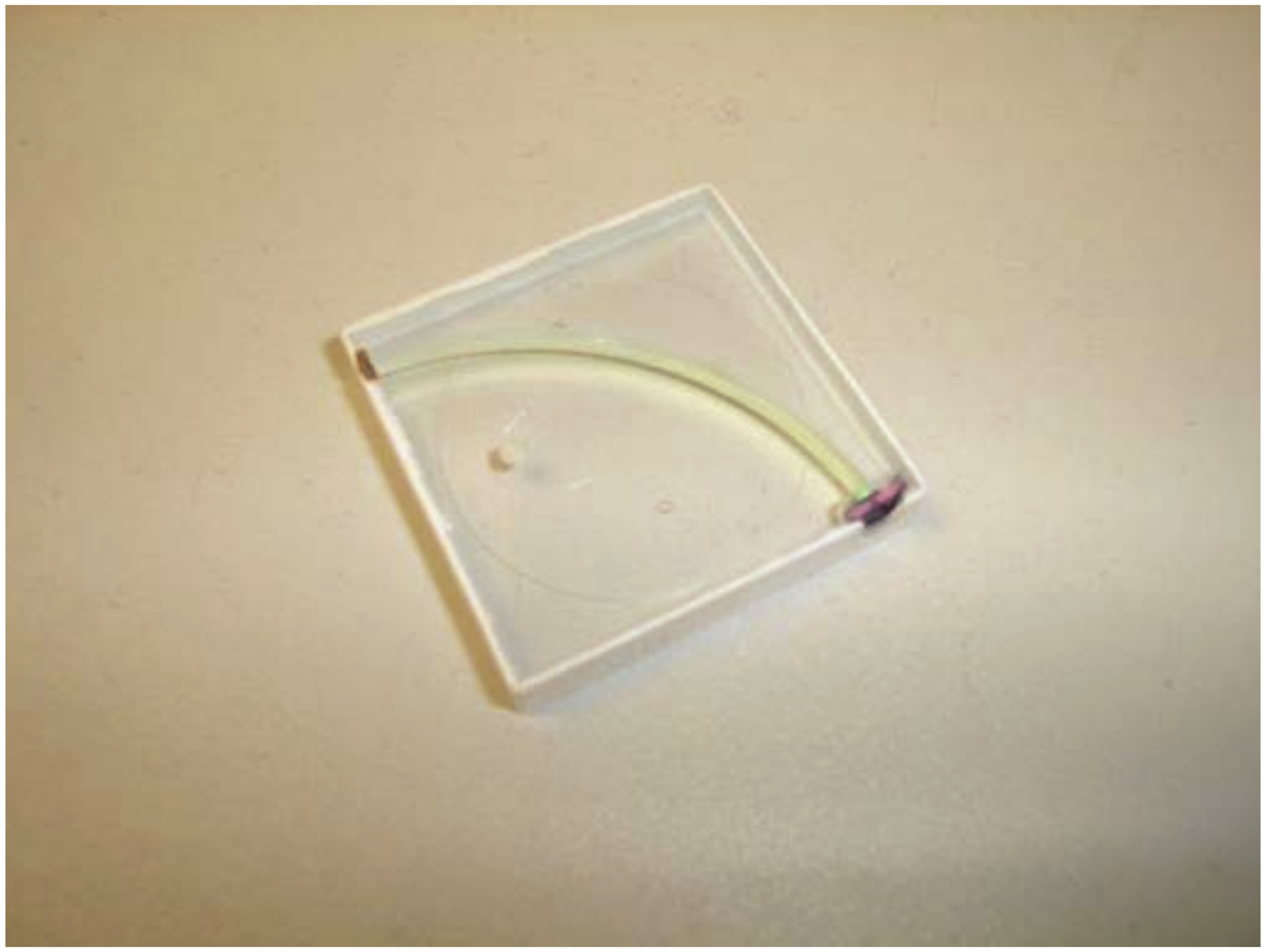}
\includegraphics[width=45mm]{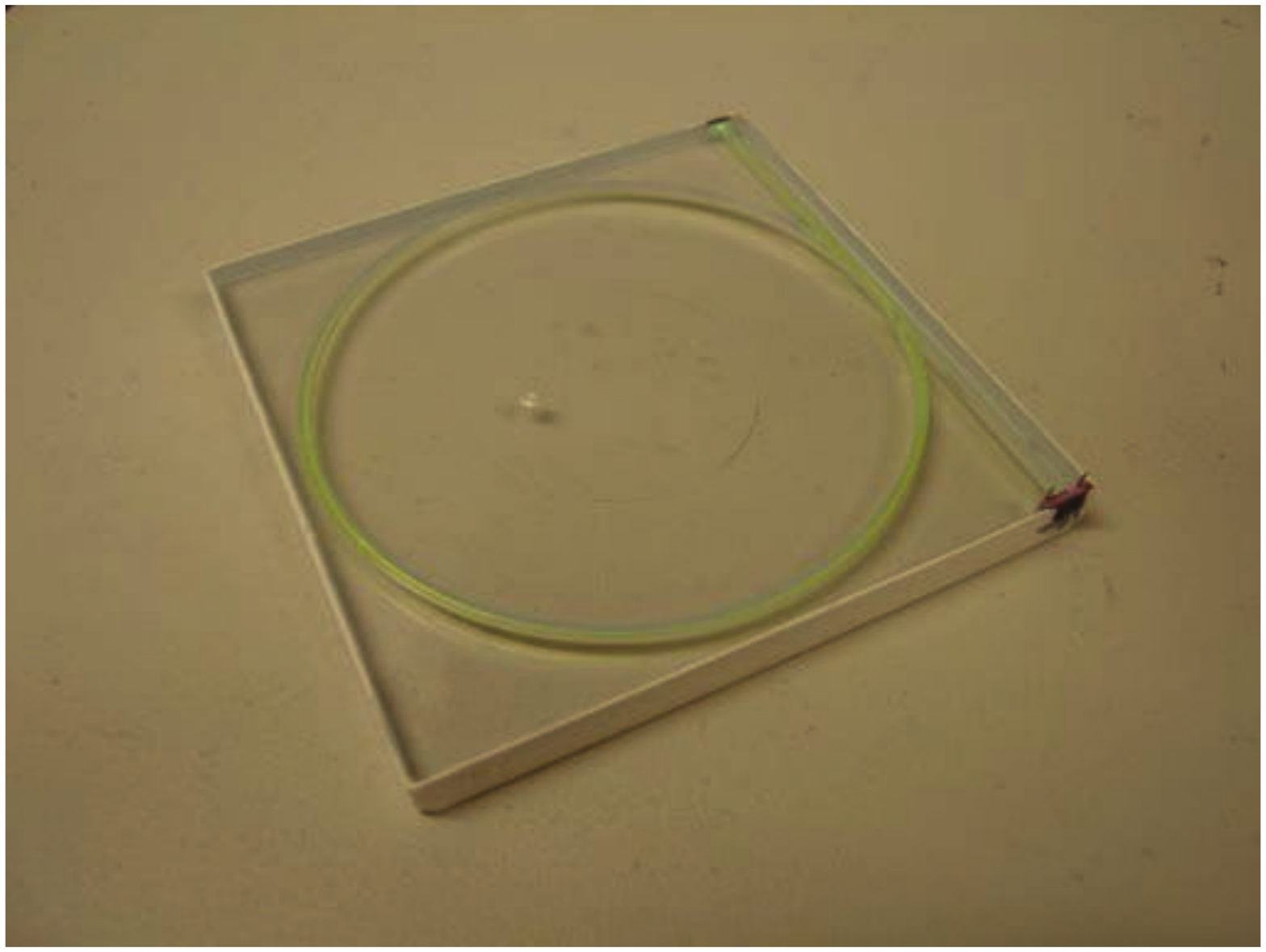}
\includegraphics[width=45mm]{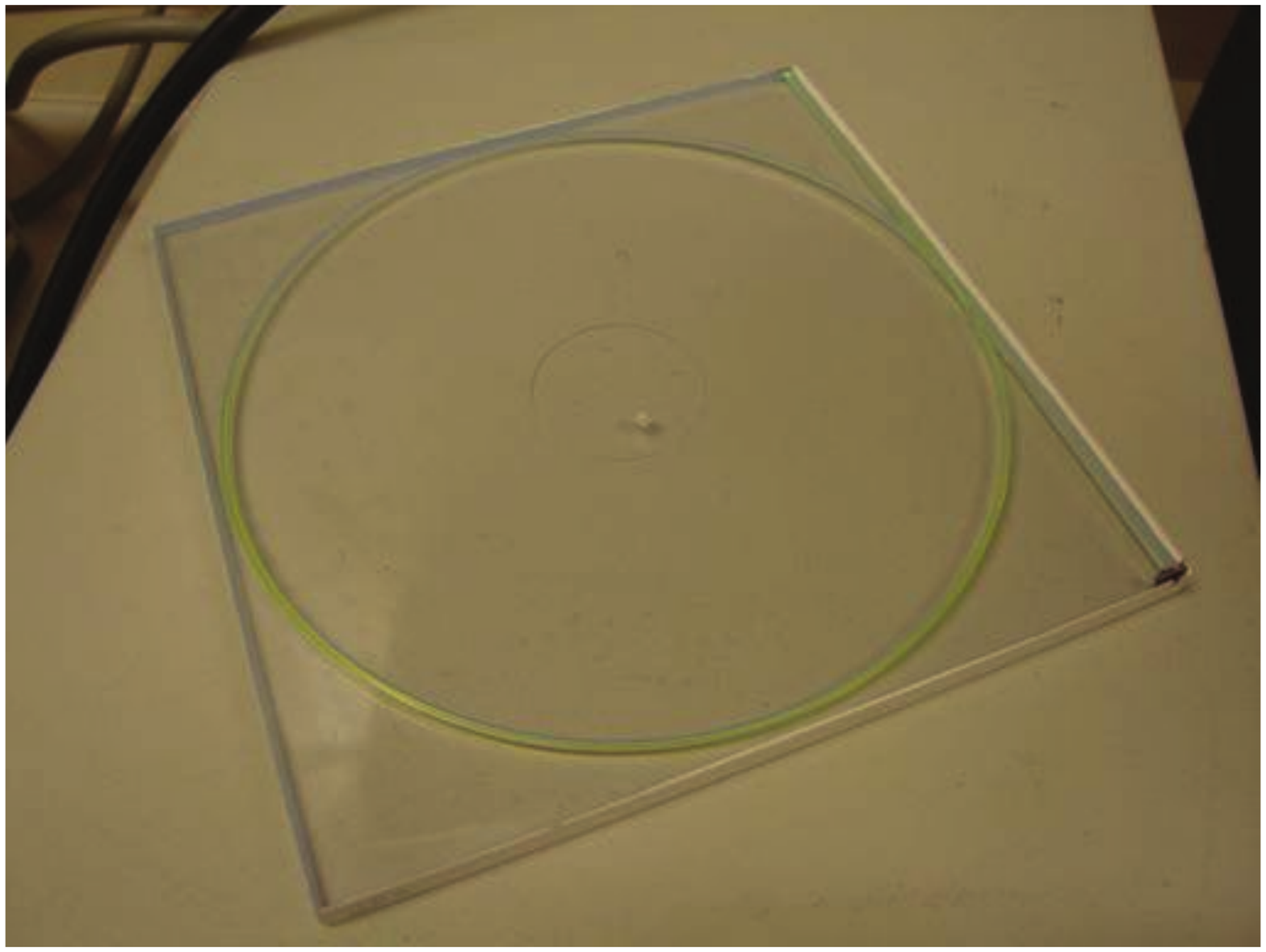}
\end{center}
\vskip -0.5cm
\caption{Readout of $\rm 3~cm \times 3~cm ~(left), \rm 6~cm \times 6~cm$ (middle), and $\rm 12~cm \times 12~cm$ tiles (right) with WLS fibers and SiPMs.}
\label{fig:tile}
\end{figure}

Figure~\ref{fig:tile} shows the fiber-SiPM readout of the three different size tiles. A rectangular-shaped groove with a cross section of 1/25'' is milled into each scintillator tile using a computer-controlled milling machine.  A Kuraray Y11 WLS fiber is inserted into the groove that collects the scintillation light. Using double cladding and coupling the fiber via an air gap to the tile
maintains the total-reflection properties of the fiber. One fiber end is pressed against a 3M reflector foil, while the other end is coupled via an air gap to the SiPM. In the $\rm 30~cm \times 30~cm$ core region the tile sizes are $\rm 3~cm \times 3~cm$ and the groove has a quarter-circle shape. A full circle is not achievable, since the bending radius becomes too small. The quarter-circle shape yields a higher light collection efficiency than a simple diagonal readout \cite{tile}. The layer core is surrounded by three rings containing $\rm 6~cm \times 6~cm$ tiles that have circular grooves, while the outer ring consists of $\rm 12~cm \times 12~cm$ tiles that also have circular grooves into which the Y11 fibers are inserted. The varying tile sizes represent a balance between shower sampling and cost. All scintillator tiles have a thickness of 5~mm, which is thick enough to ensure a sufficient signal-to-noise separation for MIPs.

\subsubsection{SiPM and Their Performance}
\label{sipm}

A SiPM is a multipixel silicon photodiode operated in the Geiger mode \cite{sipma, sipmb, sipmc}. The photosensitive area is $\rm 1.1~mm \times 1.1~ mm$ containing 1156 pixels, each $\rm 32~\mu m\times \rm 32~\mu m$ in size. SiPMs are operated with a reverse bias voltage of $ \sim 50$~V, which lies a few volts above the breakdown voltage, resulting in a gain of $ \sim 10^6$. Once a pixel is fired it produces a Geiger discharge. The analog information is obtained by summing the signals from all pixels. Thus, the dynamic range is limited by the total number of pixels. Each pixel has a quenching resistor of the order of a few $\rm M\Omega$ built in, which is necessary to break off the Geiger discharge.  Photons from a Geiger discharge in one pixel may fire neighboring pixels yielding inter-pixel cross talk. For stable operations we selected detectors with an inter-pixel cross talk of less than $35\%$ and with moderate dark current  caused by pile-up from thermal noise-induced signals. The pixel recovery time is of the order of 100~ns. However, recovery times as low as 20~ns are achievable by reducing the resistance. For short recovery times, the light pulse of a tile is sufficiently long that a pixel may fire a second time. This is a disadvantage as the SiPM saturation depends on the signal shape. Thus, we use relatively high quenching resistors. Furthermore, the sensors are unaffected by magnetic fields as tests in magnetic fields up to 4 T confirm \cite{groll}.

More than 10,000 SiPMs have been produced by the MEPhI/PULSAR group and have been tested at ITEP. The tests are performed in an automated setup, where 15 SiPMs are simultaneously illuminated with calibrated light from a bundle of Kuraray Y11 WLS fibers excited by a UV LED. During the first 48 hours, the SiPMs are operated at a bias voltage that is about 2~V above the normal operation voltage. This procedure allows us to reject detectors with unstable currents caused by long discharge. Next, the gain, noise and relative efficiency with respect to a reference photomultiplier are measured as a function of the reverse-bias voltage. The reverse-bias voltage working point is chosen such that a signal from minimum-ionizing  particle (MIP), provided by the calibrated LED light, yields 15 pixels in order to ascertain a large dynamic range and to have the MIP signal well separated from the pedestal.

\begin{figure}
\begin{center}
\includegraphics[width=66mm]{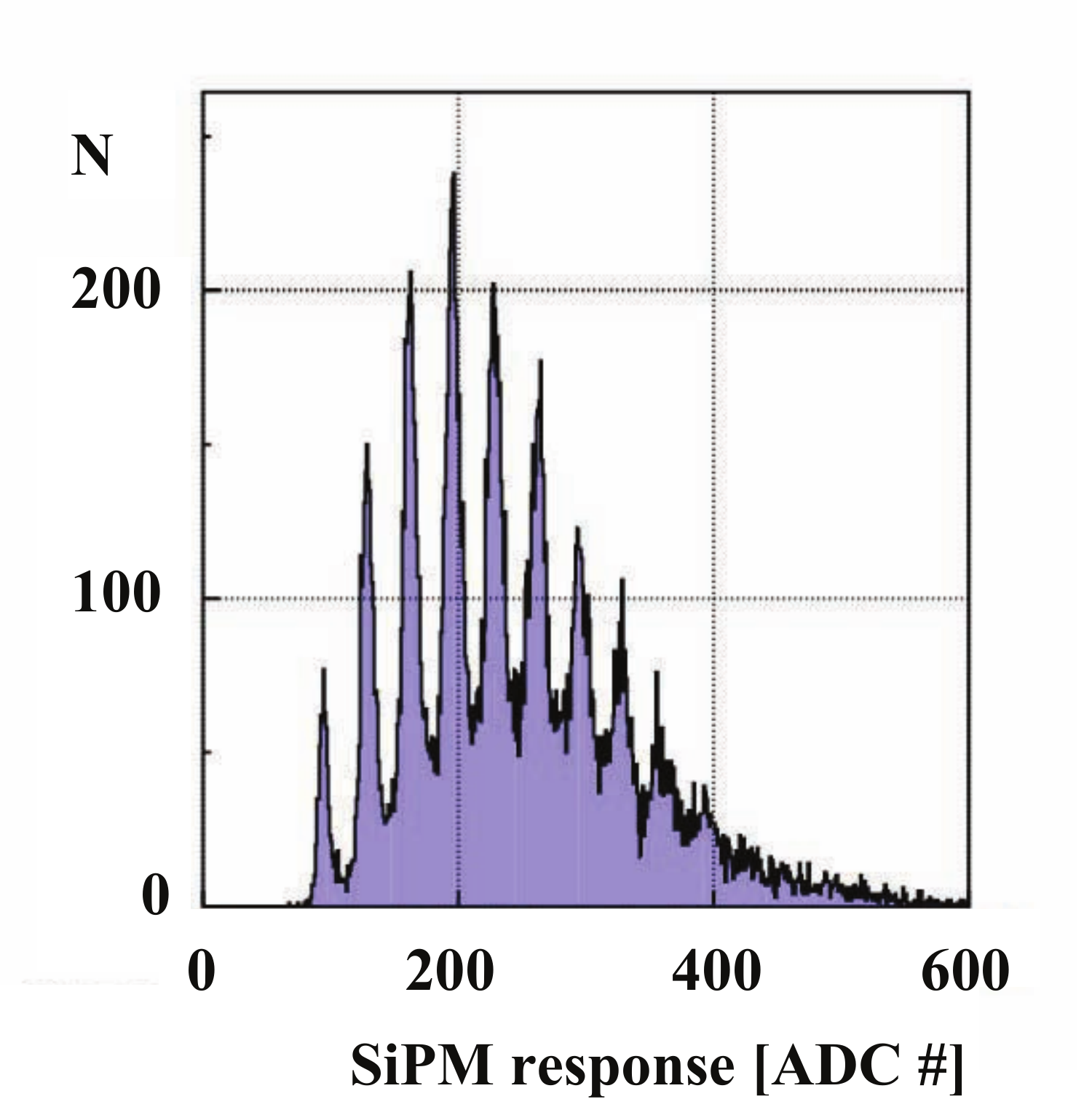}
\includegraphics[width=70mm]{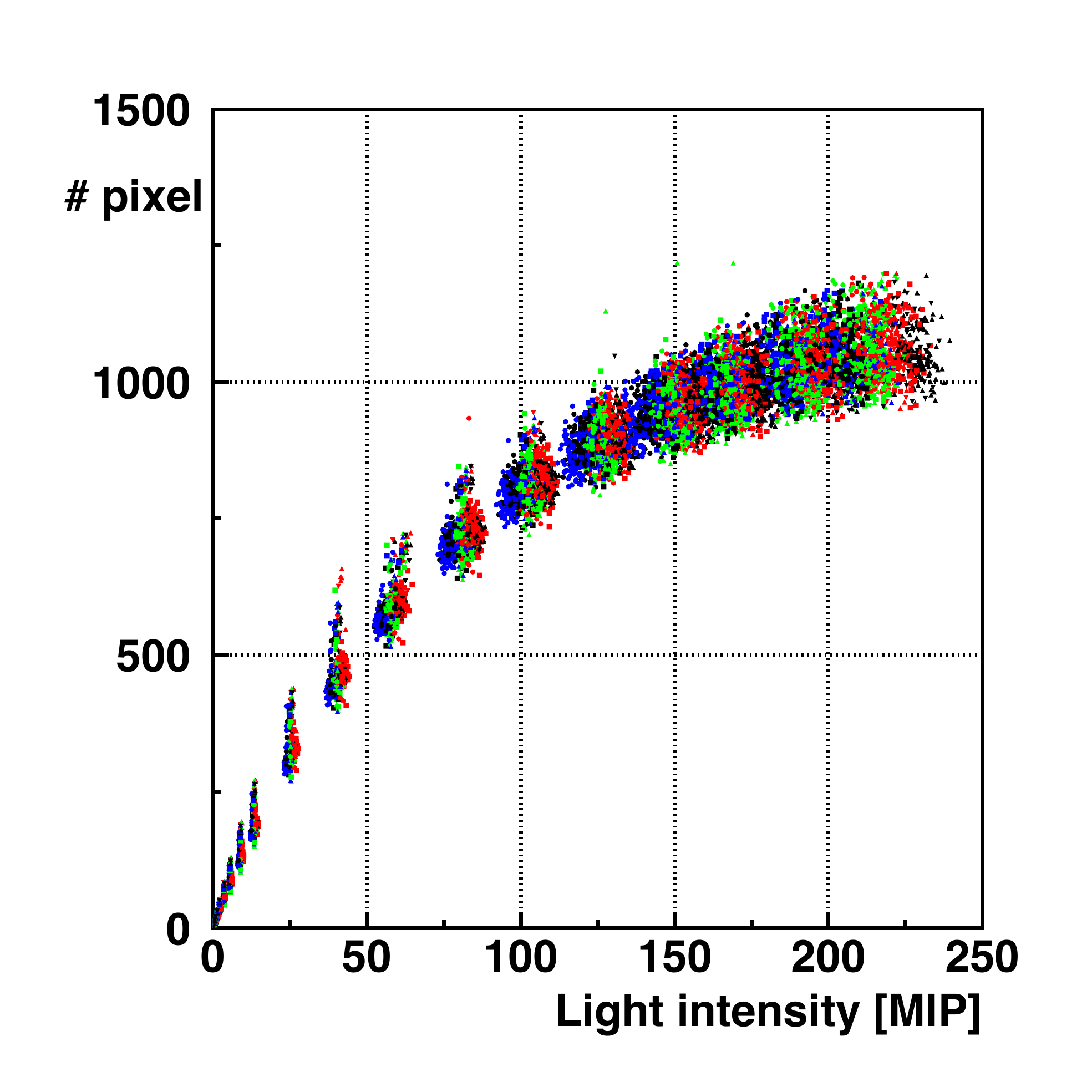}
\caption{A typical SiPM spectrum for low-intensity light (left) showing the pedestal in the first peak and up to eight fired pixels in the successive peaks. The response function for SiPMs of fired pixels versus input light (right). The curves are taken as a set of twenty measurements at increasing light intensities and can be fit with a sum of two exponential functions.}
\label{fig:sipm-pixel}
\vskip 0.0cm
\end{center}
\end{figure}

At the working point, we measure several SiPM characteristics. With low-light intensities of the LED, we record pulse height spectra that are used for the gain calibration.  A typical pulse height spectrum is
shown in Figure~\ref{fig:sipm-pixel} (left plot), in which up to nine individual peaks corresponding to different numbers of fired pixels are clearly visible.

\begin{figure}
\begin{center}
\includegraphics[width=45mm]{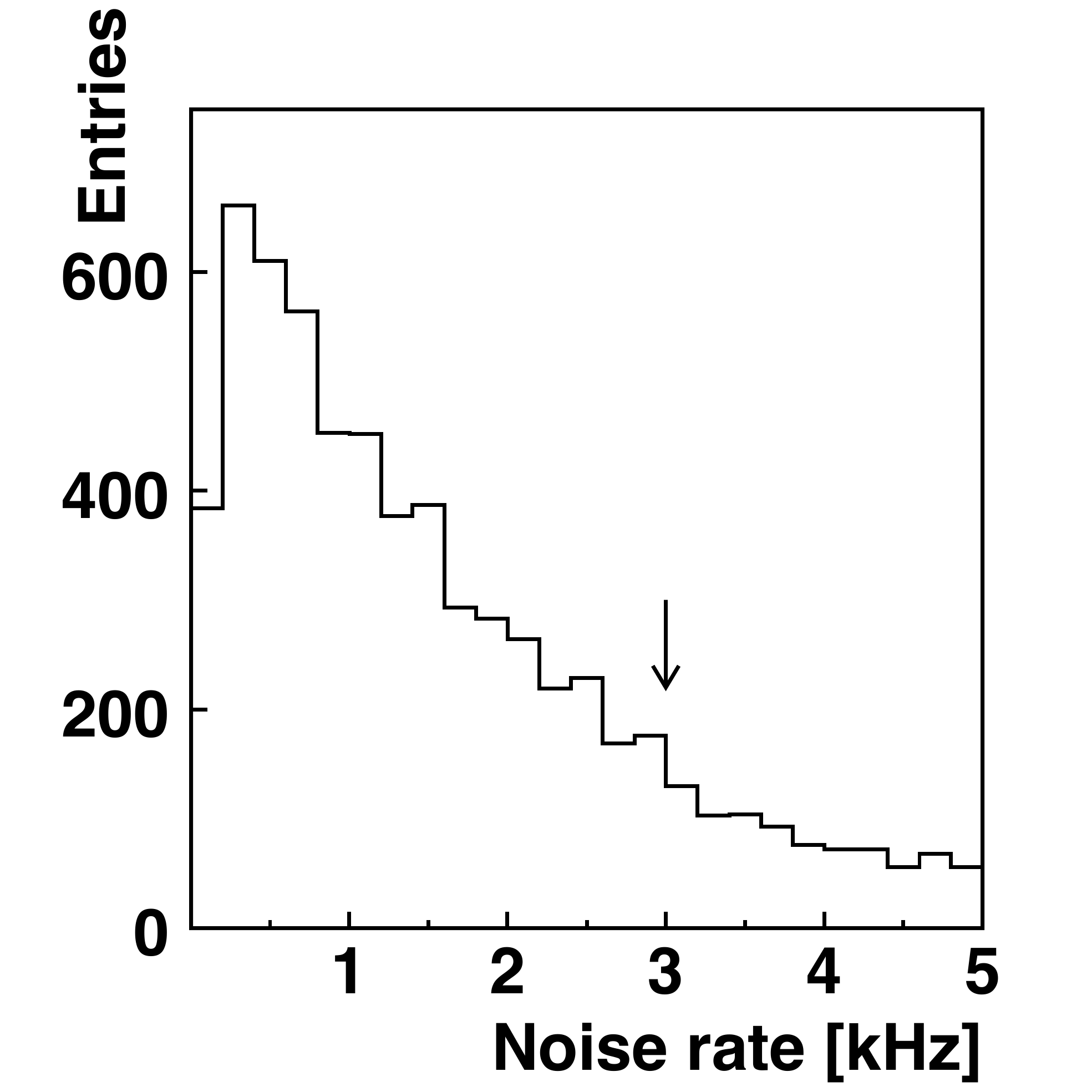}
\includegraphics[width=45mm]{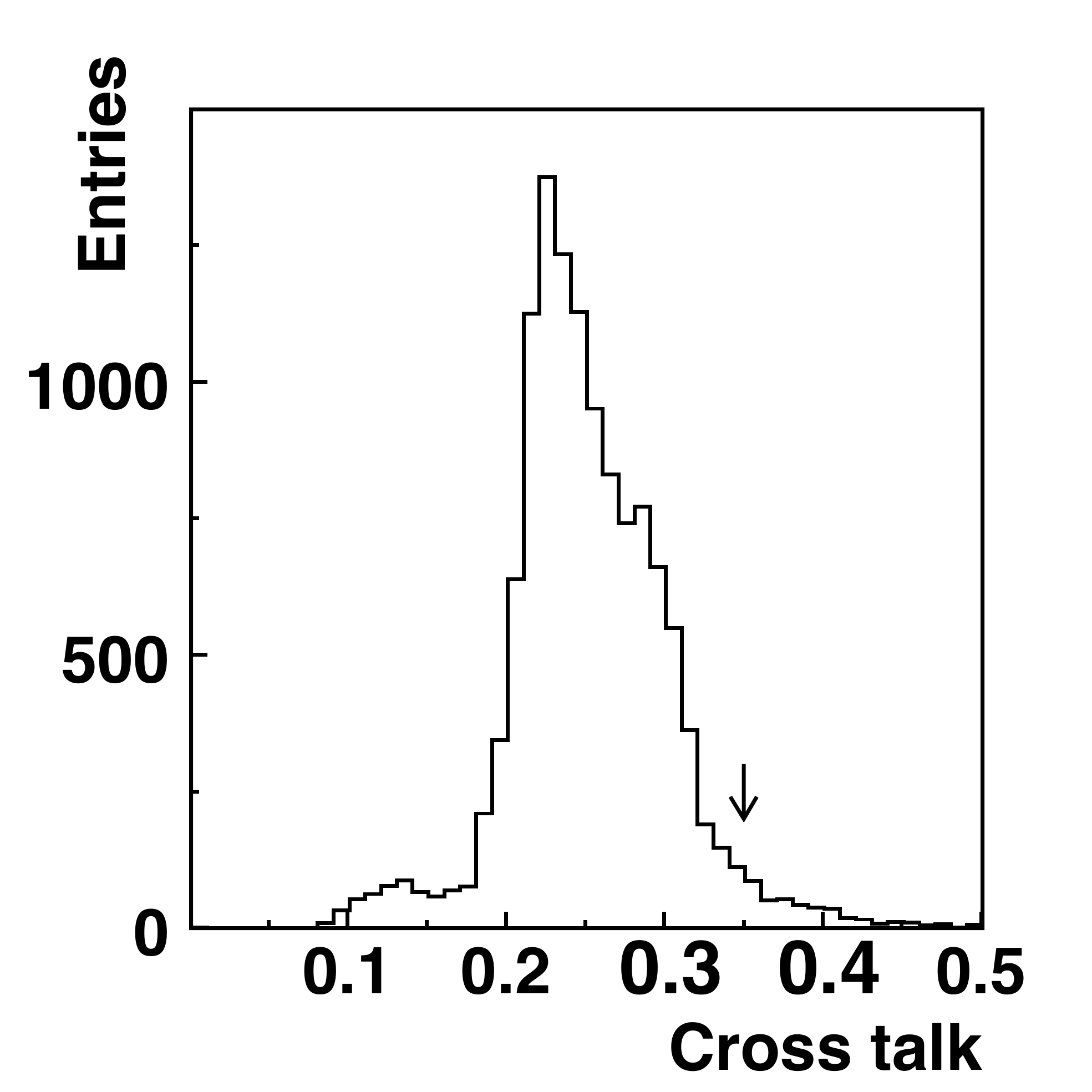}
\includegraphics[width=45mm]{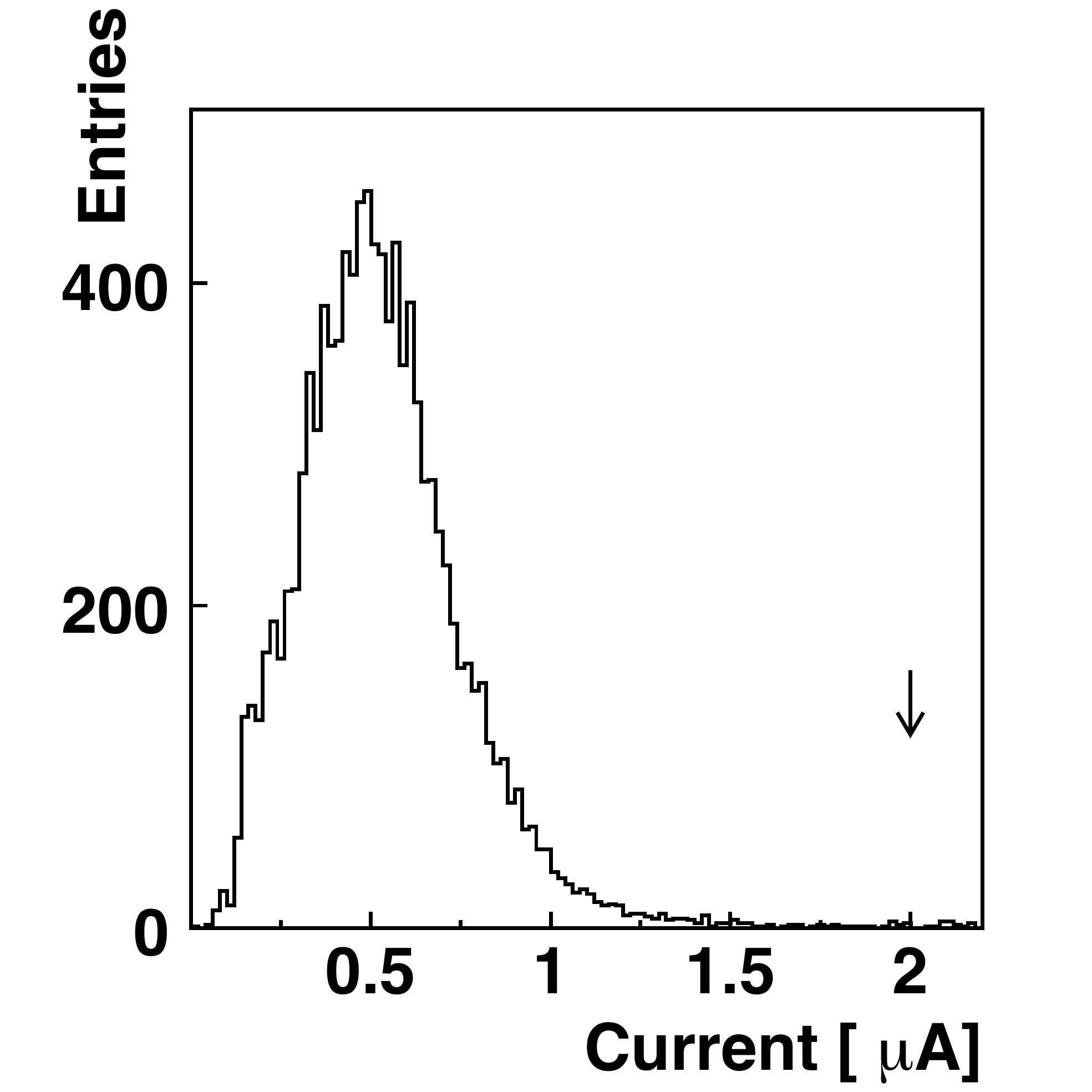}
\caption{The distribution of SiPM noise at half a MIP threshold (left), cross talk (middle), and current (right). Arrows show the maximum values on noise, cross talk and current permitted in the prototype.}
\label{fig:noise}
\end{center}
\end{figure}

This excellent resolution is extremely important for calorimetric applications, since it provides self-calibration and monitoring of each channel. We record the response function of each SiPM over the entire dynamic range (zero to saturation). Figure~\ref{fig:sipm-pixel} (right plot) shows the number of pixels fired versus the light intensity in units of MIPs for different SiPMs. 
The shape of the response function of all SiPMs is similar and individual curves are generally within $15\%$ of one another.  In addition, we measure the noise rate at a threshold of 0.5 MIPs, the inter-pixel cross talk, and the SiPM current as shown in Figure~\ref{fig:noise}. Arrows in the figures indicate the requirements for our detector selection. 

\begin{figure}
\begin{center}
\includegraphics[width=140mm]{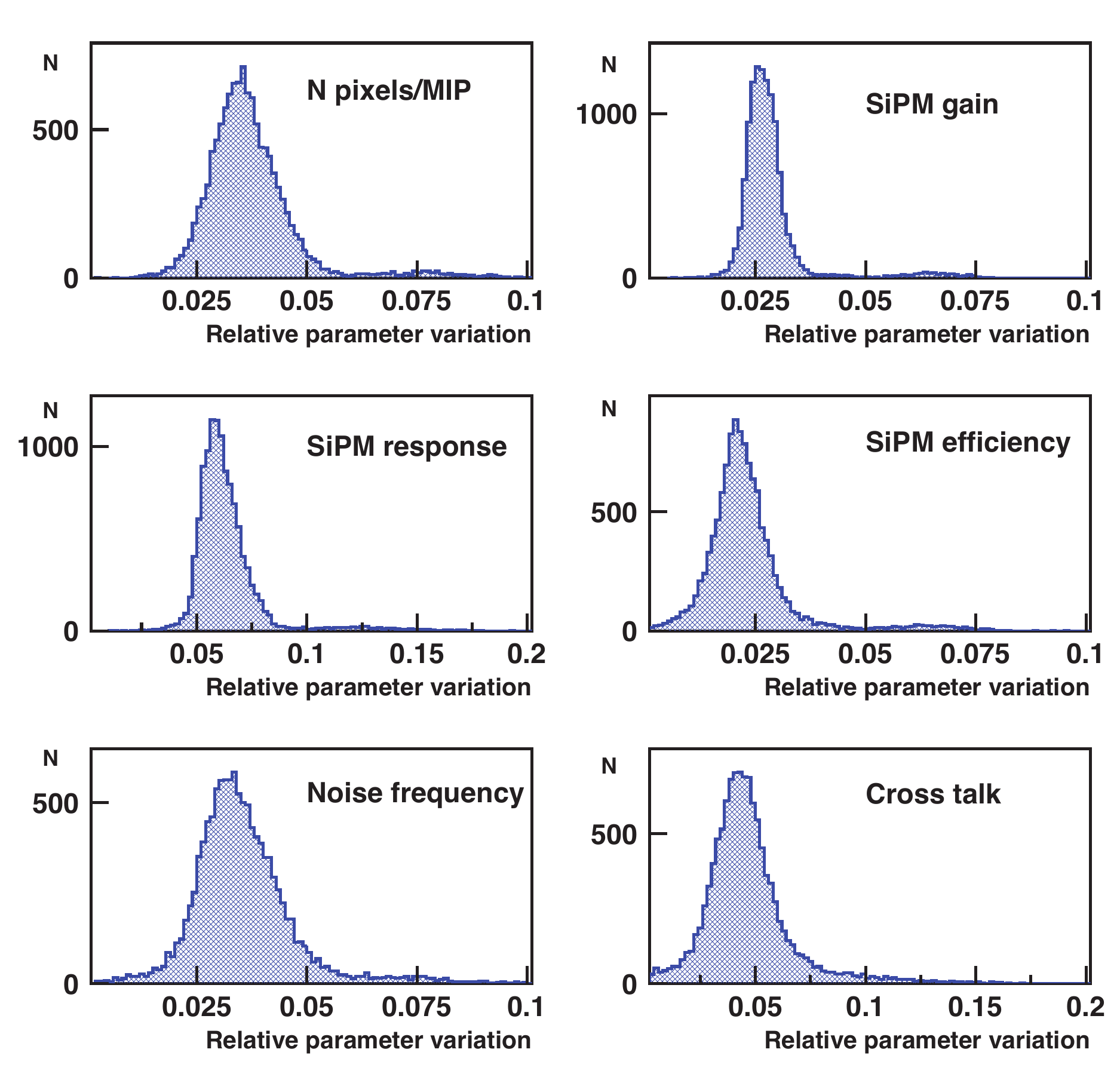}
\caption{Distributions of the relative variation of SiPM parameters (number of pixels per MIP, SiPM gain, SiPM response, SiPM efficiency, noise frequency and cross talk) for a 0.1~V change in bias voltage.}
\label{fig:sipm-var}
\vskip 0.0cm
\end{center}
\end{figure}

The relative variation of SiPM parameters for a $\rm 0.1~V$ change of the bias voltage is also measured and is shown in Figure~\ref{fig:sipm-var}. Thus, a voltage change of 0.1~V  modifies most SiPM parameters by 2--3$\%$. The largest effects of about $5\%$ are found for the cross talk and the SiPM response.

We use SiPMs with noise rates of less than 3~kHz at half a MIP threshold.\footnote{Here, a threshold of 0.5~MIPs corresponds approximately to 7.5 fired pixels.} Additional requirements on other parameters reduce the yield only slightly. The main noise sources are the SiPM dark current caused by pile-up from thermal noise-induced signals and cross talk. Both sources depend on temperature (T), fluctuations in the bias voltage $\rm (\Delta V)$ and the readout electronics. A decrease in temperature by $2^{\circ}$C leads to a decrease of the breakdown voltage by $\rm 0.1~V$, which is equivalent to an increase of the bias voltage by the same amount. In addition, a decrease in temperature leads to a decrease of SiPM dark rate.

\subsubsection{Manufacturing of the  Scintillator Tiles}
\label{scintillator}

The scintillator tiles have been produced by the UNIPLAST plant in Vladimir, Russia. The scintillator material is p-terphenyle plus POPOP dissolved in polystyrene (BASF130). Three molds have been fabricated to produce the three different tile sizes. The prototype uses 3000 small tiles, 3848 medium tiles and 760 large tiles. The four sides of each tile are matted by a chemical treatment providing a white surface that serves as a diffuse reflector. Thus, no reflector foils are needed between tiles allowing us to place one tile next to the other without gaps. 

\subsubsection{Tile Assembly and Quality Control}
\label{tile-assembly}

Before inserting the Y11 fibers into the groove they are heated to $\rm 90^\circ C$ to avoid the development of micro cracks on the surface when bent which would reduce light transport efficiency. The fibers are cut to the correct length with a diamond cutter. This procedure yields a flat surface and requires no additional polishing. A small groove is machined into the tile near the edge into which the SiPM is inserted, in order to place the SiPM at a well-defined position with respect to the fiber. The WLS fiber is coupled to the SiPM via an air gap that can vary between $\rm 50~\mu m$ and $\rm 100~\mu m$. This  ensures an optimal illumination of the SiPM pixels. 

\begin{figure}
\begin{center}
\includegraphics[width=80mm]{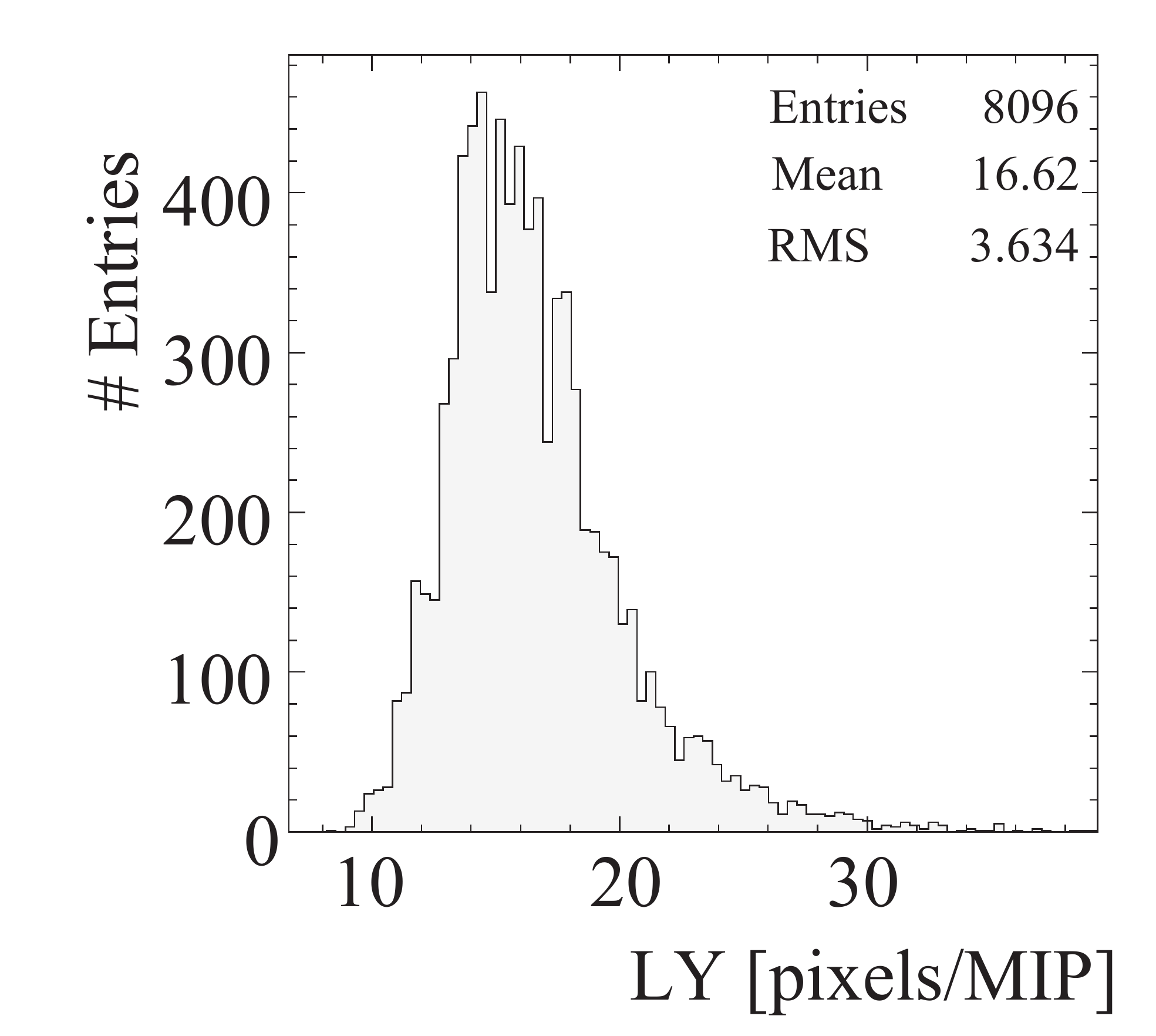}
\caption{Light yield of 8100 scintillator tiles measured at ITEP on a test bench. For the prototype, tiles with a light yield closest to 15 pixels/MIP were selected.}
\label{fig:ly}
\end{center}
\end{figure}

After inserting the Y11 fibers and mounting the SiPMs the light output of all tiles was measured at ITEP. The signal yield in the SiPM depends on the light output of the scintillator, the coupling of the scintillator-fiber and fiber-SiPM, as well as the efficiency of the SiPM. The latter depends both on $\rm \Delta V$ and on T. For each tile the MIP response is measured using triggered electrons from a $^{90}$Sr source. The trigger signal was obtained from a scintillator below the tile. The $ ^{90}$Sr spectrum collected with the trigger is very similar to the MIP spectrum measured at the ITEP test beam. The mean value of the MIP response varies from tile to tile because of the different light collection efficiencies. Figure~\ref{fig:ly} shows the measured distribution. The average light yield is 16.6 pixels with an RMS spread of 3.6 pixels. From this distribution 7608 tiles have been selected that have light yields closest to 15 pixels. Too low a yield reduces the separation of the MIP from the noise, while too high a light yield reduces the dynamic range for energy detection.

%


\subsection{Cassette Design}
\label{design}

The basic cassette structure consists of a 1~cm wide and 8~mm thick aluminum frame with front and rear covered by 2~mm thick steel sheets. Viewed in the reverse beam direction,
each cassette contains a $\rm 115~\mu m$ thick sheet of 3M reflector foil glued to the steel sheet; the scintillator tiles; another sheet of 3M reflector foil; a support plate made of FR4 holding cables and optical fibers, and a $\rm 50~\mu m$ thick mylar foil for electrical insulation (see Figure~\ref{fig:layer-xsect}). The individual contributions to the nuclear interaction length $\lambda_n$, the radiation length, and the Moli\`ere radius $R_M$ for an AHCAL layer are summarized in Table~\ref{tab:layer}.

\begin{figure}
\begin{center}
\includegraphics[width=140mm]{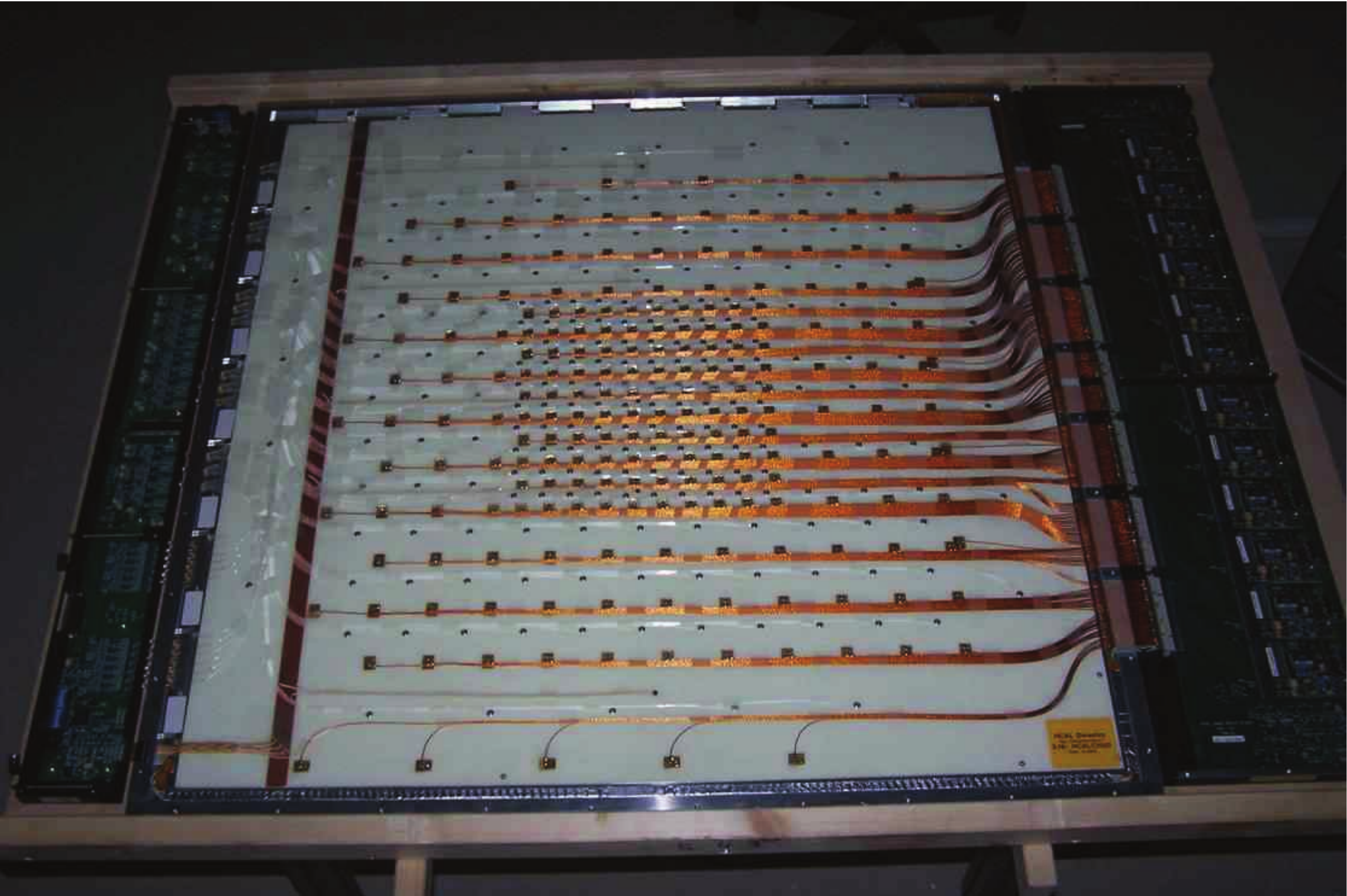}
\caption{Photograph of a fully assembled module.}
\label{fig:layerXsection}
\end{center}
\end{figure}

The SiPMs are connected to the front-end electronics via $50~\Omega$ micro-coax cables that carry both signal and bias voltage (on the shield). This design choice is based on the successful experience with the SiPM operation in a small test prototype \cite{minical}. It also provides more flexibility for different front-end electronics designs than a specially adapted printed circuit board. The latter design choice will be considered in future prototype versions. 
The FR4 plate is used to fix the positions of cables and calibration fibers, thus minimizing forces from bending or weight that may act on solder connections or SiPMs. 
Viewed from the beam direction (entering from the top in Figure~\ref{fig:layerXsection}), the front-end boards are mounted with seven~robust  two-row 64-pin connectors on the right-hand side of the cassette.  On the left-hand side, precision holes house the LEDs and PIN diodes of the  Calibration and Monitoring Board (CMB).

\subsection{Cassette Assembly}
\label{cassette-assembly}
\begin{figure}
\begin{center}
\includegraphics[width=65mm]{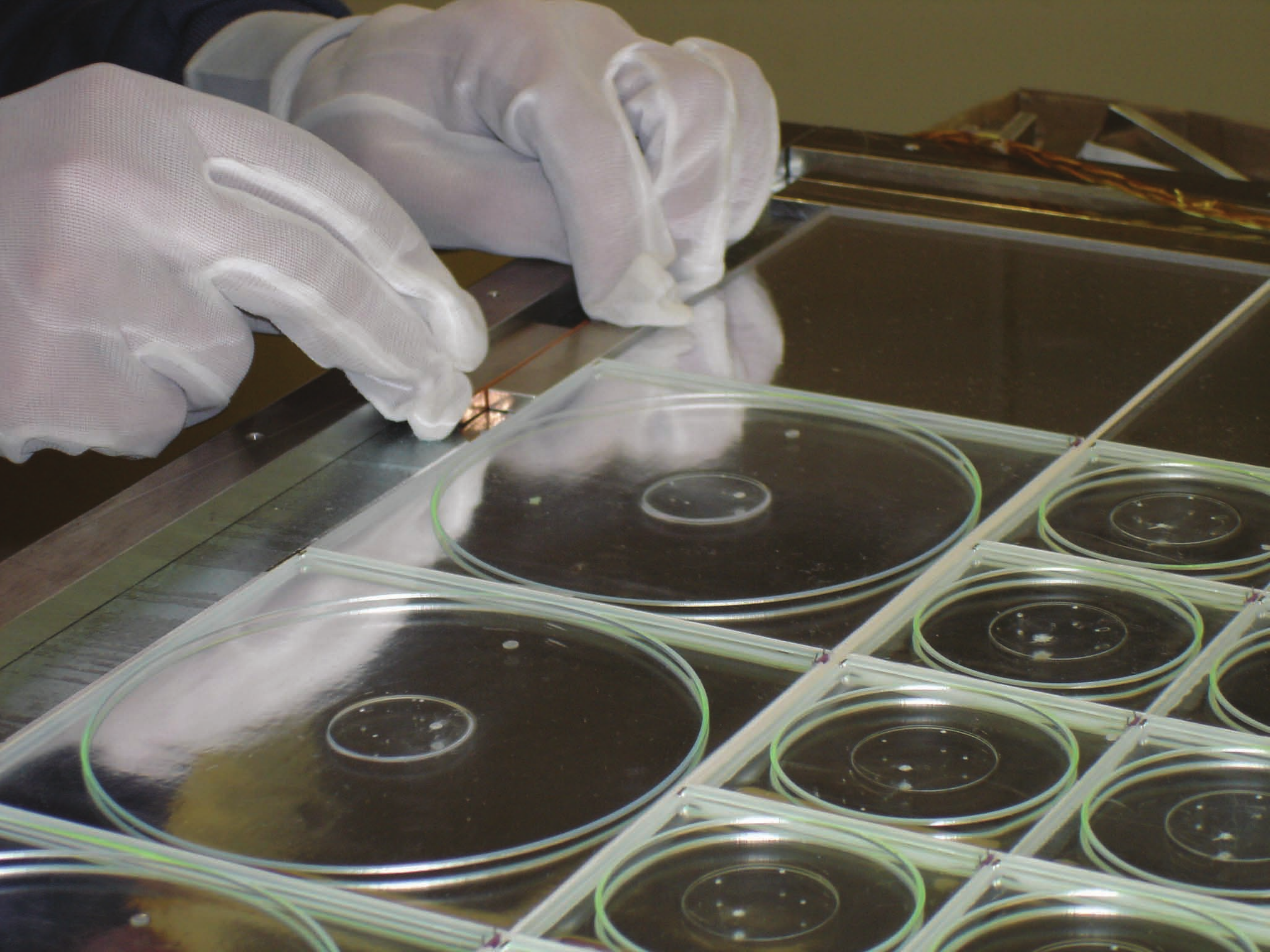}
\includegraphics[width=65mm]{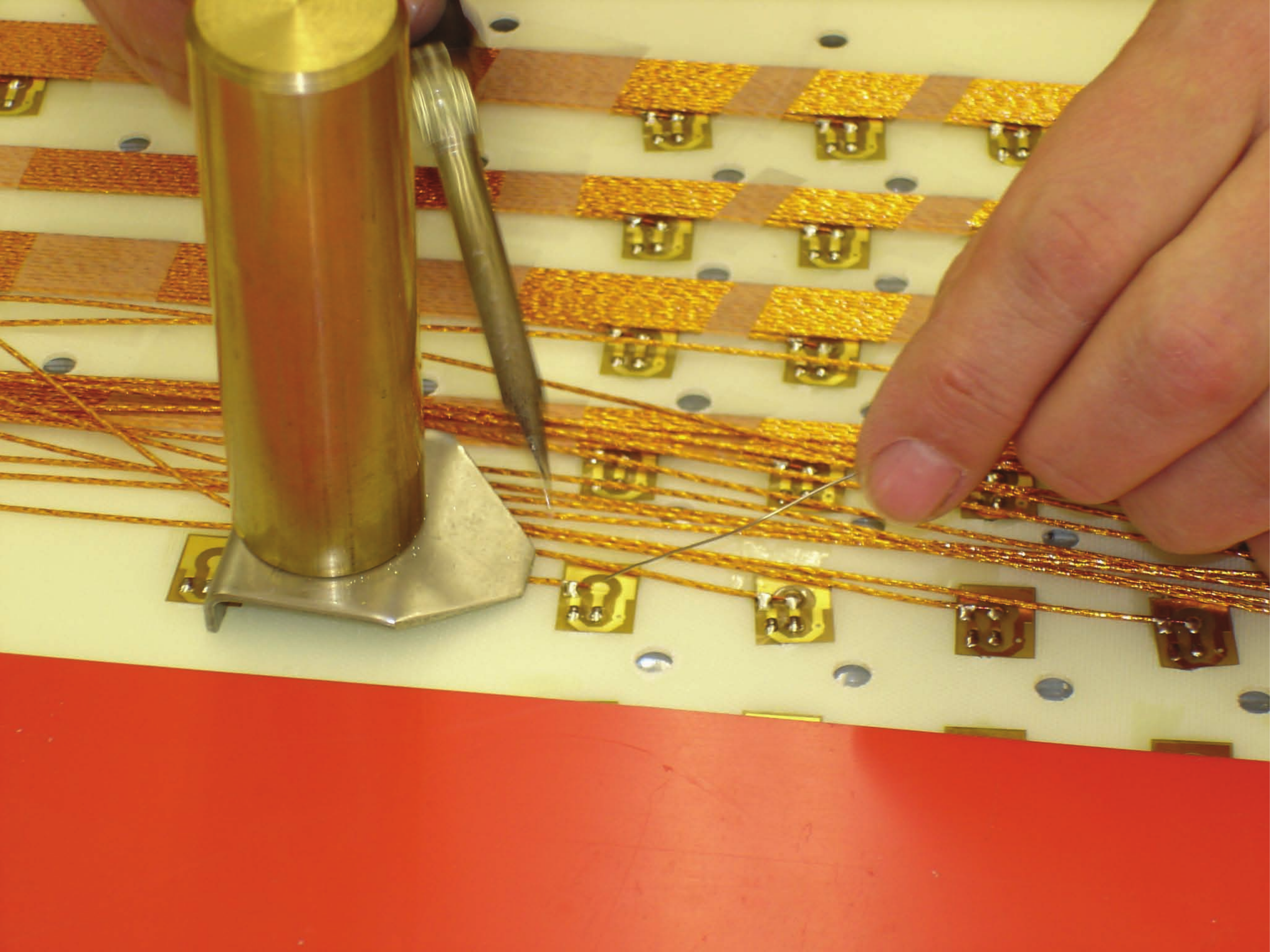}
\end{center}
\caption{Photograph of the wedges fixing the tiles inside a cassette (left) and photograph of the flexfoils used in a cassette (right).}
\label{fig:wedges}
\end{figure}

The cassette assembly was accomplished at DESY in cooperation with Russian groups. First, the rear steel plate and the internal FR4 plate were covered with 3M super-radiant reflector foil. The cassette frame was mounted on the assembly table with the rear plate on the bottom. Tiles were laid out and fixed in their final position by means of several pairs of wedges on one horizontal side and one vertical side of the cassette, see Figure~\ref{fig:wedges} (left). This method has the disadvantage that tile size inaccuracies, which have been found to mainly originate from the chemical matting process, may accumulate and need to be compensated by thin spacers in particular places. To overcome this problem in the future we will consider positioning the tiles individually with respect to the electronics structure.

The FR4 plate has two properly positioned circular holes per tile, one to connect the SiPM, the other to inject UV light from the optical fiber into the scintillator.
In order to protect the SiPM from mechanical stress perpendicular to the tile layer, the SiPM pins are soldered to a small flexible PCB  of 0.3~mm thickness, see Figure~\ref{fig:wedges}~(right). We can easily adjust for  residual tolerances in the position relative to the FR4 plate and extend the coaxial cable shield to the SiPM pins. 
After gluing the PCB foil to the FR4 plate the coaxial cable is soldered to it.

In the next step, we position the optical calibration system fiber bundles and connect the individual fibers to small conical metal mirrors for light injection into each tile. 
We have optimized the optical connection to the tiles, as well as the gluing of the  bundles and their connection to the LED in order to equalize the light intensity in the 18 channels connected to one bundle. 

In the final step we install the temperature sensors and calibration signal lines. We test each photosensor with dark pulses before closing the cassette to ensure that the entire channel from the SiPM to the ADC works. Occasionally, we found a damaged SiPM  and simply replaced it. The holes in the FR4 plate are sufficiently large to perform this operation from the top, without disassembling the entire tile.

%


\section{The Readout  System} 
\label{readout}

\begin{figure}[th]
\begin{center}
\includegraphics[width=130mm]{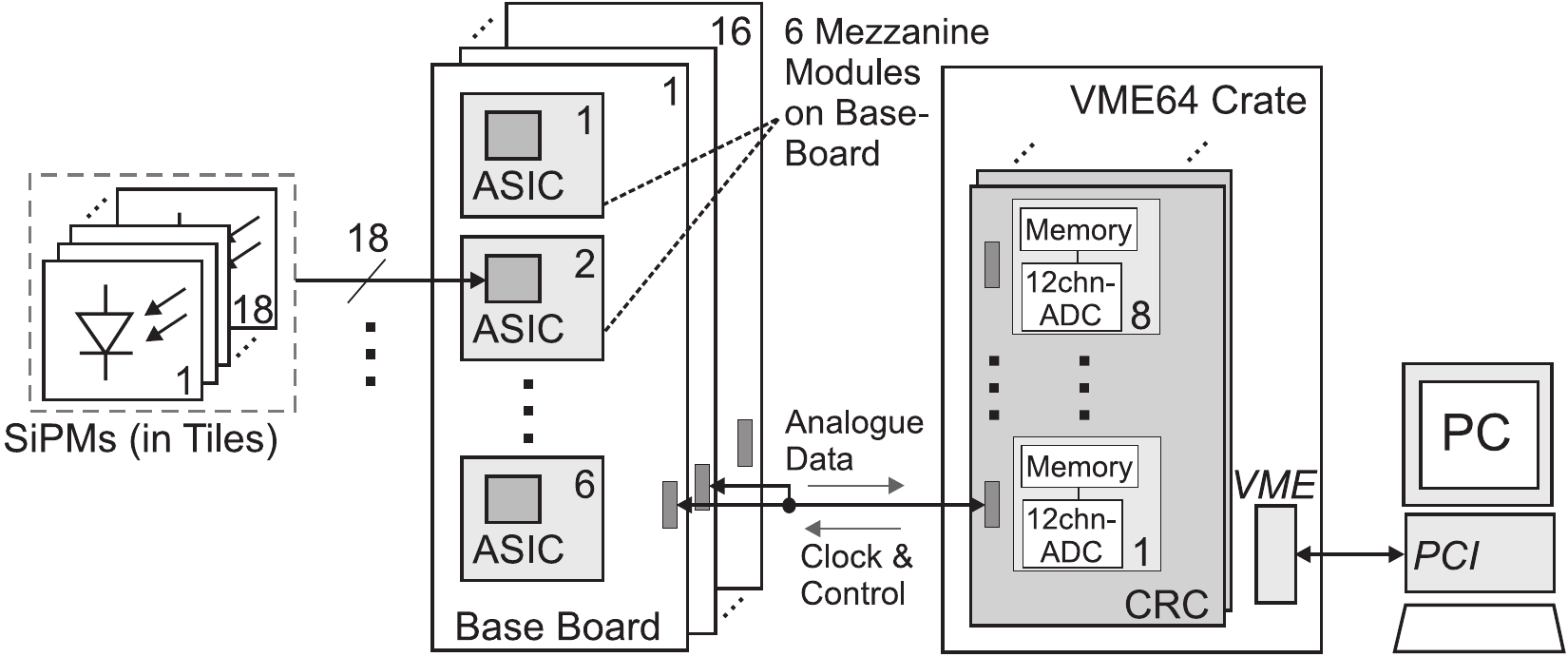}
\end{center}
\caption{Schematic view of the readout and data acquisition.}
\label{fig:daq}
\vskip 0.2cm
\end{figure}

The AHCAL readout concept is based upon the same architecture as that of the ECAL prototype \cite
{ecal}. A schematic view of the entire readout system is shown in Figure~\ref{fig:daq}.
Adopting the same design for the back-end data acquisition of VME CALICE readout cards (CRC~\cite{crc}) as for the ECAL provides a considerable simplification, since the number of readout channels of both prototypes is similar. Thus, in a combined test beam setup the same data acquisition system may be used. The front-end electronics are matched to the SiPM needs. A single ASIC houses
an 18-fold multiplexed chain of pre-amplifier, shaper, and sample-and-hold circuit (ILC-SiPM~\cite{flcphy3}). The signals from twelve ASICs are fed into one of the eight input ports of the CALICE readout card and are digitized by 16-bit ADCs. 

We operate the readout system in two different modes, either to detect individual pixels or to measure the full dynamic range. In calibration mode with LED signals, we use high gain and a fast shaping time of $\rm 40~ns$ (peak), while in the physics mode we use low gain and a shaping time of $\rm 180~ns$ providing sufficient latency for the particle beam trigger.

\begin{figure}[th]
\begin{center}
\begin{center}
\includegraphics[width=130mm]{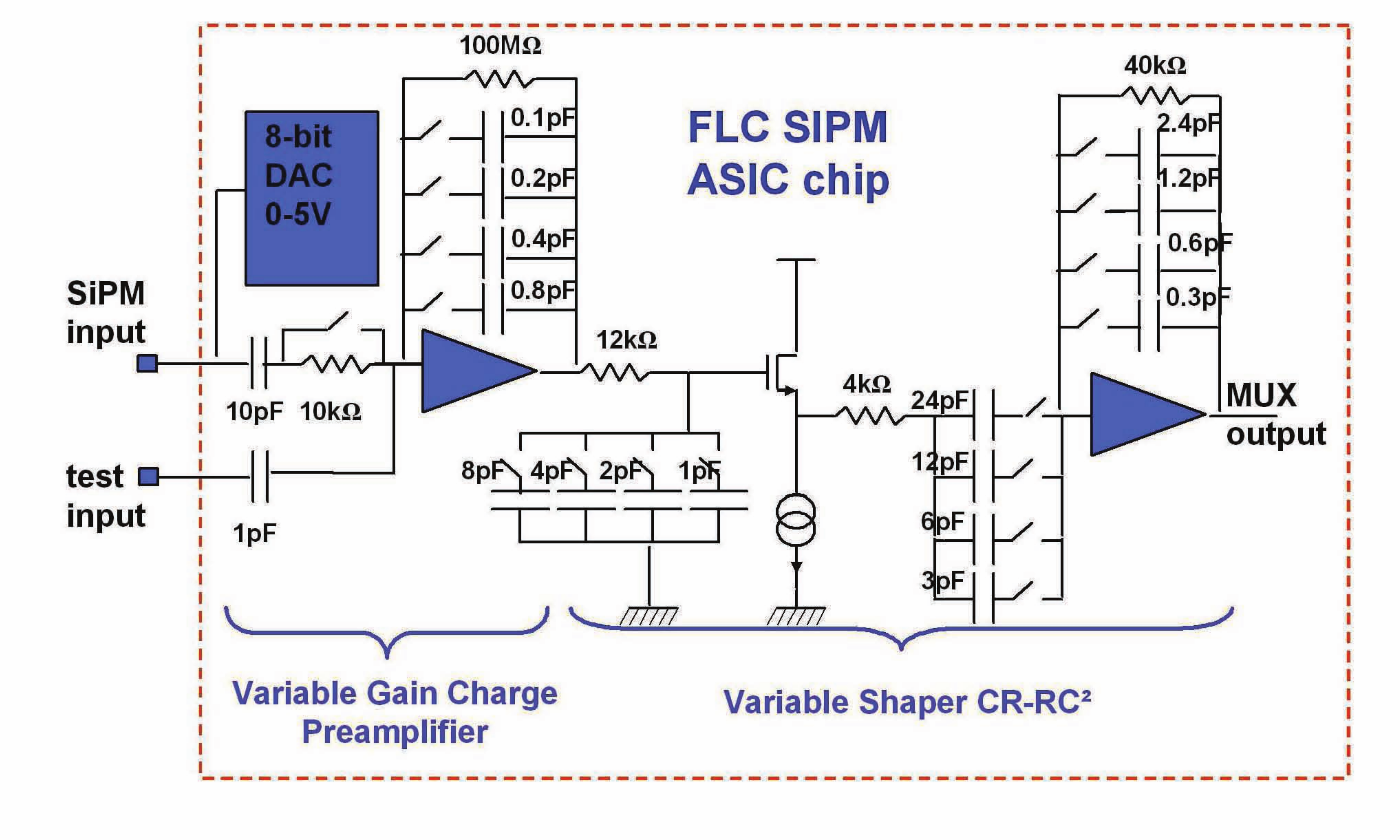}
\end{center}
\caption{Block diagram of the ASIC chip.}
\label{fig:asics}
\end{center}
\end{figure}

\subsection{The Very-Front-End ASICs}
\label{asics}

The 18 channel ASIC chip is based on the readout chip for the Si-W ECAL prototype and uses AMS $ \rm 0.8 \mu m $ CMOS technology \cite{cmos}. A schematic view of a single channel of the  ILC-SIPM ASIC chip is depicted in Figure~\ref{fig:asics}. 
The integrated components allow us to select one of sixteen fixed preamplifier gain factors from 1 to $\rm 100~mV/pC$, and one of sixteen $\rm CR-RC^2$ shapers with shaping times from 40 to 180~ns. A $\rm 10~k \Omega $ resistor may be added at the preamplifier input to further delay the signal peaking time at the expense of a $40 \%$  increase in noise. After shaping, the signal is held at its maximum amplitude with a track and hold method and is multiplexed by an 18 channel multiplexer to provide a single analog output to the ADC. The total power consumption of the chip is around 200~mW for a 5~V supply voltage. At the end of 2004 about 1000 ASICs were produced  and were packaged in a QFP100 case.

\subsubsection{Slow and Fast Shaping, Bias Adjustment and Input Coupling of the SiPM}

\begin{figure}[th]
\begin{center}
\includegraphics[width=70mm]{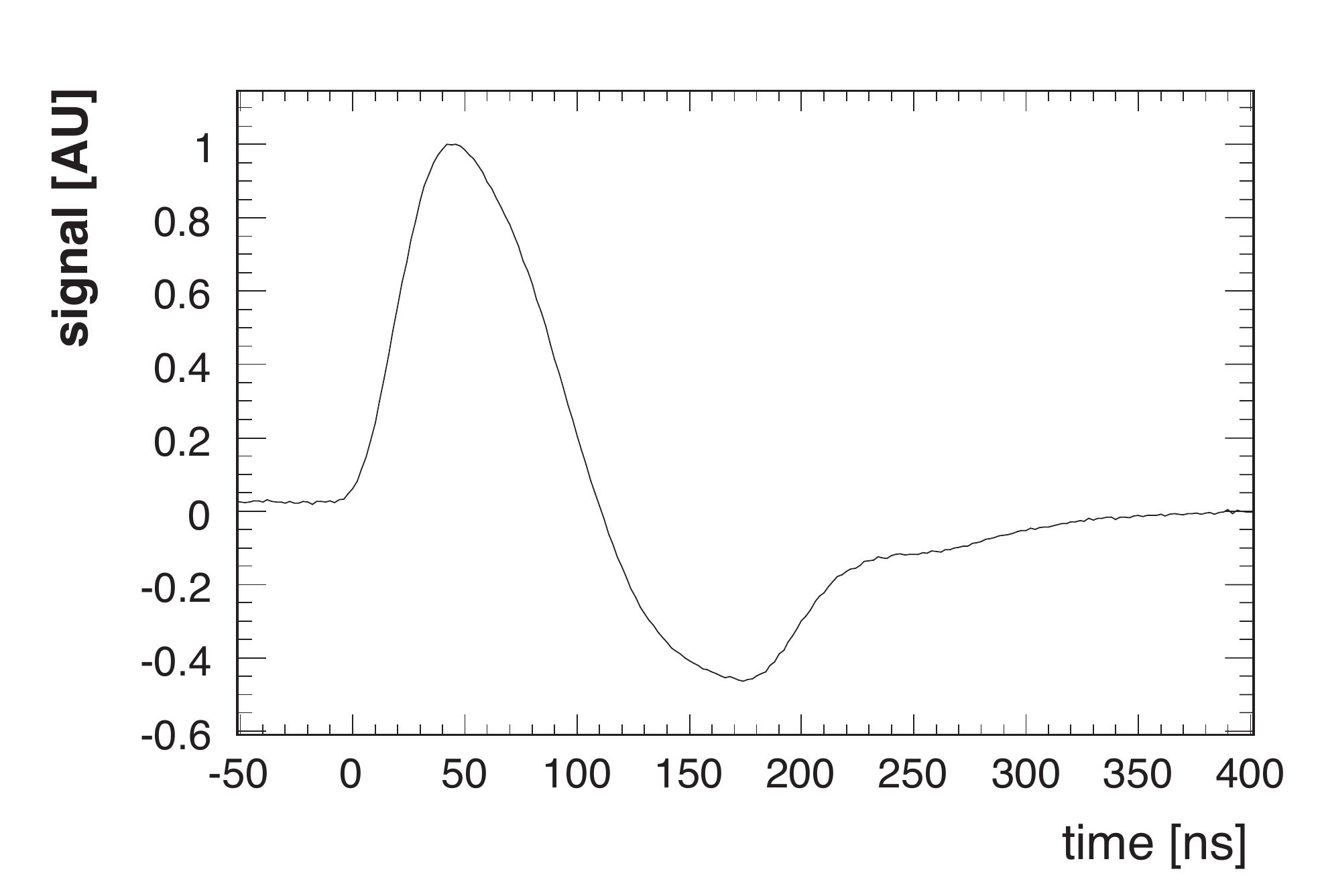}
\includegraphics[width=70mm]{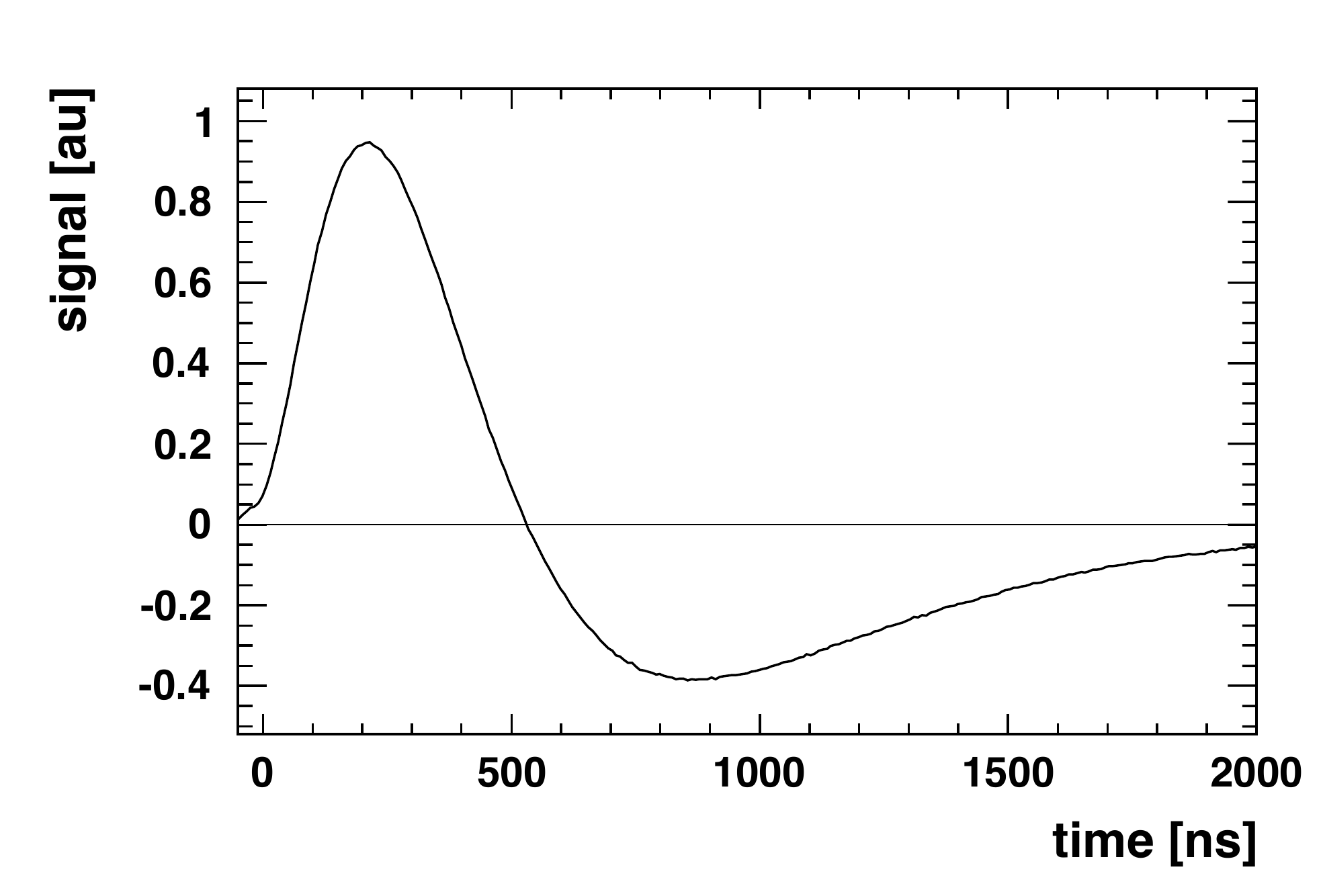}
\caption{Pulse shapes in the calibration mode (left) and the physics mode (right).}
\label{fig:pulse}
\end{center}
\end{figure}

The ASIC is operated in two different modes due to different timing requirements for physics events and calibration data. In physics events, the signals from each cell can vary between half a MIP and full saturation of the SiPM. In addition, the signals need to be delayed until the trigger decision has been reached in a test beam environment. We, therefore, need the longest shaping time of 180~ns in the physics mode. On the other hand, we obtain calibration data in dedicated runs with LED light to determine the SiPM gain. Here, we minimize the shaping time and use the highest amplification from the preamplifier in order to achieve the optimum signal-to-noise ratio. The pulse shapes in the calibration mode and the physics mode are shown in Figure~\ref{fig:pulse}. In the calibration mode, the measurements are performed without the input resistor.

\begin{figure}[th]
\begin{center}
\includegraphics[width=130mm]{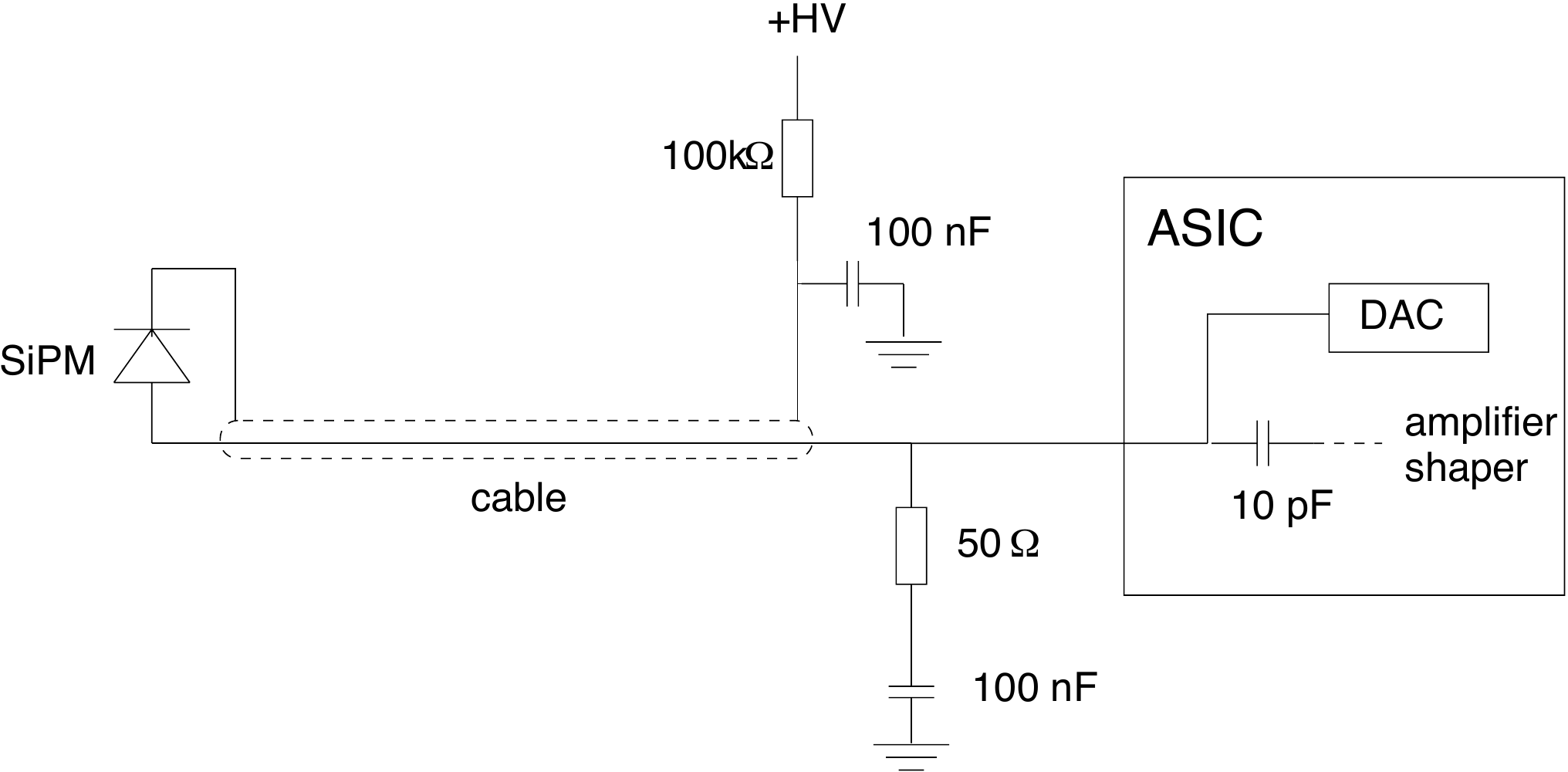}
\caption{Coupling diagram of the SiPM to the ASIC chip.}
\label{fig:coupling}
\end{center}
\end{figure}

\begin{figure}[th]
\centering
\includegraphics[width=100mm]{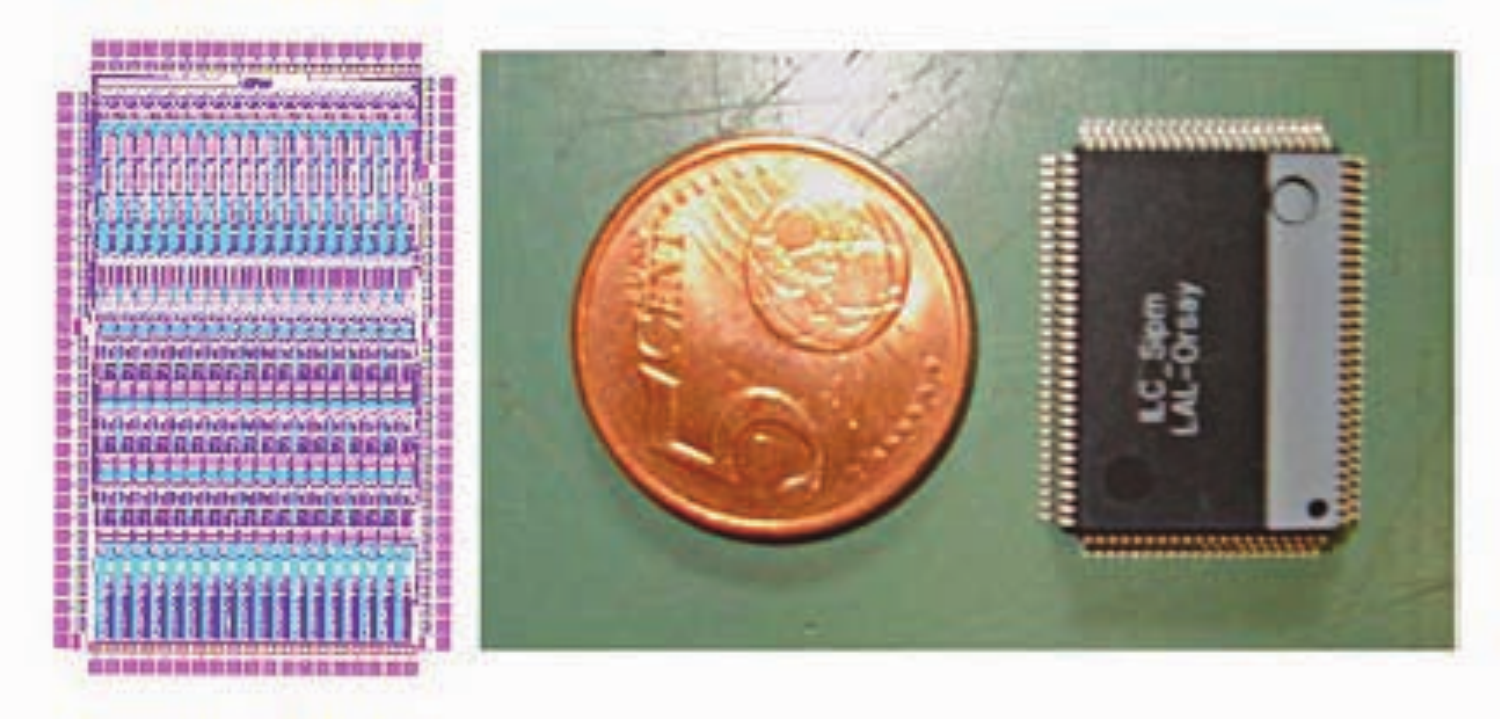}
\caption{Schematic diagram (left) and photograph (right) of the ASIC chip.}
\label{fig:chip}
\vskip 0.2cm
\end{figure}

The SiPM is directly connected to the chip as shown in Figure~\ref{fig:coupling}. In order to adjust the reverse bias voltage of the SiPMs individually, an eight-bit DAC is placed directly at the preamplifier input. 
Except for high voltage decoupling and cable matching of the $50~ \Omega$ components no external circuitry is needed.  The layout of the ASIC chip itself and a photograph are depicted in Figure~\ref{fig:chip}. 




\subsubsection{Linearity, Gain  and Noise Performance}

\begin{figure}[th]
\includegraphics[width=75mm]{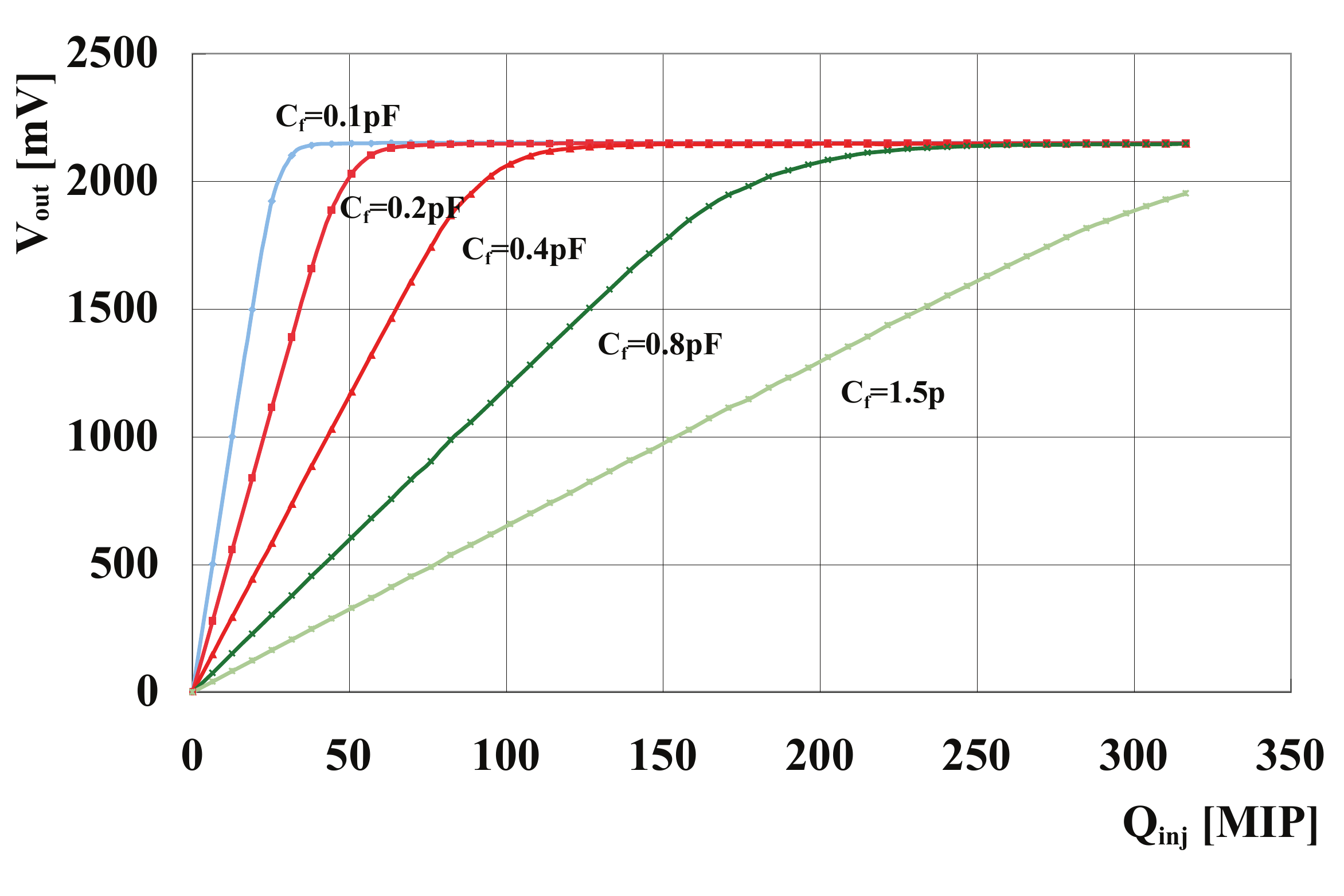}
\includegraphics[width=75mm]{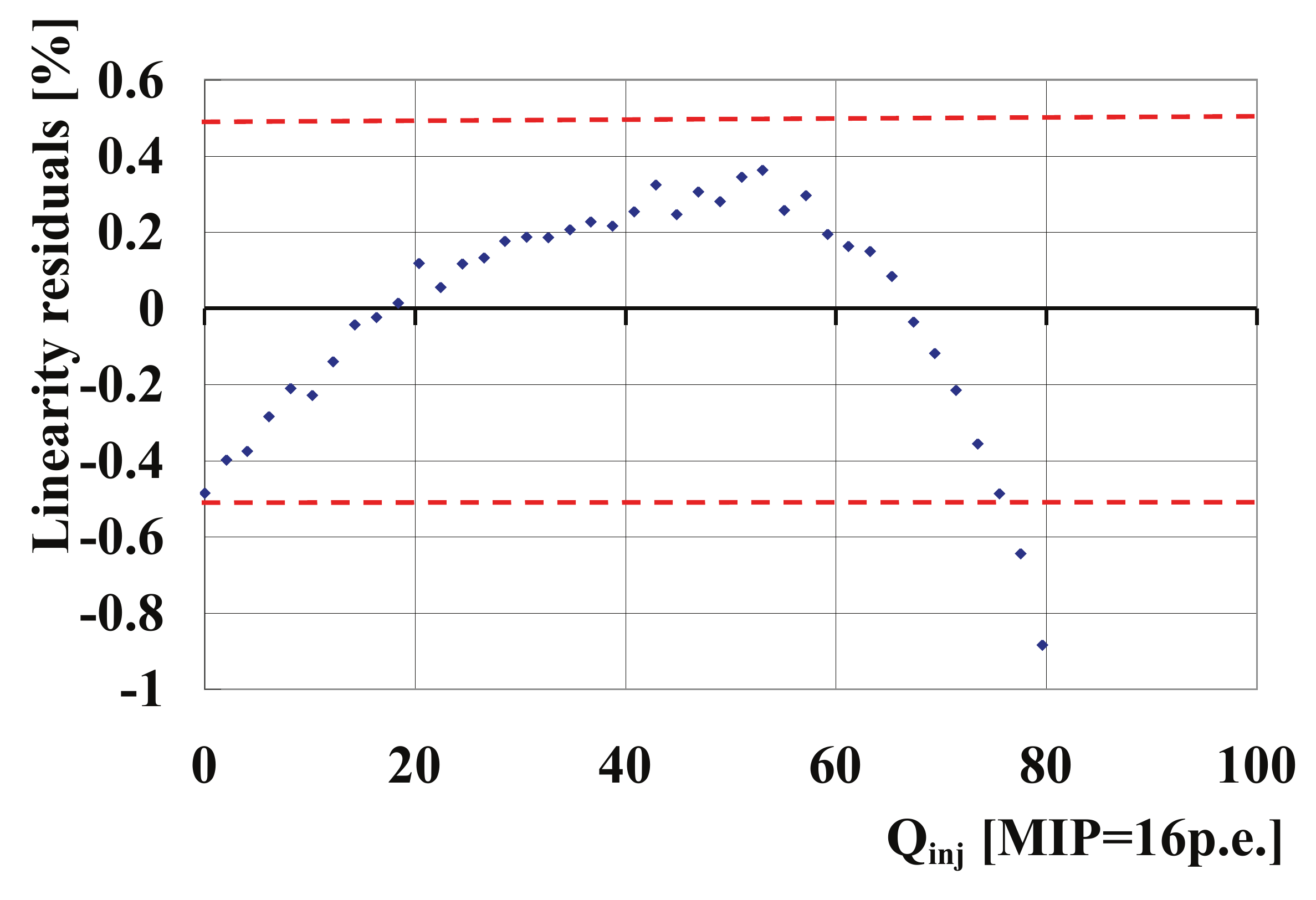}
\caption{Linearity measurements of preamplifier and shaper for input capacitance of $\rm C_f =0.1~pF, ~0.2~pF,~0.4~pF,~0.8~pF~and~1.5~pF$ (left) and residuals for operation in the physics mode (right). The dashed (red) lines show a $\pm 0.5\%$ deviation from linearity.}
\label{fig:linamp}
\vskip 0.1cm
\end{figure}

We measured linearity, gain, and noise for both operation modes. Figure~\ref{fig:linamp} (left) shows charge injection measurements for different preamplifier gain settings and shaping times. Figure~\ref{fig:linamp} (right) shows the deviations from linearity for operation in the physics mode. Based on these measurements we chose gain settings in the physics mode such that a linearity better than 3\% was achieved for input charges up to 190~pC. This ensured that SiPMs with pixel gains up to $10^6$ could be recorded up to full SiPM saturation~\footnote{A gain of $19^6$ corresponds to a charge of 160~fC}. The calibration mode covered a range up to 10~pC with a linearity of $1\%$ or less~\cite{beni_thesis}.

We measured a gain of $92~\frac{\mathrm{mV}}{\mathrm{pC}}$ in calibration mode and $8.2~\frac{\mathrm{mV}}{\mathrm{pC}}$ in physics mode. We determined the chip noise in the calibration mode to be $1.72~\mathrm{mV}$  ($1.52~\mathrm{mV}$),  when the input signal is (not) connected yielding a signal-to-noise ratio of 8.6 per fired single pixel. For a SiPM with $10^6$ pixel gain and  an average of 15 pixels per MIP we expect a signal-to-noise ratio of 12.6 for a MIP signal in physics mode ~\cite{beni_thesis}.
 
During commissioning of the CERN~2006 test beam, we observed typical signal-to-noise ratios of four for single pixels and eleven for minimum ionizing muons~\cite{marius_thesis}. The large difference in the single-pixel signal-to-noise ratio between the measurement and the expectation mainly results from the very short shaping time. The  signal-to-noise ratio measured in the test beam, however, is smaller than the expectation, because most SiPMs have a gain lower than $10^6$ and due to the short shaping time not all the signal charge is collected.



\begin{figure}[th]
\includegraphics[width=87mm]{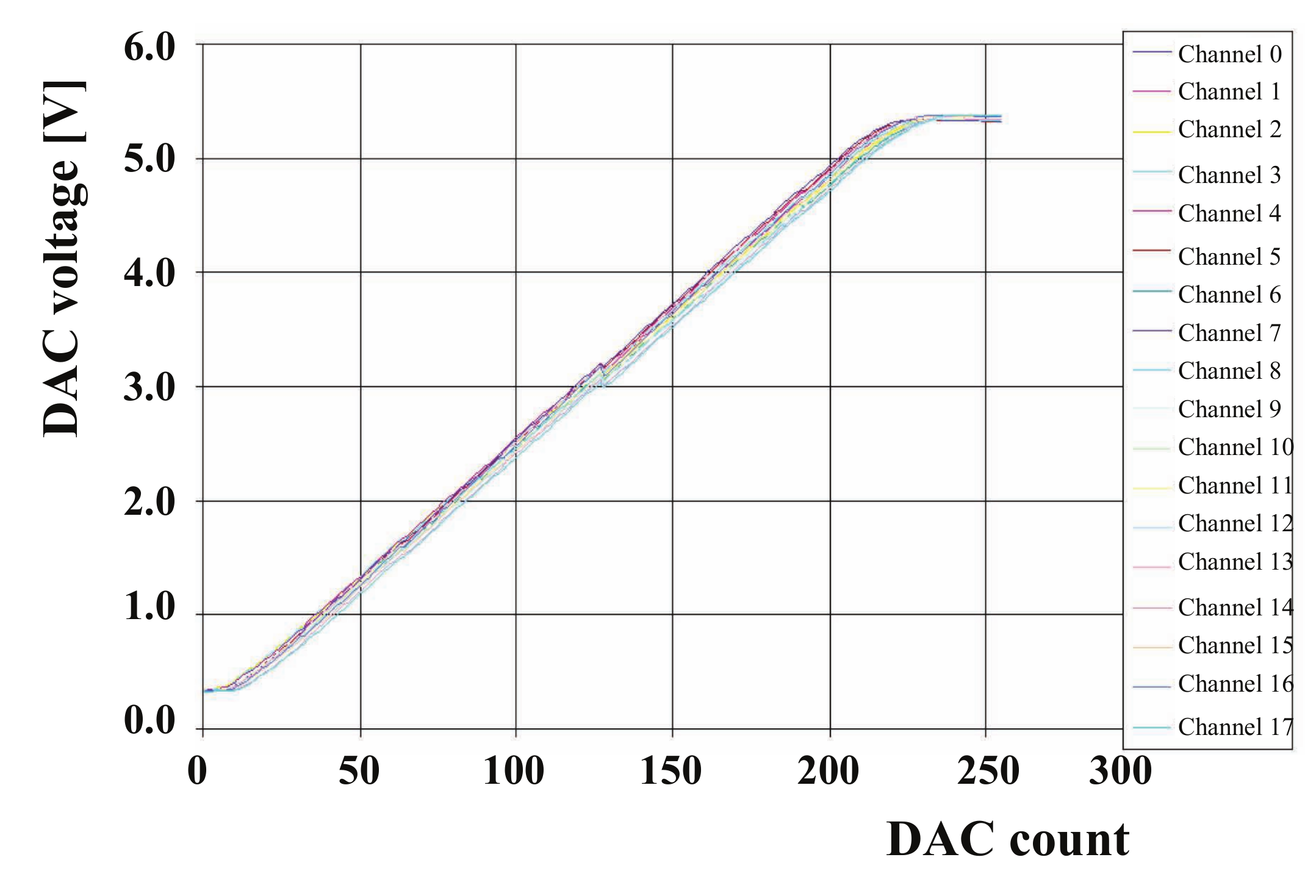}
\includegraphics[width=62mm]{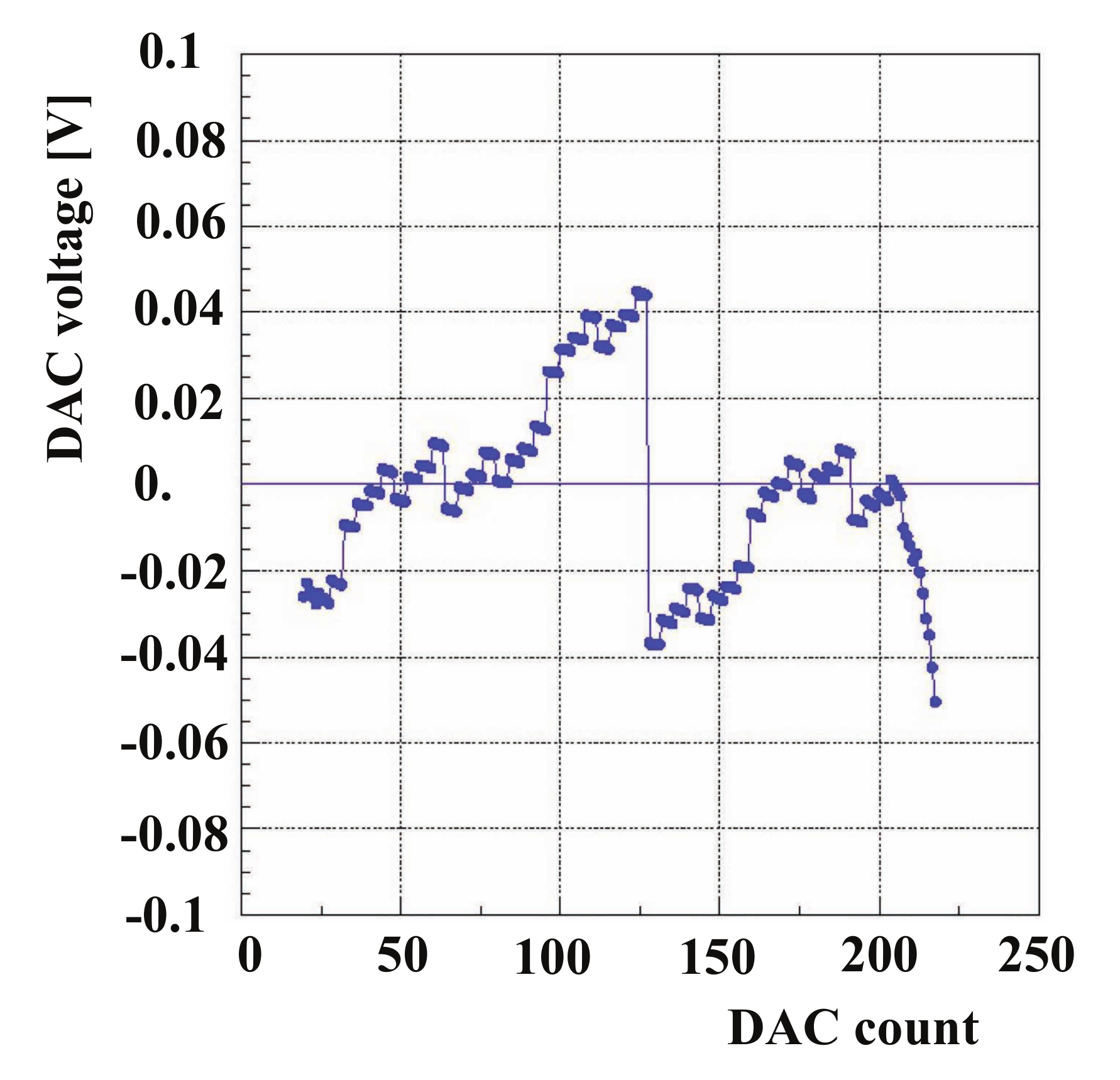}
\caption{Measurements of the DAC linearity of 18 channels (left) and residuals for channel zero (right).}
\label{fig:lindac}
\end{figure}

The linearity of the 18 channels of the DAC is shown in Figure~\ref{fig:lindac} (left). The non-linearity of $\sim 2 \%$ essentially results from a mismatch in current mirrors. The total range of $\rm 4.5~V$ specifies the voltage selection range of SiPMs grouped together and connected to a given base board in a single cassette. The deviation from linearity is shown in Figure~\ref{fig:lindac} (right) for channel zero as an example. The non-linearities of the other channels are similar in size. 

\subsection{The Very-Front-End Readout Boards}
\label{boards}

\begin{figure}[htb]
\vspace{-0.0cm}
\setlength{\unitlength}{1cm}
\begin{center}
\includegraphics[width=130mm]{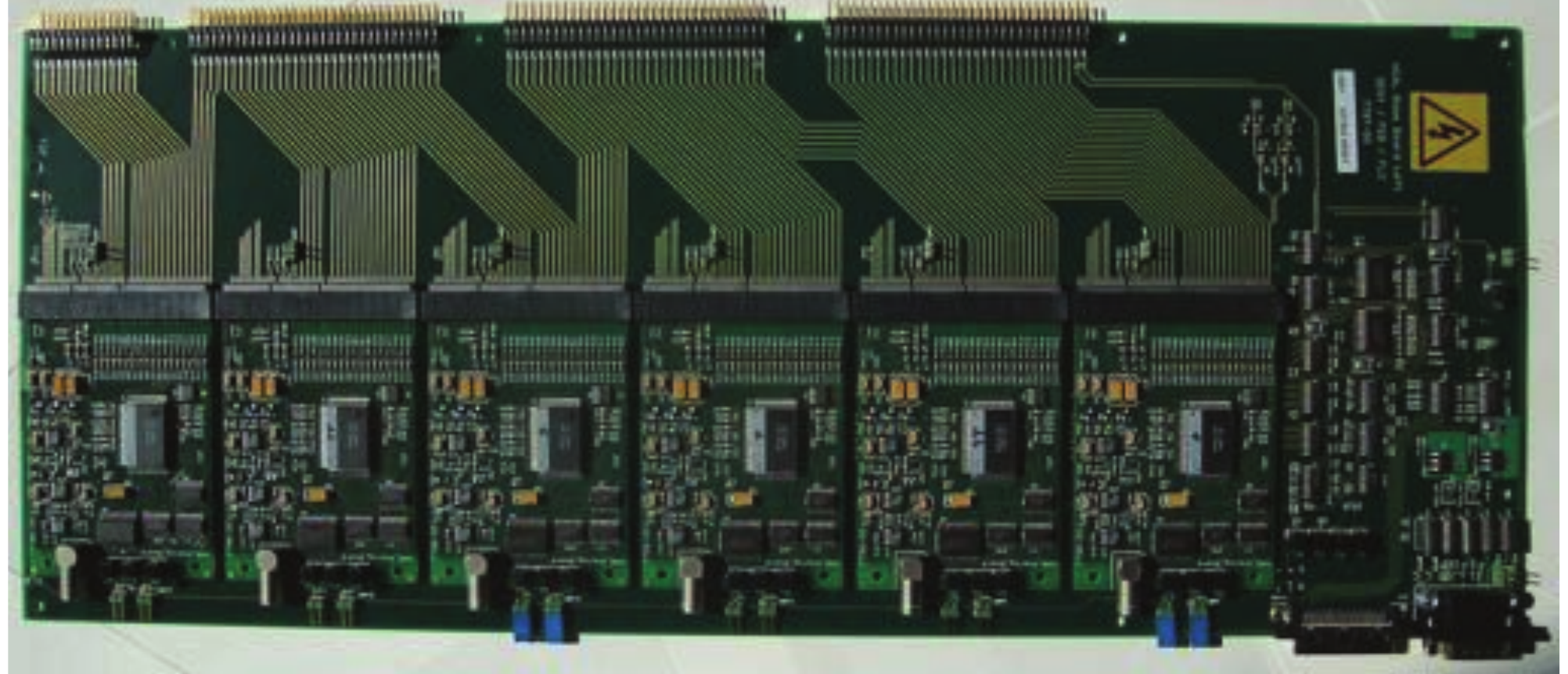}
\caption{The VFE AHCAL base board with six mezzanine boards. The connection to the SiPMs is
via coaxial cables.
\label{fig:VFE_boards}}
\end{center}
\end{figure}

As shown in Figure~\ref{fig:VFE_boards}, we use two different types of Very-Front-End (VFE) readout boards to interface the SiPM outputs to the data acquisition system (DAQ). The first is a mezzanine board, called AHCAL analog board (HAB) that carries one VFE ASIC with 18 SiPM input channels. The second is a base board, called AHCAL base board (HBAB) that carries up to six HABs. Each HAB provides an output driver (with a gain of two) for the processed analog SiPM signal, contains control and configuration electronics that is set remotely by the DAQ, and provides  the correct bias voltage for each SiPM. The HBABs, with  dimensions of 47.5~cm$\rm \times$19~cm, house all control functions and configuration data from the DAQ to the ASICs. Two HBABs read out the 216 SiPM channels of one AHCAL layer. The modular design of the readout and use of connectors allow us to replace individual boards in case of malfunction. 

Both types of boards are six-layer designs with a dedicated power-ground system. All critical signals, especially the analog outputs of the VFE ASICs and the fast LVDS control signals, are routed differentially with a 120~$\Omega$ line impedance that matches the impedance of the cable to the DAQ. The external trigger line from the DAQ used by the ASICs to hold the current input signal is length-balanced in order to achieve synchronous sampling of all ASICs. Without the SiPM connected, the VFE noise (at most 1.5~mV RMS) is dominated by the ASIC noise.

\section{Off-Detector Readout Electronics}

The off-detector electronics distributes the sample-and-hold signals required by the VFE electronics within a latency of $\rm 180\pm 10~ns$, drives the VFE signal sequence to multiplex the analog signal readout, digitizes the signals, and stores the data. The CRCs (see Figure~\ref{fig:daq}) performing this task contain eight front-end (FE) sections that are fanned into a single back-end (BE) section providing the interface to VME. Each FE section controls twelve VFE readout ASICs needed to read out a single layer of the AHCAL. It is equipped with twelve 16-bit ADCs to digitize the SiPM signals. The data volume within each FE is 512 bytes per event. The BE collects the digitized signals and stores them into an 8~MByte memory. Since the typical event size is 4~kByte per CRC, about 2000 events can be stored before readout is required.

The CRCs also handle the trigger control and distribution. The trigger logic built in the BE allows for significant complexity, since we define the trigger condition via software. We set a trigger busy signal to prevent further triggers until the digitization of the VFE analog data is completed. We use the rising edge of the trigger signal to synchronize the entire system and derive all timing information including the time-critical sample-and-hold signal from this edge.

The system performs as expected meeting all requirements. The sample-and-hold is distributed on the derived 160~MHz clock within a minimum of 160~ns. This allows us to implement the rest of the latency of 180~ns as a software delay in the FE FPGA. Time jitter on this signal comes from the 6.25~ns rounding errors of the clock. The CRC noise without the AHCAL  is only 1.4~ADC bins on average, while the VFE electronics has a noise of 18 ADC bins. The trigger counter synchronization is very reliable. 
Further details are given in \cite{ecal}.

The dynamic range for measuring hadron showers is determined by the number of pixels per MIP and the total number of pixels in the SiPM. For an average of 15~pixels per MIP and 1156 pixel the dynamic range of the photosensor is of the order of 70~MIPs. The dynamic range of the ADC is around 100~MIPs, which is somewhat higher to ensure that that the hadron energy measurements are not affected by non-linearities of the ADCs at high ADC bins.

\section{Online Software and Data Processing}

The online software is designed to satisfy the electronics requirement of
an instantaneous event rate of up to 1 kHz and an accepted average event
rate of 100 Hz. We store information on the hardware configuration and beam properties
together with the data in the run data files to simplify subsequent data processing. We save
the data as C++ objects and serialize them explicitly. The software is designed to use a simple memory management by avoiding virtual functions or pointers within the data classes.
We have implemented variable-length data objects using free memory
immediately following the object. The online software uses fast accessor
classes to write objects and to locate existing objects within the event
record. During data taking, a copy of the complete event record is written
to disk once the online processing of the event is completed.

The CALICE DAQ system is optimized for maximal acquisition speed. In a first processing step, the conversion step, the raw data are tested with respect to their integrity and are converted into LCIO \cite{lcio}, the standard data format for  ILC-related studies. This step also acts as an event builder.  During the conversion, a database is filled with information of the detector configuration, including temperature recordings, voltage settings, calibration constants and other specific information of the run, such as machine-related parameters. In a second processing step, the reconstruction step, zero suppression is applied and  calibrated calorimeter hits are produced as the main output of this step. The size of the processed events is typically reduced by a factor of three compared to that of the raw data. In addition, the DAQ has the capability to include relevant data from other detectors, such as tracks from a tracking chamber and hits from Cherenkov counters. This is important information needed in a test beam to locate the particle position and determine the particle type. 
Digitization algorithms developed for the test beam data are easily portable into a full detector simulation. Further details about the management and processing of CALICE  data are given in \cite{eudet}.

The full system achieves a readout rate of 120~Hz when not limited by the beam intensity or spill structure. The average readout rate during high-intensity spills is approximately 90~Hz. The trigger rate during spills is  $\sim$600~Hz and is limited by the rate of reading out the trigger counters for synchronization purposes.
A high readout frequency helps to check for synchronization errors.  
The system recorded up to 10 million events per day and ran for several beam periods of months. In total, about 300 million events have been collected, filling approximately 30~TBytes of disk space.
Further details are given in \cite{ecal}.

\section{The Calibration and Monitoring System}   
\label{monitoring} 

Since the SiPMs are sensitive to changes in temperature and operation voltage, we need to monitor them during test beam operations. The monitoring system needs sufficient flexibility to perform three different tasks. First, we utilize the self-calibration properties of the SiPMs to achieve a calibration of the ADC in terms of pixels that is needed for non-linearity corrections and for directly monitoring the SiPM gain. Second, we monitor all SiPMs during test beam operations with a fixed-intensity light pulse. Third, we cross check  the full SiPM response function by varying the light intensity from zero to the saturation level \cite{calibration}.

We built a monitoring system that distributes UV light from an LED to each tile via clear fibers. We use UV light to attain efficient light collection by the WLS fiber. The pulses are 10~ns wide in order to match real signals as closely as possible. If the signal width becomes too long, the probability for a pixel to fire twice increases rapidly. The LED light itself is monitored with a PIN photodiode to correct for fluctuations in the LED light intensity that is adjustable from the DAQ system. By varying the voltage, the LED intensity covers the full dynamic range from zero to saturation ($\rm \sim 70~MIPs$).

In addition, we use high-precision power supplies to supply the reverse-bias voltage to the SiPM. 
Finally, we have placed five thermosensors in each cassette that are read out regularly by a slow control system storing a time stamp and the temperature recordings in a database \cite{slow}.

\subsection{The Light Distribution System}
\label{light-distribution}

\begin{figure}
\begin{center}
\includegraphics[width=80mm]{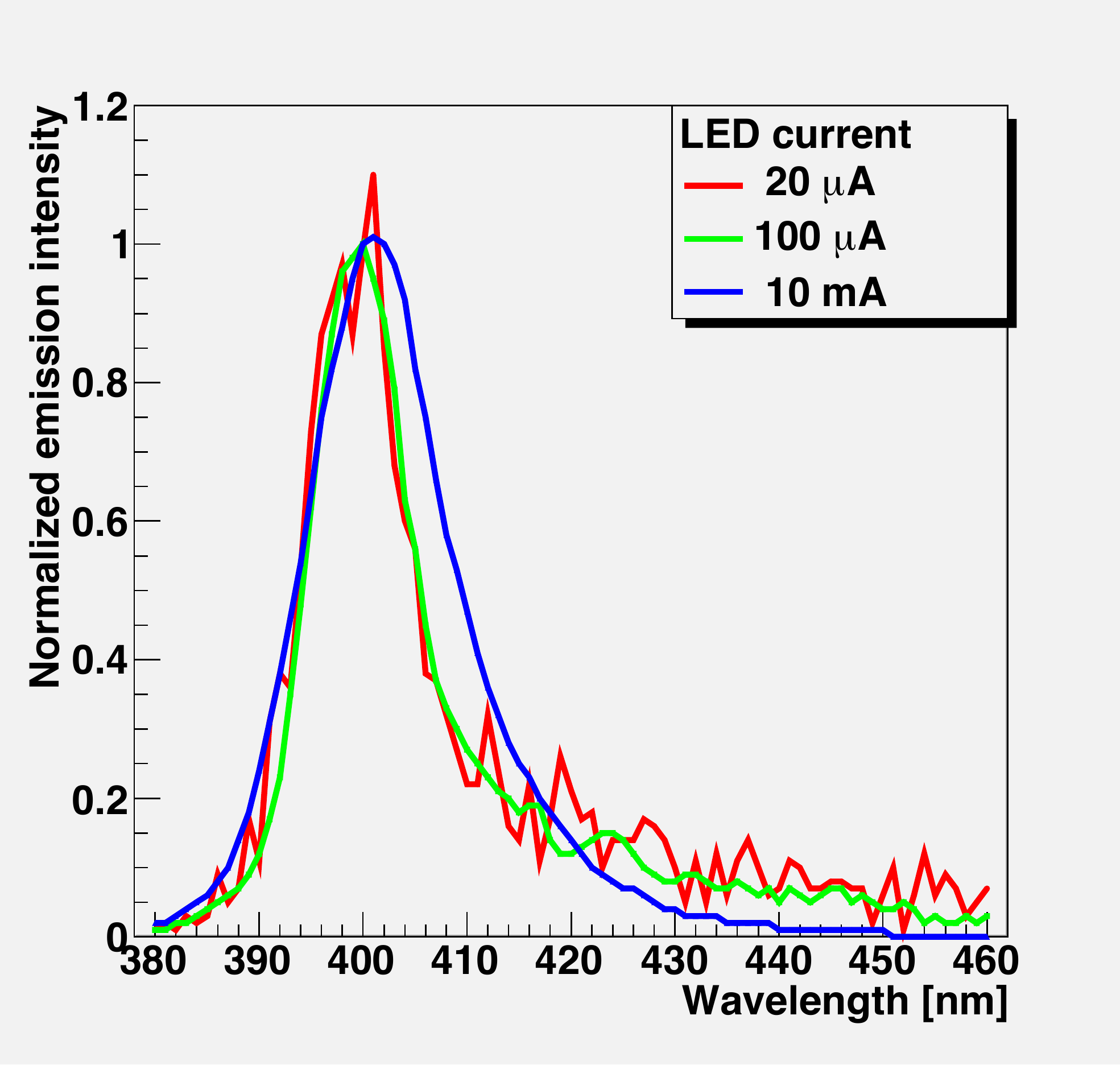}
\end{center}
\caption{The emission spectrum of the UV LED.}
 \label{fig:led}
\end{figure}

Since space constraints set a limit on the total number of LEDs plus electronics in a cassette, we couple 19 clear fibers to one LED of which 18 fibers are routed to the scintillator tiles and one fiber is routed to a PIN photodiode. The LEDs have to produce fast, bright and uniform light signals. A standard UV LED with an emission maximum at 400~nm fulfills all our requirements. Figure~\ref{fig:led} shows the emission spectrum of the UV LED. Measurements of the light emission pattern in a dedicated laboratory setup show that the light variation is less than a factor of two between brightest and dimmest illuminated fiber. This is sufficient to perform gain monitoring as low as two pixels
and to measure the saturation curve at least in a region where the maximum number of firing pixels can be extrapolated from a fit. For low LED intensities we observe prominent pixel peaks in all SiPMs. When the LED intensity is increased the pixel peaks broaden and become less well distinguishable.

We use 10~cm to 90~cm long, clear, single-clad fibers with a diameter of  0.75~mm to transport the LED light into the tiles. We have performed aging tests of the fibers by pulsing the LED at high rate and measuring the light intensity at the end of a fiber. This test, which is equivalent to a ten year operation, shows no deterioration. To ascertain a straight alignment of the fibers in front of the LED, the 19 fiber ends are glued together inside a 1.4~cm long metal sleeve with 0.5~cm diameter. The metal sleeve is inserted into a hole drilled into the aluminum frame of the cassette. An LED is inserted into this hole from the other side. The LEDs are soldered first to the LED driver electronics located on the CMB, which is attached to the cassette frame pushing the LEDs towards the fiber bundles. On the other end, each fiber is glued to a conical aluminum mirror that is 0.7~cm in diameter and is glued to a FR4 plate reflecting the LED light onto the scintillator tile at an angle of $90^\circ$ (see Section \ref{cassette-assembly}). The PIN photodiodes are also mounted on the CMB and are inserted into a hole in the cassette frame. The fiber is inserted from the other side and is centered in the hole by a special adapter. All couplings consist of thin air gaps. This construction is rather robust providing a reliable and uniform light distribution after moving the cassettes into place.

\subsection{Calibration and Monitoring Electronics}
\label{monitoring-electronics}

The CMBs contain all the electronics needed to drive the LEDs, set the gain of the preamplifiers for the PIN photodiode signals, and read out the temperature sensors. The LED light amplitudes are tunable from low intensities triggering individual pixels to high intensities producing far over 70~MIP signals. The pulse widths are adjustable within a few ns. Here, we use a typical pulse width of 10~ns. The trigger for the pulse start is obtained from the DAQ and is synchronized with the readout, while an analog signal from DAQ, (V-calib), is used to control the amplitude of LED light. We use a common steering of the amplitude and the pulse-width for all LEDs.

\begin{figure}
\begin{center}
\includegraphics[width=130mm]{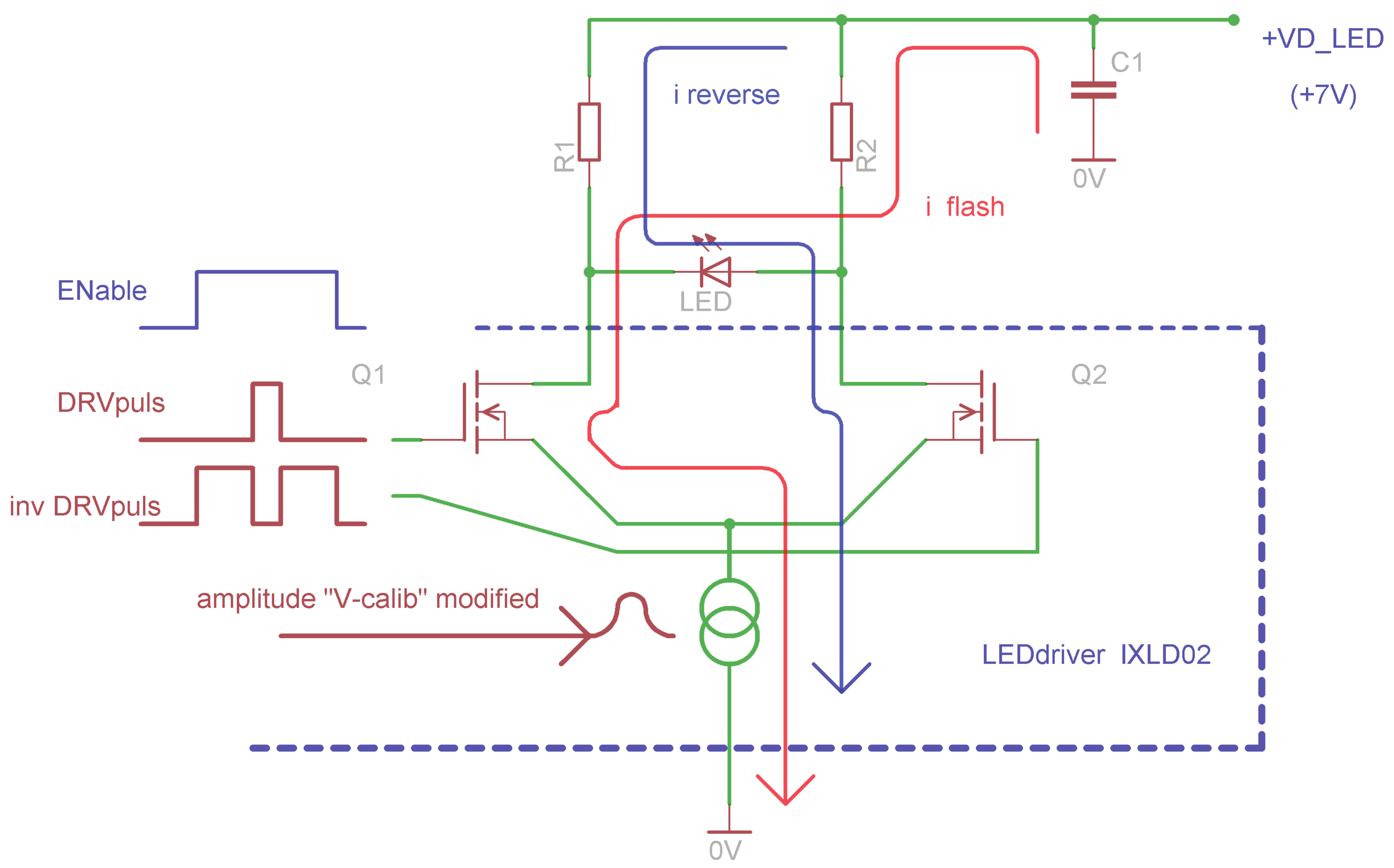}
\end{center}
\caption{Layout of the LED driver.}
 \label{fig:led-driver}
 \vskip 0.4cm
\end{figure}

\begin{figure}
\begin{center}
\includegraphics[width=90mm]{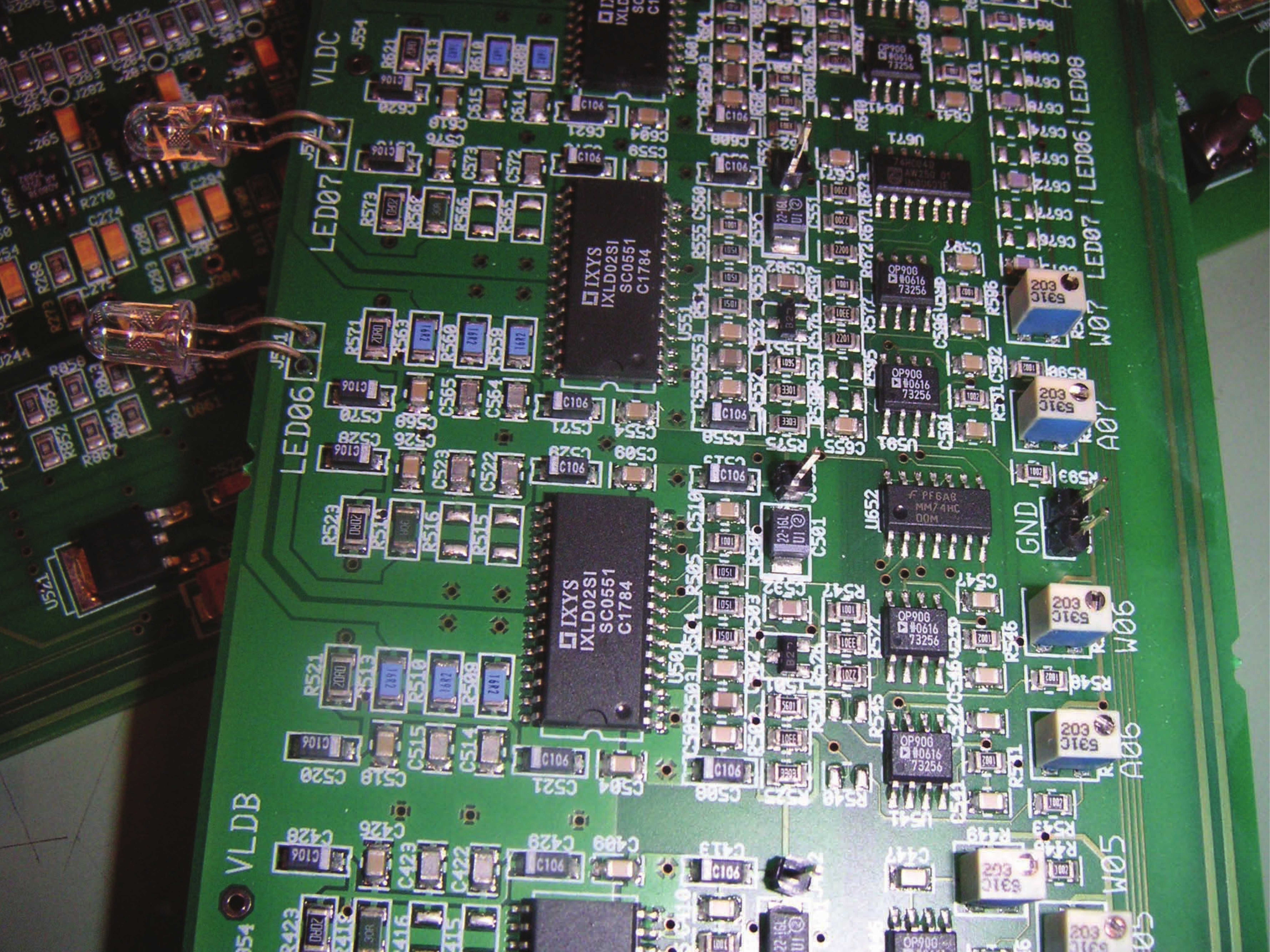}
\end{center}
\caption{Photograph of the calibration and monitoring board.}
 \label{fig:cmb}
\end{figure}

Figure~\ref{fig:led-driver} displays the schematic of the LED driver. Our design is particularly optimized for the generation of nearly rectangular pulses, $i.e.$ pulses with a fast rise time and fast fall time ($\rm \sim 1~ns$). We have checked that the LED light emission characteristics do not change with the increase of the LED current in the range $\rm 10 ~\mu A $ to 10~mA. In order to reduce light fluctuations among the LEDs we have selected a subsample with a similar light-intensity profile for one cassette. A photograph of the CMB is depicted in Figure~\ref{fig:cmb}.

Since PIN photodiodes have a gain of one, we need an additional charge-sensitive preamplifier for the PIN photodiode readout. The rest of the readout chain is the same as that of the SiPMs. The presence of a high-gain preamplifier directly on the board in the vicinity of the power signals for LEDs, however, has turned out to be a source of cross talk. 

Another task of the CMB consists of reading out the temperature sensors via a 12 bit ADC. The temperature values are sent to the Slow Control system via a CANbus interface. One CMB operates seven temperature sensors, two sensors directly located on the readout board and five sensors distributed across the center of the cassette. The sensors consisting of integrated circuits of type LM35DM produced by National Semiconductor are placed in a $\rm 1.5~mm$ high SMD socket. Their absolute accuracy is $\rm <0.6^\circ C$. A microprocessor of the PIC 18F448 family in association with a CAN controller interface, PCA82C250, provides the communication of the CMB with the Slow Control system. We have not observed any noise pickup in the CMB.

During test beam operations, we take sets of special runs to scan the SiPM response function, perform gain calibrations with low-intensity LED light,  determine the electronics intercalibration between operation in calibration mode and  physics mode, and record the LED reference signals and pedestals between spills. 
LED amplitudes for readout in high-gain mode and low-gain mode have been optimized individually for each module to provide sufficient points for the gain calibration and to record the response curves. In addition, all twelve LEDs in one module (one CMB) are manually tuned to deliver the same amount of light to ensure that all SiPM signals show single-pixel peaks. All adjustments have been performed after complete module assembly. First tests were obtained using the full data acquisition chain in a special test installation at the DESY electron test beam. After tuning of the LED monitoring system the efficiency for successful gain calibrations on all modules is around $98\%$. This includes a few broken LEDs and noisy SiPMs for which the single photoelectron peak spectrum cannot be fitted to extract a gain value.

\section{Calibration Procedure}

We calibrate each cell to the MIP scale by
\begin{equation}
E_i\left[\mathrm{MIP}\right] = \frac{A_i\left[\mathrm{ADC}\right]}{C_i^{\mathrm{MIP}}}\cdot f_i^{-1}\left(\frac{A^g_i\left[\mathrm{ADC}\right]}{C_i^{\mathrm{pixel}}}\right),
\label{eq:calib}
\end{equation}
where $A_i\left[\mathrm{ADC}\right]$ ($A^g_i\left[\mathrm{ADC}\right]$ ) is the pedestal subtracted energy measured for physics (calibration) events in units of ADC bins for cell $i$, $C_i^{\mathrm{MIP}}$ is a conversion factor of ADC bins into MIPs, the function $f_i$ is the SiPM response function describing the SiPM output signal as a function of the incoming light intensity, and $C_i^\mathrm{pixel}$ is a conversion factor of ADC bins into units of a single avalanche in a SiPM pixel  that includes the intercalibration constant.

\begin{figure}
\includegraphics[width=75mm]{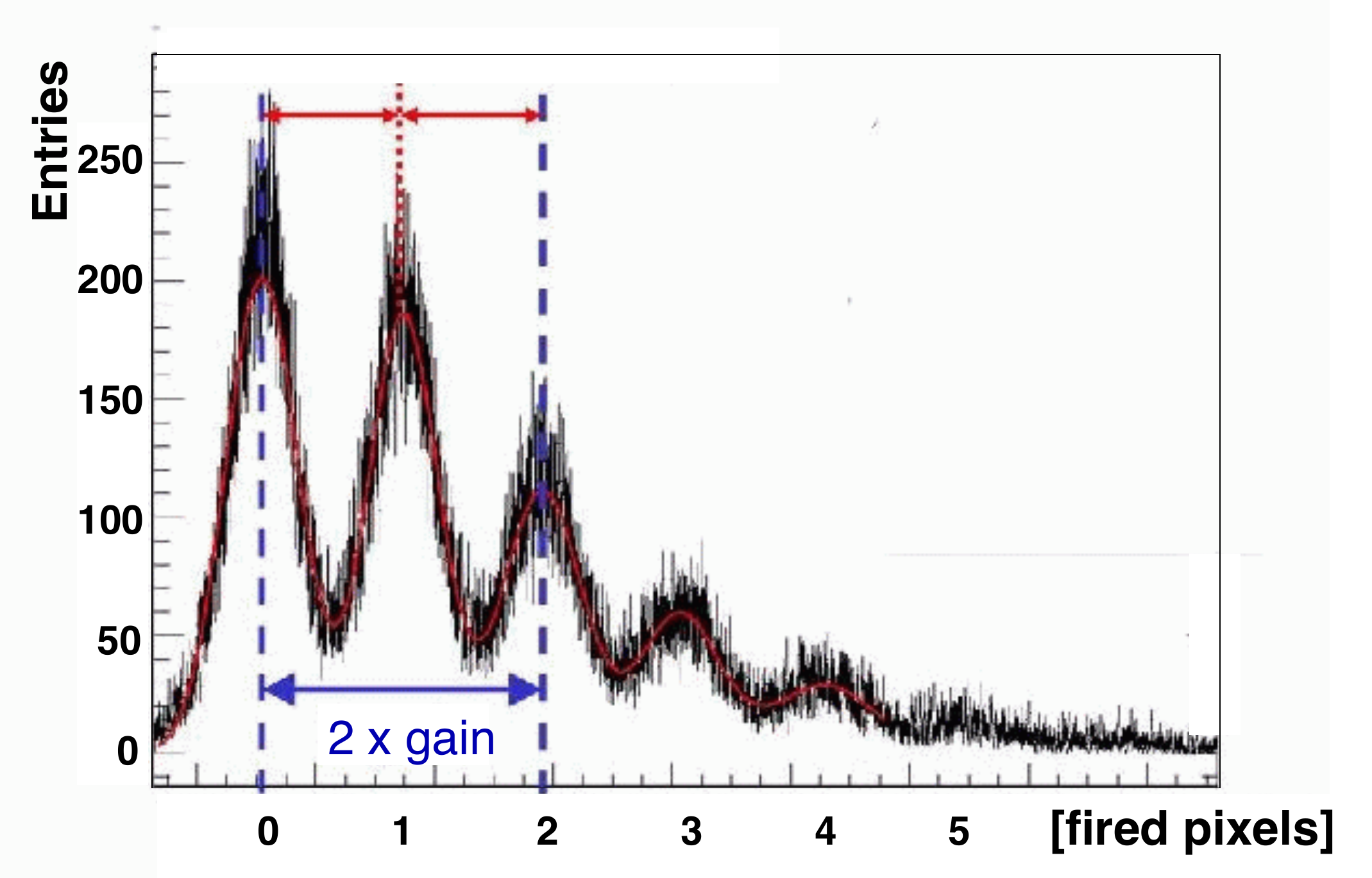}  
\includegraphics[width=74mm]{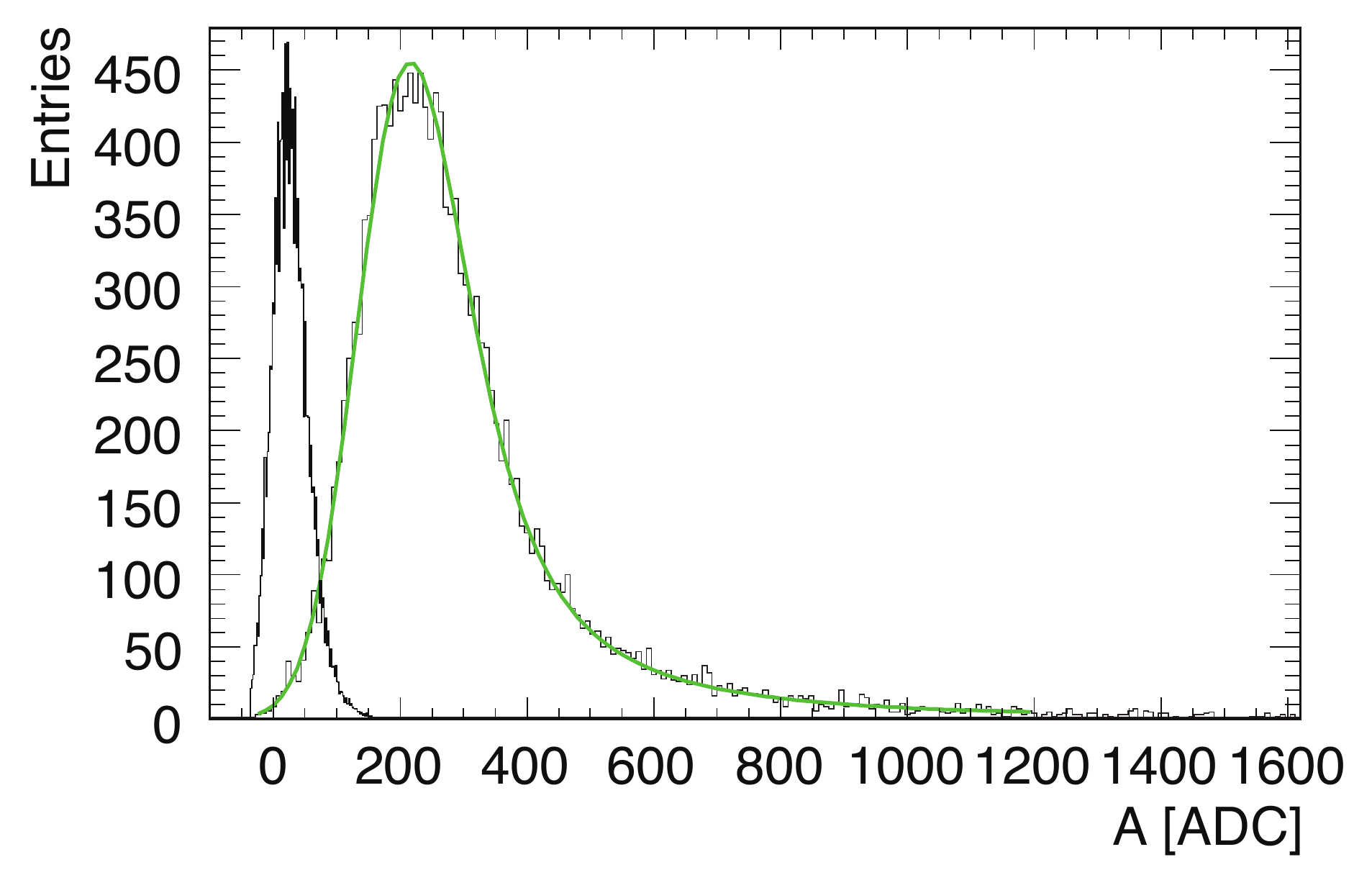}
  \caption{Left: Distribution of single pixel peaks measured in the gain calibration. The superimposed curve is a fit with multiple Gaussian functions. Right: The MIP distribution of muons measured in a cell in the physics mode. The superimposed curve is a fit to a Landau distribution convolved with a Gaussian function. The noise spectrum is shown as a reference.} 
  \label{fig:calib}
\end{figure}

We extract the MIP conversion factors $C_i^\mathrm{MIP}$ of each cell using the most probable response to  the approximately minimum ionizing muons. A typical MIP spectrum together with the pedestal spectrum of the same channel is shown in Figure~\ref{fig:calib} (right). In the summed shower energy, we consider only cells in which the energy exceeds half a MIP. This eliminates most of the electronic noise, while keeping a high efficiency for single particles.

We measured the response function $f_i$ of each SiPM  on the test bench before installation (see Figure~\ref{fig:sipm-pixel} right). The inverse function $f^{-1}_i$ is used to correct for non-linearities in the response. In future, these measurements could be replaced by regularly updated in-situ measurements providing improved non-linearity corrections. A typical calibration spectrum is depicted in Figure~\ref{fig:calib} (left). The inverse response function $f_i^{-1}\left(A_i\left[\mathrm{ADC}\right]\right)$ is nearly one for small measured energies increasing approximately exponentially towards higher energies. In the range of our operation the maximum correction factor does not exceed three. 
The conversion from the MIP scale to the GeV scale is achieved using test beam data of known energies.


\section{Commissioning and Initial Performance} 
\label{commissioning}

Commissioning started with modules without absorber plates in a DESY electron beam. In a dedicated setup we have tested up to four modules with 3~GeV electrons. We have determined a first MIP calibration and have measured the in situ light yield by combining the MIP calibration and gain calibration. Furthermore, we have tested the non-linearity correction with additional absorber material in front of the module under study. These initial studies were performed to establish the readout, calibration and monitoring of the  7608~SiPMs used in the AHCAL prototype.

\subsection{Test of Single Modules}
\label{beam-test}

\begin{figure}
  \begin{center}
\includegraphics[width=140mm]{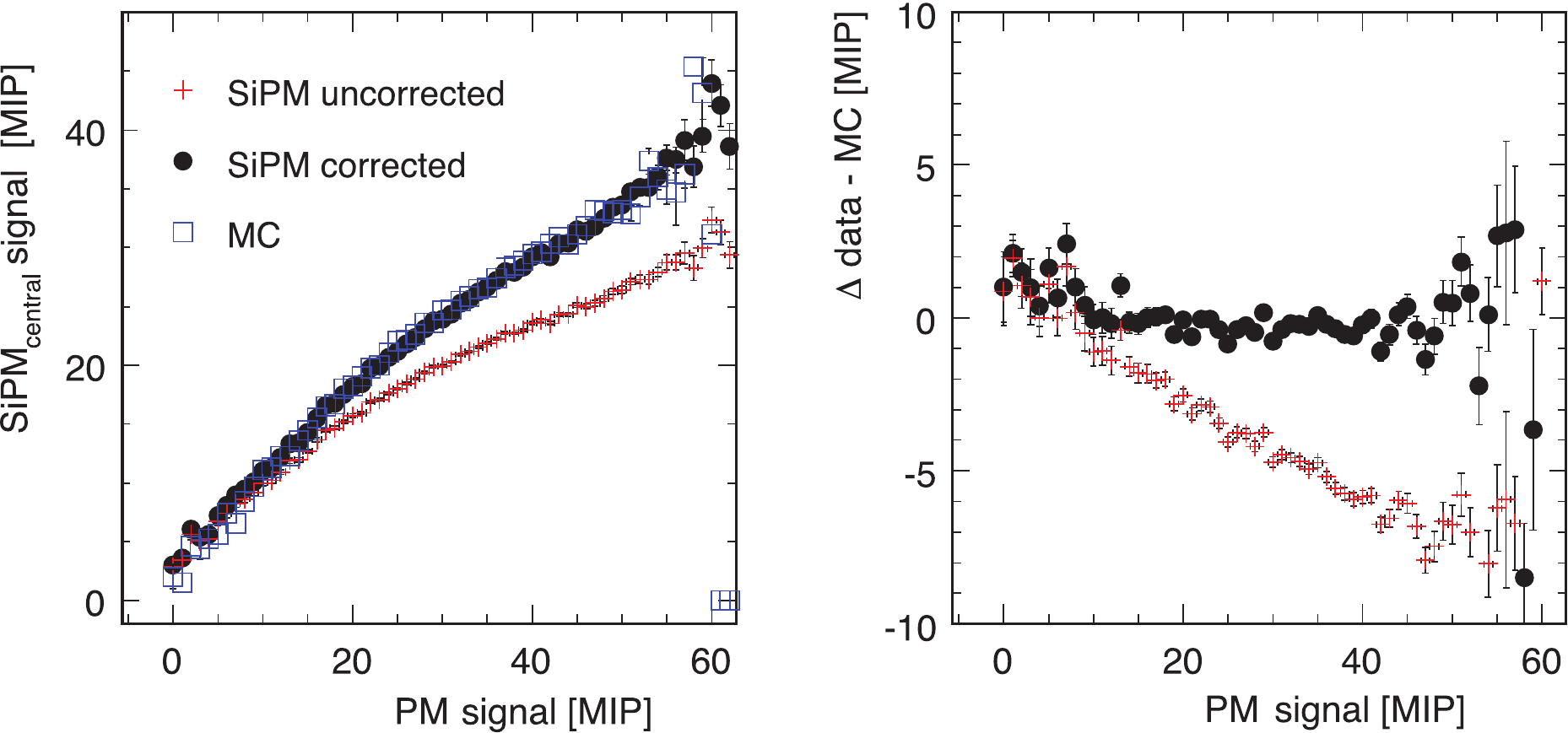}
  \end{center}
  \caption{{ Left: Comparison of SiPM readout and PM readout for a 5~GeV $e^+$ beam showering in a 5~$X_0$ thick lead absorber for uncorrected data (crosses), corrected data (solid points) and a GEANT3 simulation (squares);  Right: Residuals of the SiPM data and simulation for uncorrected (crosses) and corrected data (solid points).
  }}
  \label{fig:shower5X}
\end{figure}

Using the initial test beam setup at DESY we have measured energies up to $\sim$40 MIPs per tile. We 
have compared the performance of the SiPM readout with that of a photomultiplier  (PM) readout using a specially equipped cassette.
For the latter setup we used long WLS fibers that were routed from the scintillator tiles to the outside of the module where the PMs were located. We positioned a lead absorber plate of variable thickness on the beam line in front of the module in order to initiate an electromagnetic shower. Figure~\ref{fig:shower5X} (left plot) shows the correlation of the energy collected by the reference scintillator-PM system to the energy recorded with the tile-SiPM system for a 5~GeV $e^+$ beam showering in a 5~$X_0$ thick lead absorber. For energies above 15 MIPs the uncorrected data (crosses) show a clear deviation from a GEANT3 simulation (squares) \cite{geant3}. The largest fraction of energies in the shower maximum lies in the range between 15 and 35 MIPs, where the simulation deviates from the data by $10-25\%$. After correcting the SiPM signals on an event-by-event basis using the measured response function (see Equation~\ref{eq:calib}), the simulation and the data show good agreement. This is depicted in Figure~\ref{fig:shower5X} (right plot) displaying the residual of the corrected data with respect to the simulation.

\subsection{Test Beam Setup at CERN}
\label{cern}

\begin{figure}[h!]
\includegraphics[width=55mm]{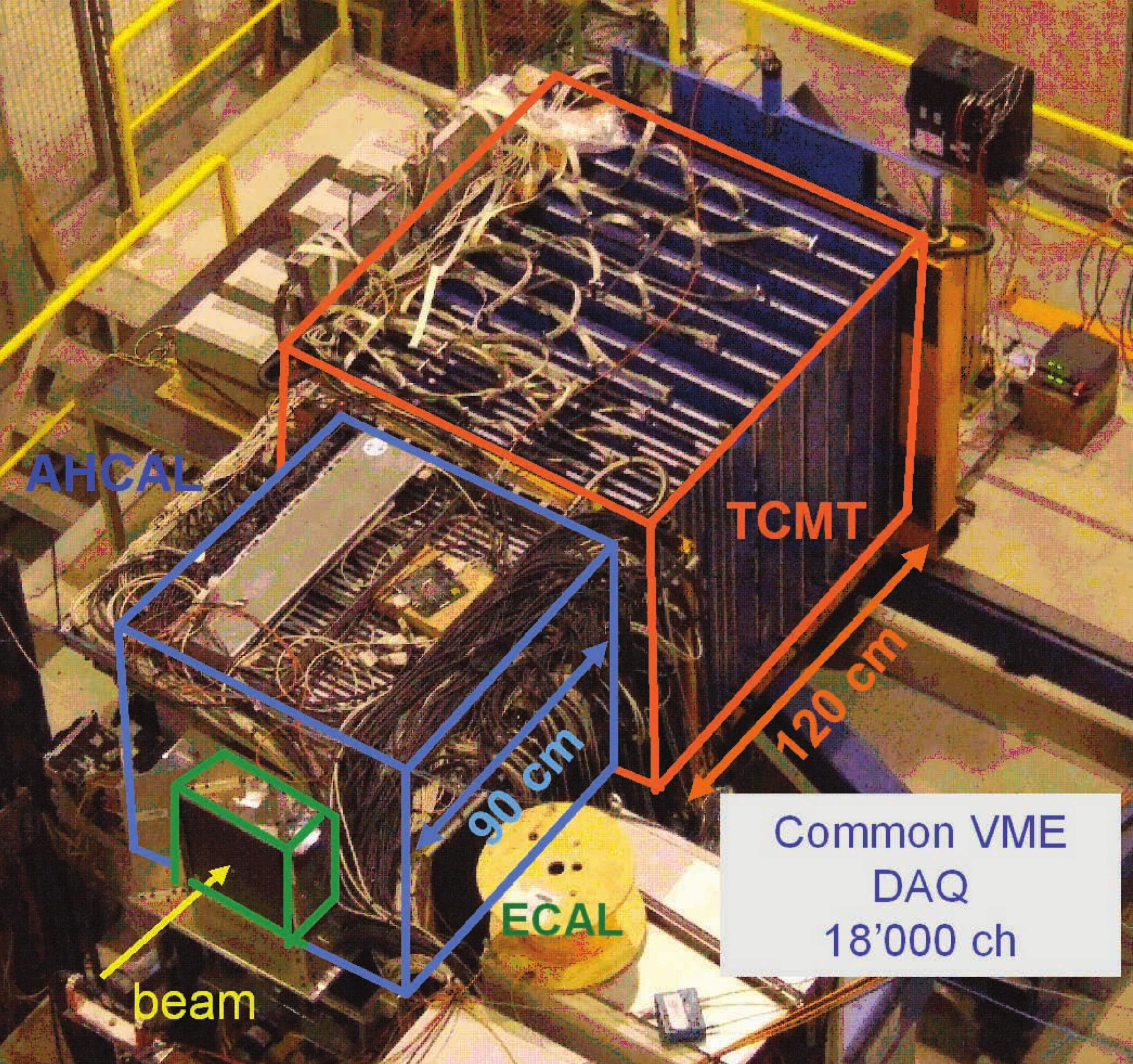}
\includegraphics[width=85mm]{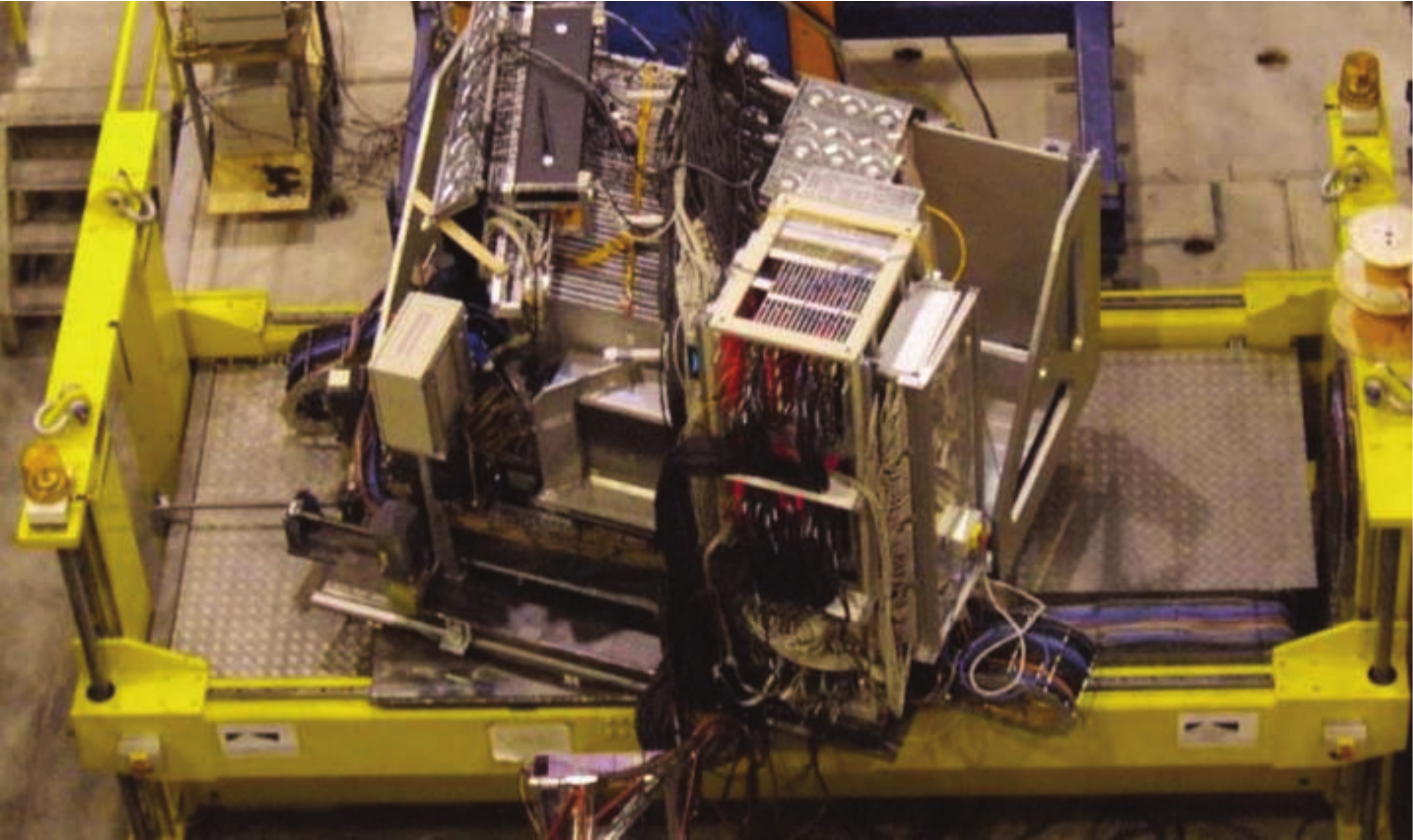}
\caption[]{{ CALICE calorimeter system setup at CERN in 2006 (left) and 2007 (right).}}
  \label{fig:setup_tb}
  \vskip 0.2cm
\end{figure}

The test beam setup at CERN in the H6 area started in 2006 with up to 23 modules installed. In 2007 the full AHCAL prototype was completed and installed. Figure~\ref{fig:setup_tb} shows the setup during 2006 and 2007 test beam. Data taking was carried out with a Si-W ECAL upstream and the TCMT downstream of the AHCAL. The 2006 data provide important tests and a validation of calibration and monitoring procedures. We can also use them to perform first studies of hadron shower shapes. The 2007 data, taken with the completed AHCAL prototype, allow us to perform the full program.

\begin{figure}[h!]
\includegraphics[width=67mm]{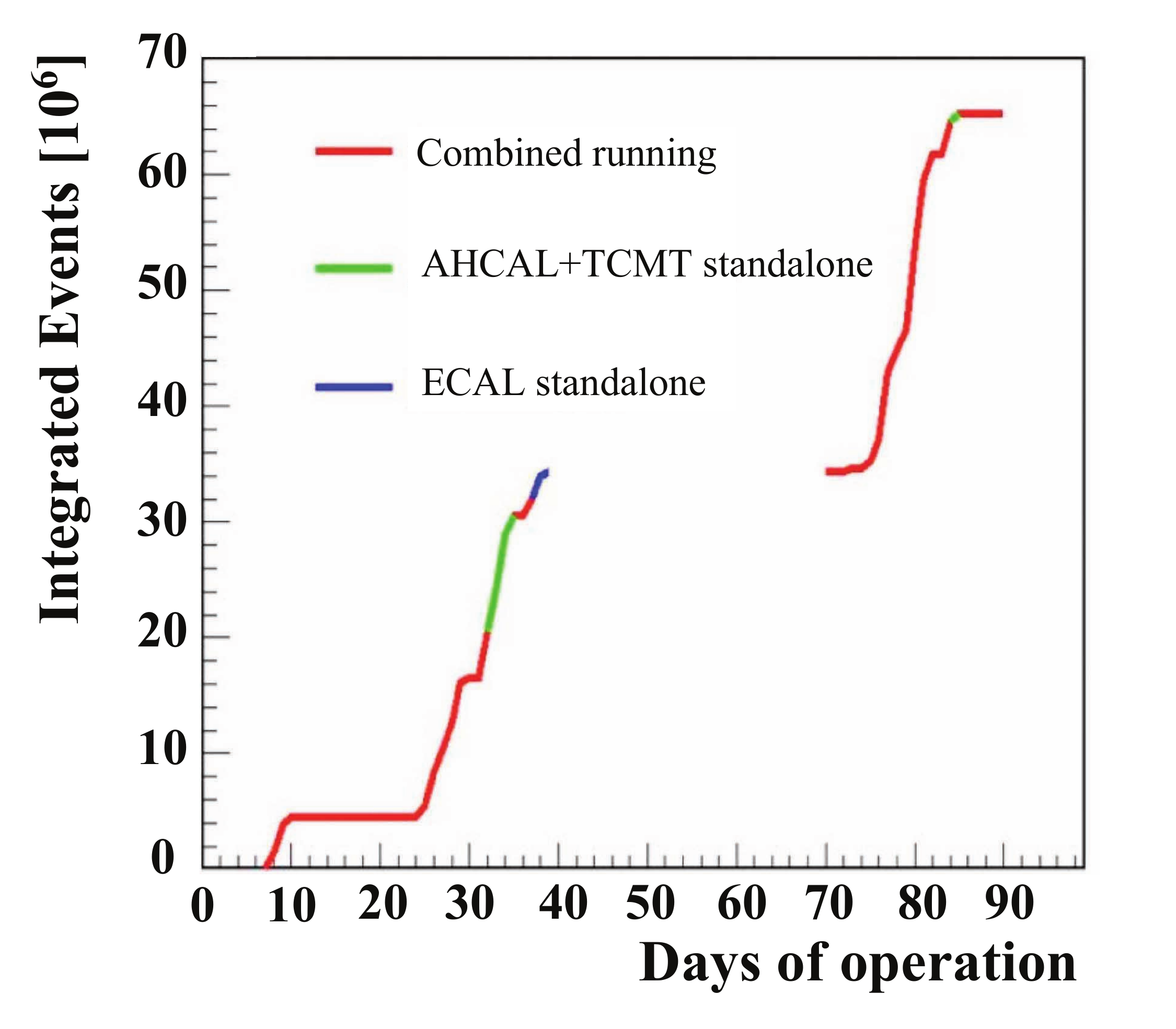}
\includegraphics[width=77mm]{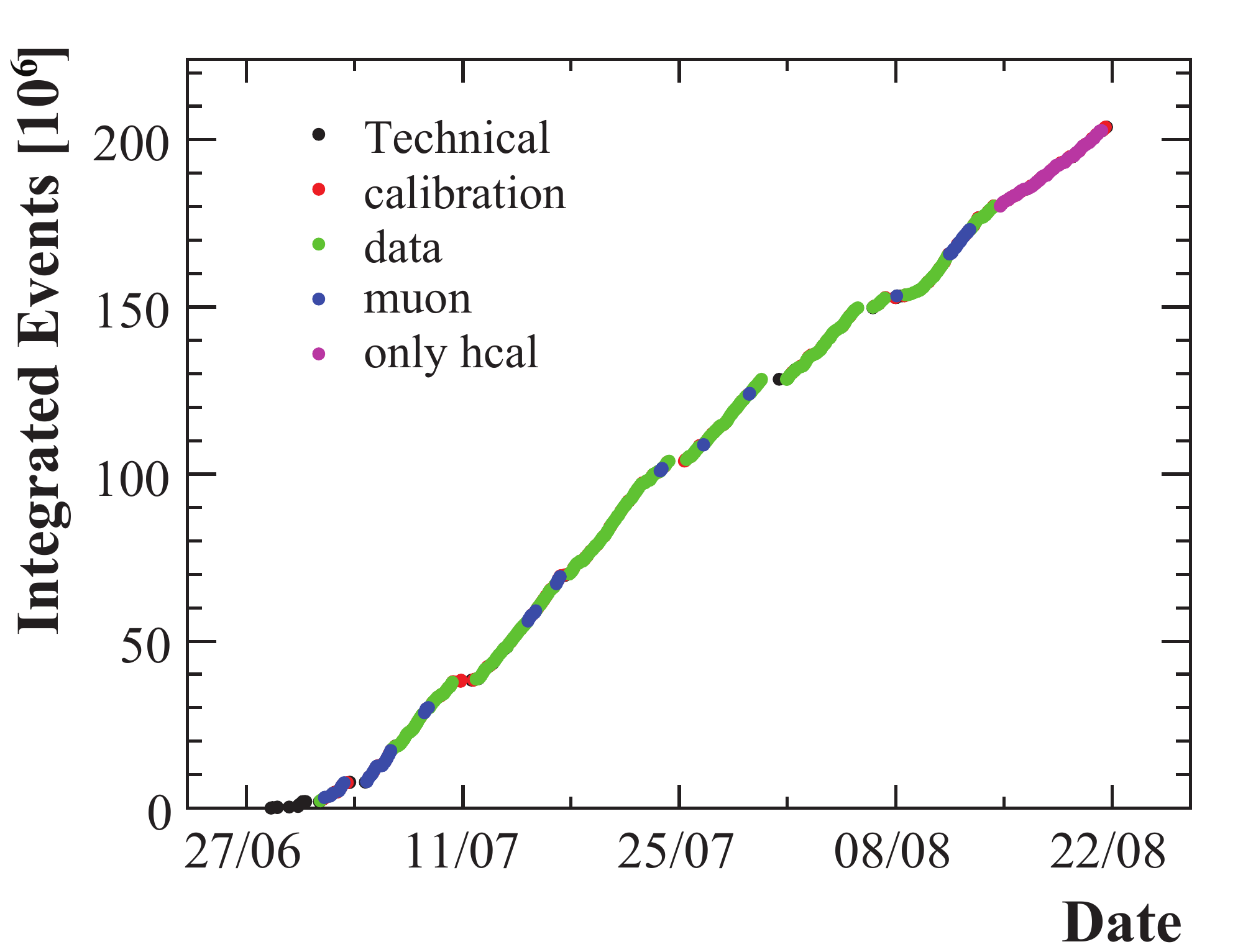}
\caption[]{{Event collection rate during test beam in 2006 (left) and 2007 (right).}}
\label{fig:setup_uptime}
\end{figure}

All detectors, beam monitoring devices and the DAQ itself showed a very high reliability. The uptime of the total system during test beam was $>95~\%$. During the two test beam periods we collected more than $2.5 \times 10^8$ events without zero suppression. We collected large samples of $\mu^\pm$, $e^\pm$ and hadrons ($h^\pm$). We
collected showers in the 6--80~GeV (6--45~GeV)  energy range for $h^\pm$ ($e^\pm$). Figure~\ref{fig:setup_uptime} (right plot) summarizes the event collection rate during the 2006 and 2007 test beam periods.

\subsection{MIP Calibration}
\label{section-muon}

The calibration of each cell is accomplished with muons from a beam that has a sufficiently broad distribution to cover the entire front face of the AHCAL. A minimum of 2000 muon events per cell is necessary to obtain a reliable fit to the pulse height spectrum that is parameterized as a convolution of a Landau distribution and a Gaussian function. For a uniform beam distribution this amounts to a total of $5 \times 10^5$ events. Since the beam has a Gaussian-shaped profile, we need approximately $4 \times 10^6$ events to achieve a good calibration in the outermost cells. With an average data acquisition rate of 100~Hz, the MIP calibration of the entire AHCAL takes about 12 hours.

\begin{figure}[!t]
\begin{center}
\includegraphics[width=90mm]{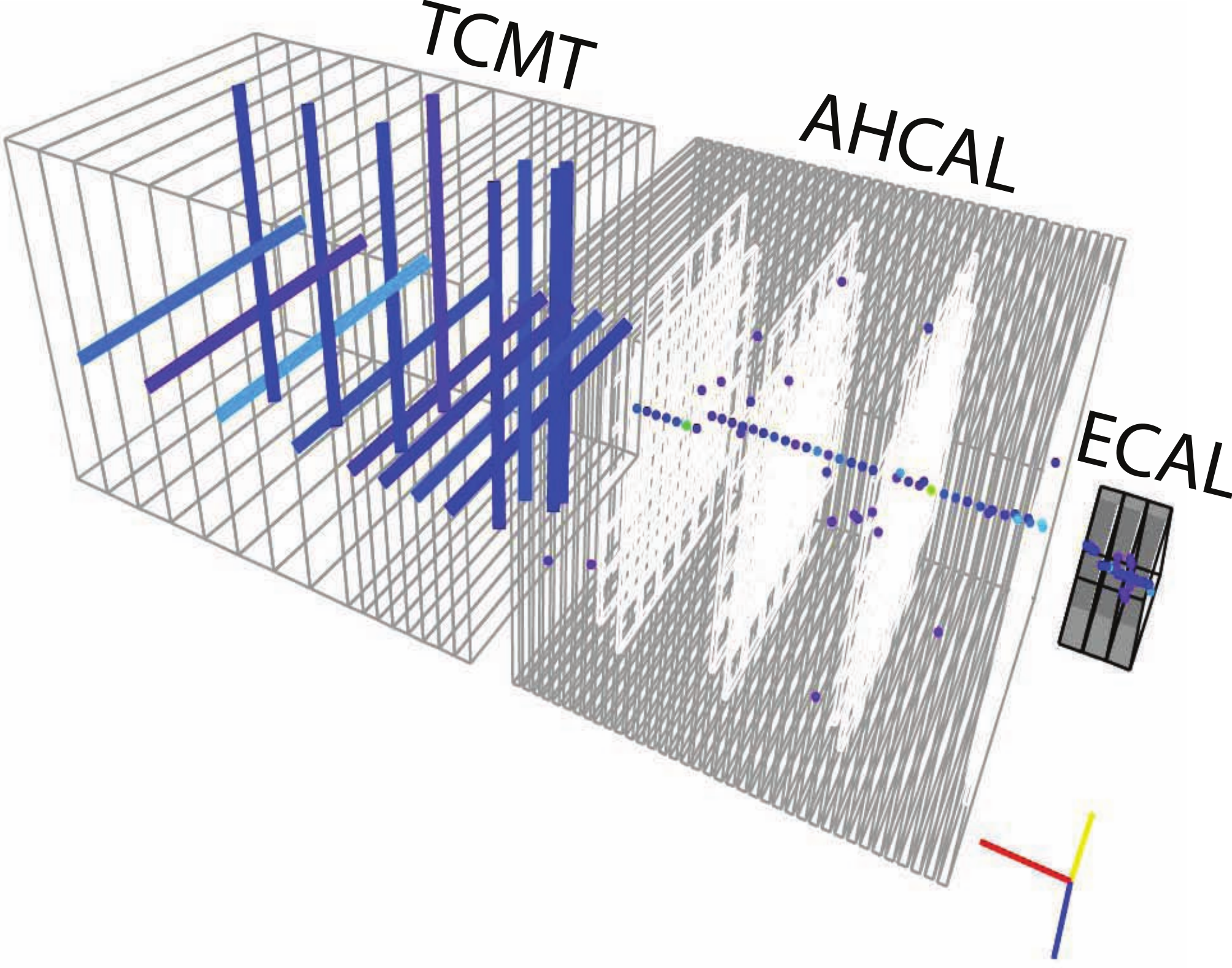}
\caption{Three-dimensional event display of a muon from the test beam penetrating the ECAL, AHCAL and TCMT. Hits in the TCMT are indicated by the colored bars.}
\label{fig:mu-ev}
\end{center}
\vskip 0.1cm
\end{figure}

In the offline processing of the data, we apply muon quality selection criteria to reject noise events and non-muon triggers. We define a muon as a track having at least $N_{\rm{ min}}$ hits within a cylinder of radius $R_{\rm{tile}}$, where $N_{\rm{ min}}$ is the minimum number of accepted hits  that have a pulse height larger than half a MIP, based on the initial calibration. The value of $R_{\rm{tile}}$ is chosen to be the same for each of the three tile sizes. Figure~\ref{fig:mu-ev} shows a one-event display of a 32~GeV/c muon that passed our track selection for $N_{\rm{ min}}$ = 16. Figure~\ref{fig:calib} (right plot) displays a typical energy spectrum in a $\rm 3~cm \times 3~cm$ calorimeter cell measured with selected muons. The fit shown represents a Landau distribution for the energy loss convolved with a Gaussian function to account for the system noise and the Poisson statistics introduced by the SiPM response. The  most probable value of the fit function is defined to be the MIP calibration factor. In order to reduce a possible bias from our track selection, we fit only the part of the spectrum that lies above the noise threshold.

\begin{figure}[!t]
\includegraphics[width=75mm]{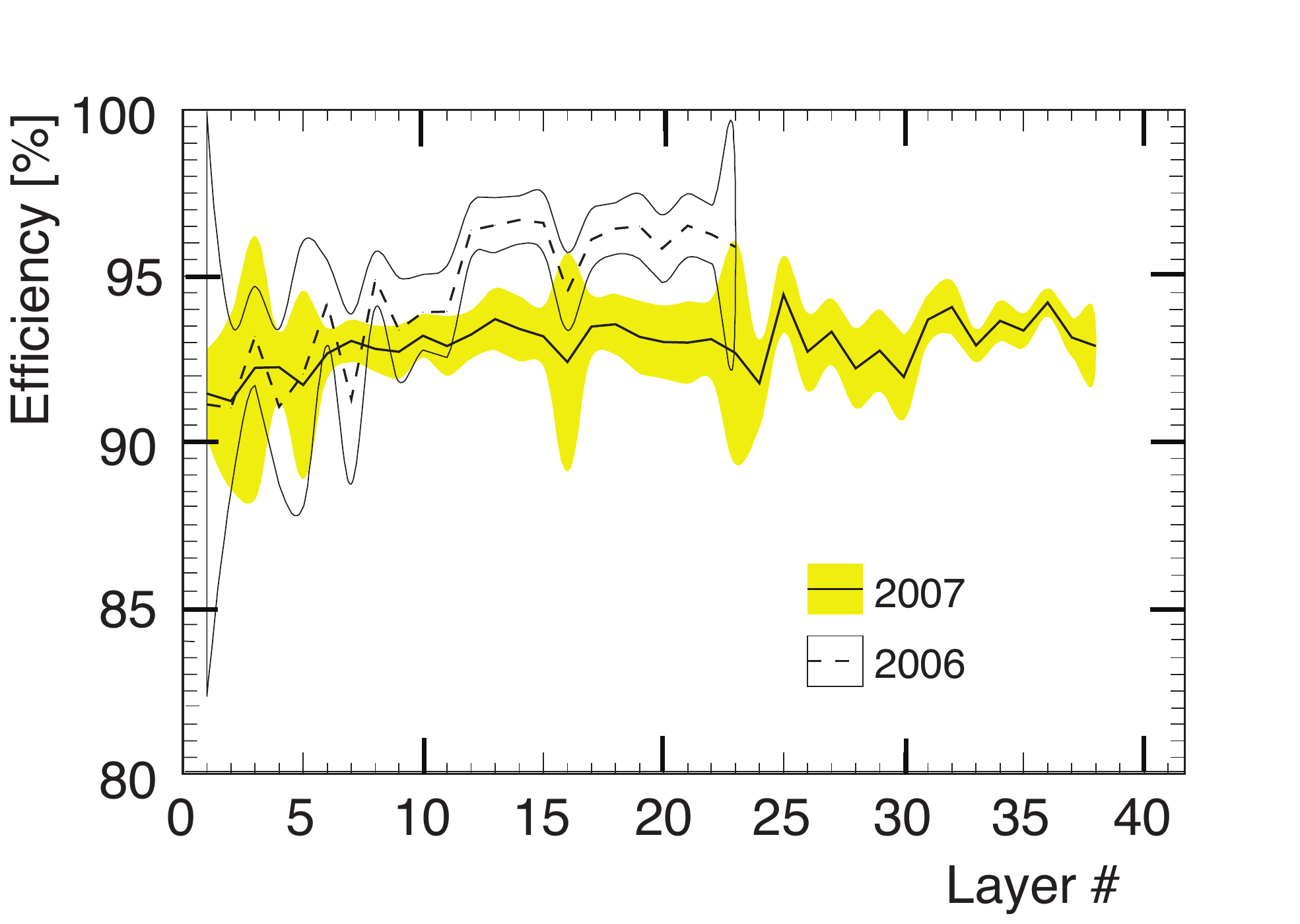}
\includegraphics[width=75mm]{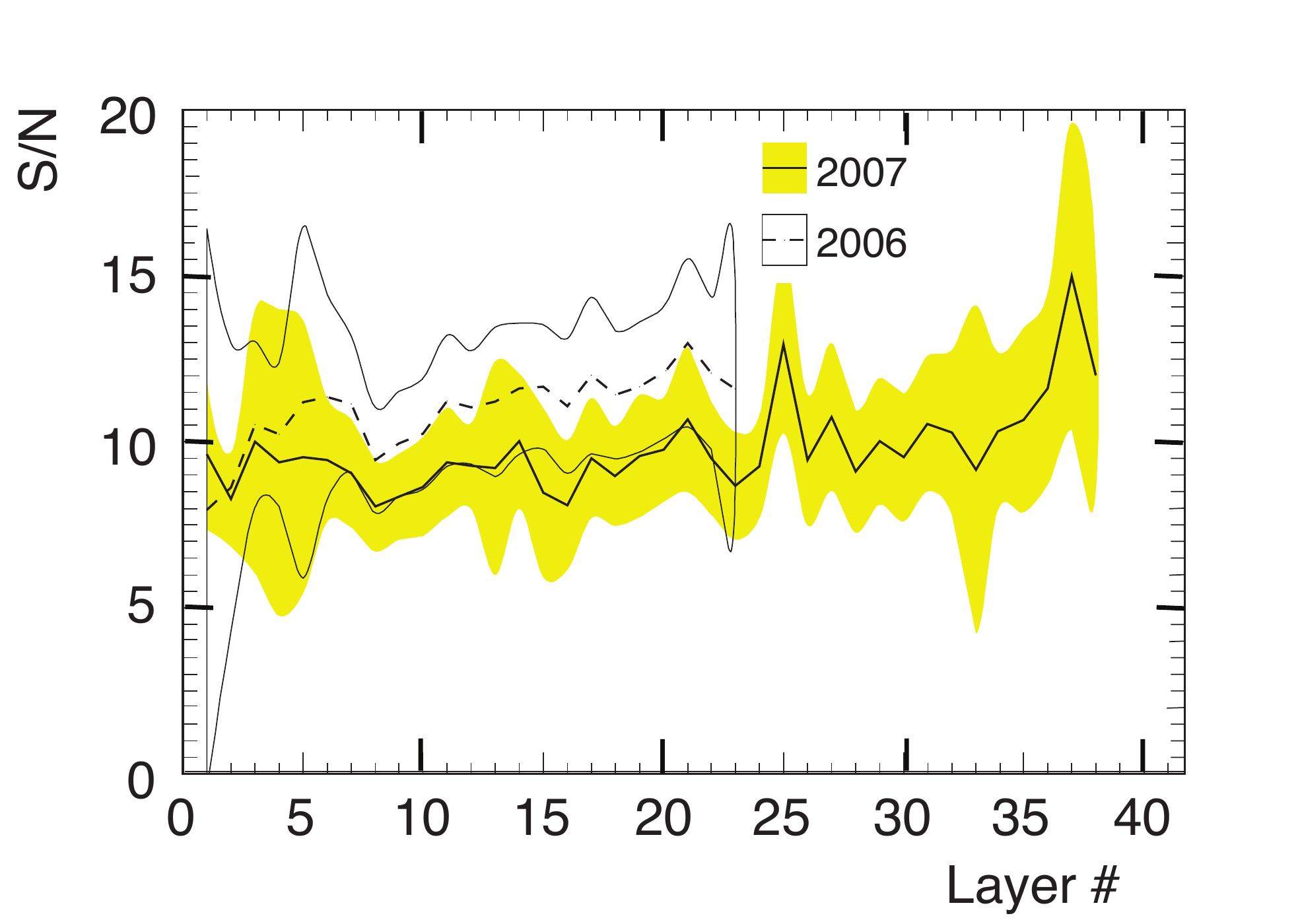}
\caption{{Average MIP detection efficiency in each calorimeter layer (left). Average signal-to-noise ratio in each calorimeter layer for a MIP signal (right). The bands indicates the spread among the tiles in one layer.}}
  \label{fig:mip_eff}
\end{figure}

\begin{figure}[!t]
\begin{center}
\includegraphics[width=90mm]{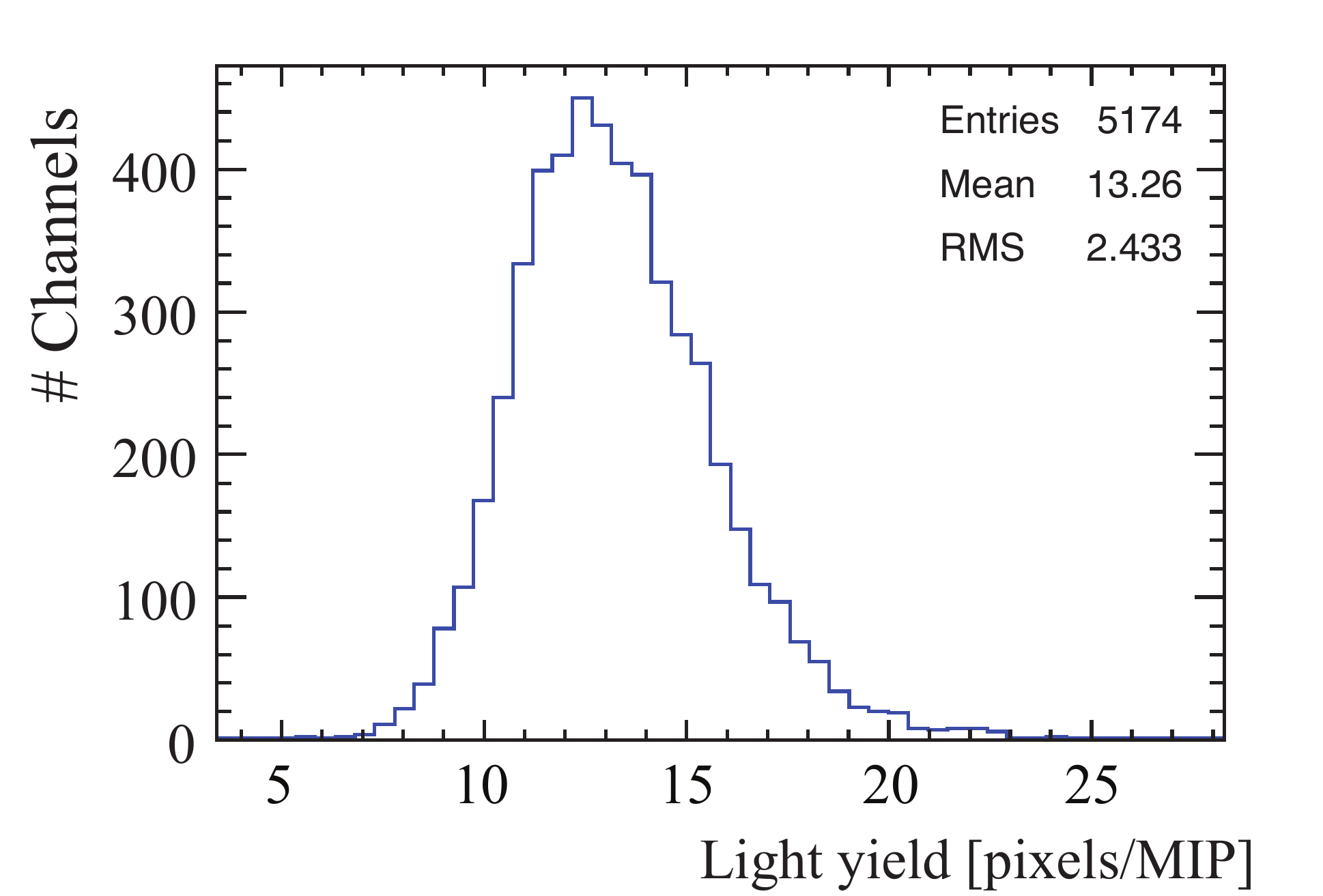}
\caption{Recorded light yields in the test beam.}
  \label{fig:ly_tb}
\end{center}
\end{figure}

Figure~\ref{fig:mip_eff} (left plot) shows the MIP detection efficiency of calibrated cells per module after imposing a minimal threshold of half a MIP. The band indicates the spread (RMS) of all cells in one module. We determine an average efficiency of $93\%$ that lies slightly below our expectation of $95\%$. The reduced detection efficiency, however, is consistent with the fact that the calorimeter was operated at a working point of 13 pixels/MIP instead of the design value of 15 pixels/MIP. Figure~\ref{fig:mip_eff} (right plot) shows the average signal-to-noise ratio for the MIP signal in each module. We measure an average signal-to-noise ratio in the entire calorimeter of $\sim$10. Figure~\ref{fig:ly_tb} shows the light yields measured in the 2006 and 2007 test beam data. The discrepancy between the measurements at ITEP and in the test beam are due to different operating conditions.

\subsection{Noise and Occupancy}
\label{noise}

\begin{figure}[h!]
\begin{center}
\includegraphics[width=72mm]{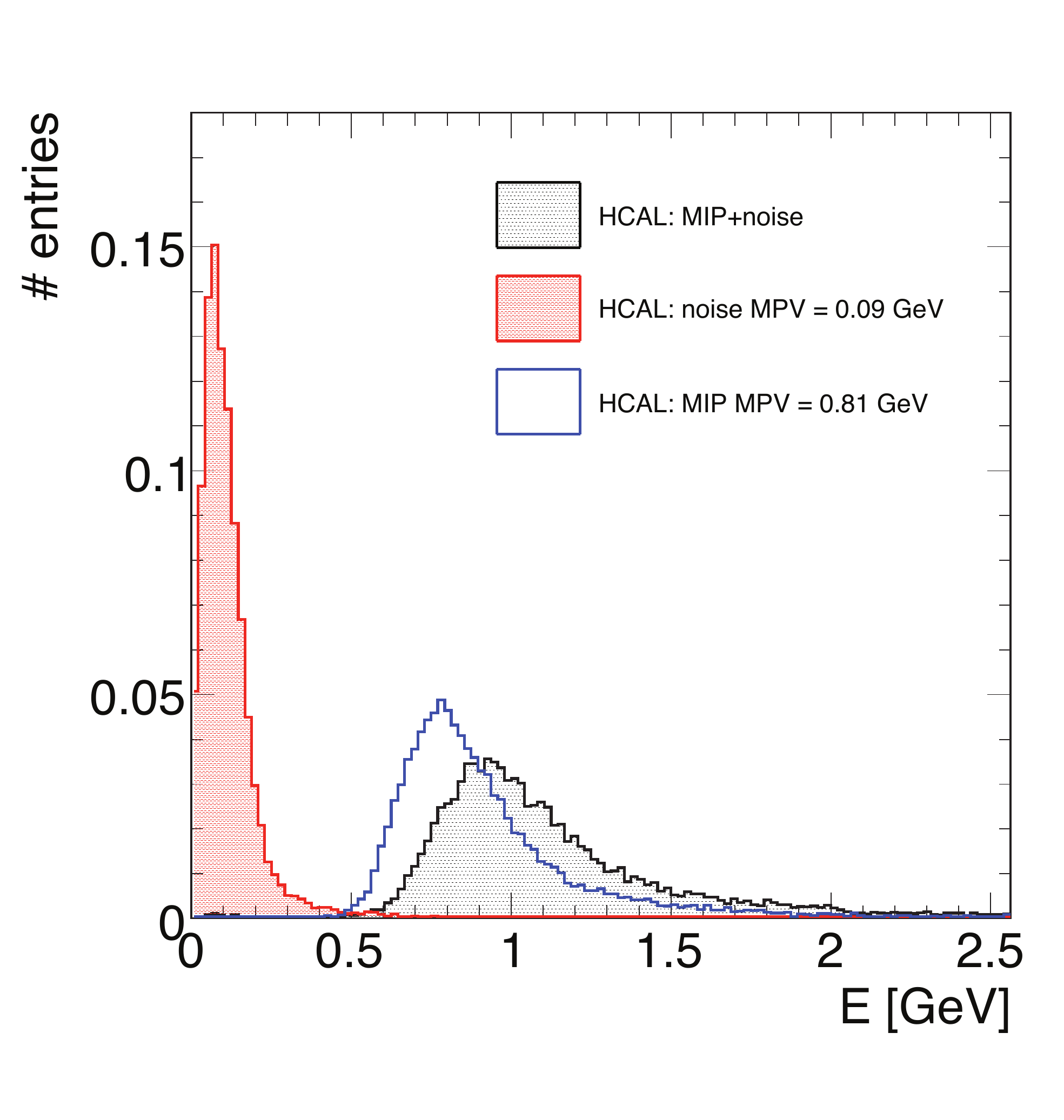}
\includegraphics[width=72mm]{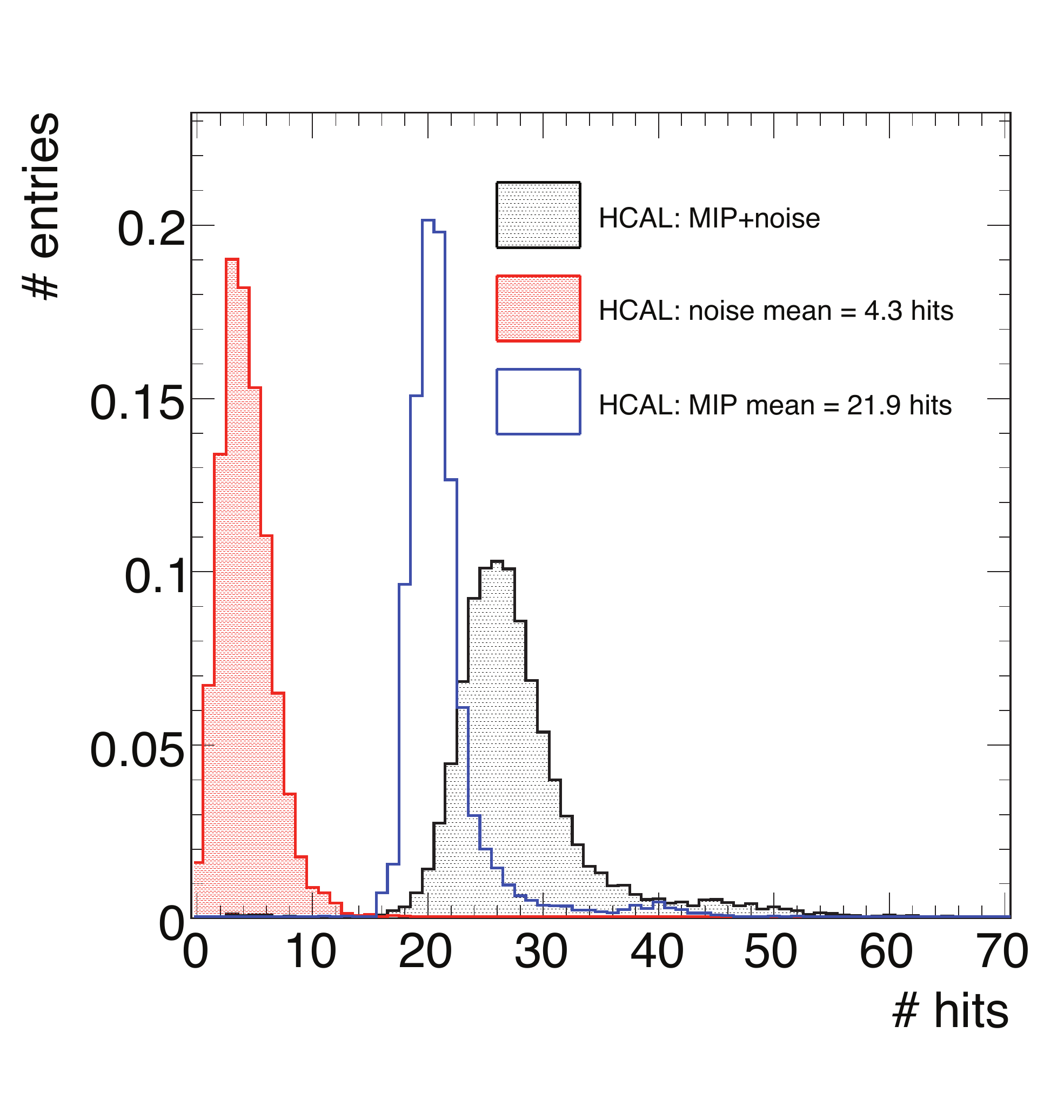}
     \caption{ Distributions of total energy (left) and number of hits (right) for noise (red solid histogram), MIP spectra (unshaded blue histogram), and their combination (dark-shaded black) in the AHCAL. The area of the distributions are normalized to one.}
\label{fig:noisehits}
\end{center}
\end{figure}

An adjustment of the bias voltage for the entire AHCAL prototype is necessary in order to set the light yield to 15 pixels/MIP and in turn to minimize the noise above threshold. If the reverse bias voltage is set too low, the number of pixels per MIP is reduced. In turn, if the threshold is lowered the noise level increases. If, on the other hand, the reverse bias voltage is set too high, the SiPM dark current increases and in turn the noise rises. The optimization procedure allows us to compensate for light yield changes due to large temperature variations (about 57~mV/K). The SiPM operation voltage specified during production is referred to as nominal voltage ($\Delta$ HV = 0 V). The SiPM noise behavior with operation voltage leads to an optimization range of $\rm \pm 1~V$. This is sufficient to compensate for differences between ambient temperature at the test beam and during the SiPM production tests. Figure~\ref{fig:noisehits} shows the distributions of total energy (left plot) and number of  hits (right plot) for  noise (random triggers), for MIP signal measured in muon-induced events, and for their combination. While the noise spectra are 
obtained by summing the entire calorimeter for random triggers, the MIP-only spectrum is obtained by selecting only cells through which the muon passed. By requiring a threshold of half a MIP in each channel the noise is negligible. It should be noted that the MIP position depends on the total number of layers while the noise depends on the volume of the AHCAL.

\begin{figure}[h!]
\vskip 0.0 cm
\begin{center}
\includegraphics[width=70mm]{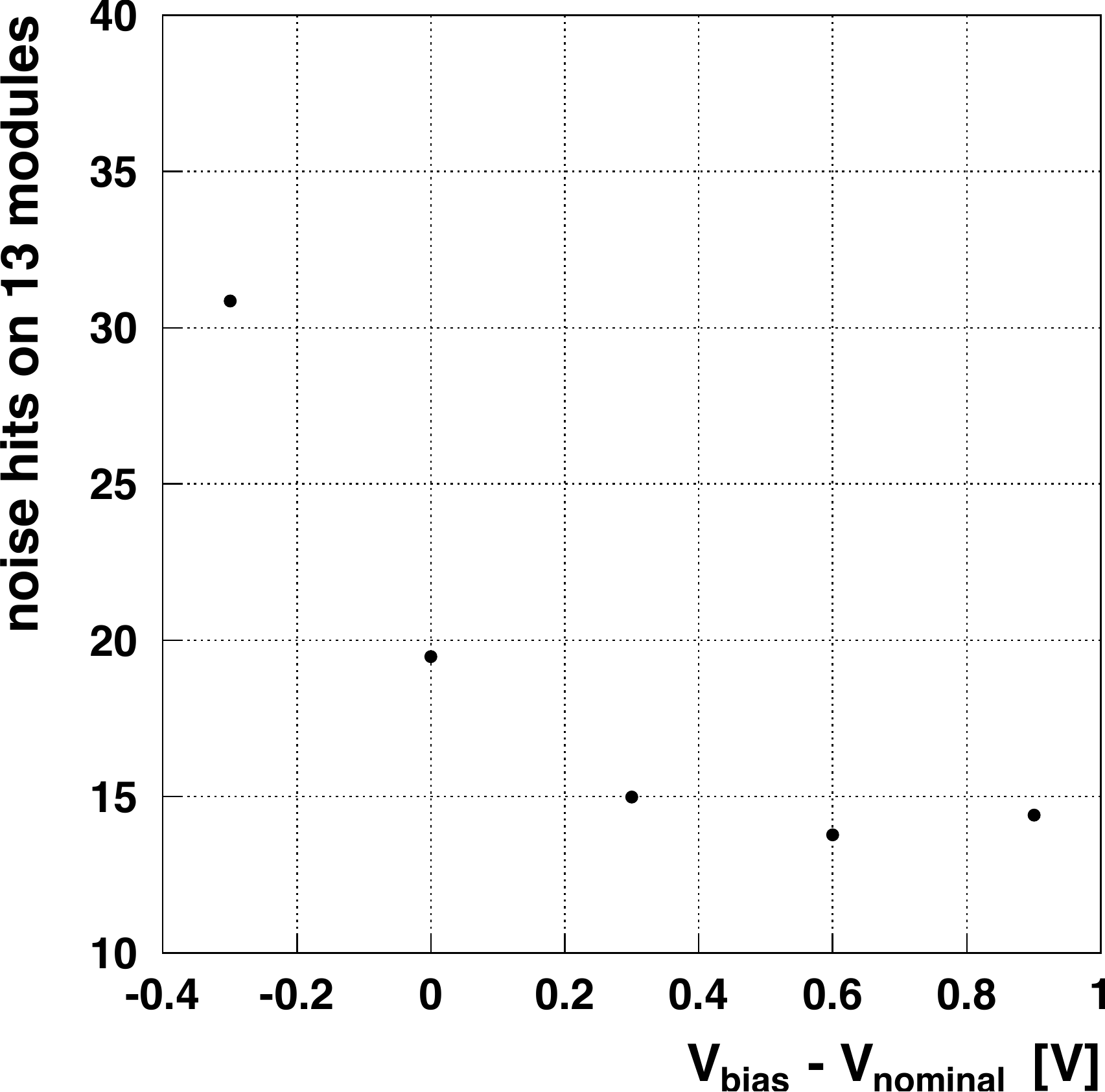}
 \caption{Noise hits as a function of reverse bias voltage with a threshold of half a MIP.}
\label{fig:noise-opt}
\end{center}
\end{figure}

An increase in the reverse bias voltage yields an increase in both gain and noise. The effective noise at half a MIP, however, reaches a minimum before rising again. This is shown in Figure~\ref{fig:noise-opt} for the 2006 data, where we optimized the effective noise. For the 2007 data, we optimized the light yield instead to achieve a homogeneous light yield in the entire calorimeter resulting in an increase in noise levels.  In 2006, the average noise level per cell was $0.9\times 10^{-3}$. In 2007, it increased to 1.3--2$\times 10^{-3}$, which is about a factor of ten higher than the design goal of $10^{-4}$. 

In the entire AHCAL prototype about 2~\% of all cells are unusable for data analysis.
Initially, modules 1 and 2 that were assembled with the very first batch of SiPMs developed high-noise problems because a coating layer on the silicon surface was missing. Dust particles caused shorts to develope frequently between the poly-silicon resistor and the aluminum bus.
The affected pixels drew high current creating sources for high noise in these devices. The problem was identified in test bench studies and the subsequent SiPM production was
modified accordingly. The SiPMs in modules 1 and 2 were exchanged between the 2006 and 2007 test beam operations. In the remaining modules, only $0.2\% $ of the channels suffer from a long-discharge behavior.  A revised quality control procedure on the test bench allowed us to achieve such a small fraction of unusable cells.

\subsection{Performance of the SiPMs in the Test Beam}
\label{sipm-performance}

\begin{figure}
\vskip 0.3cm
\begin{center}
\includegraphics[width=140mm]{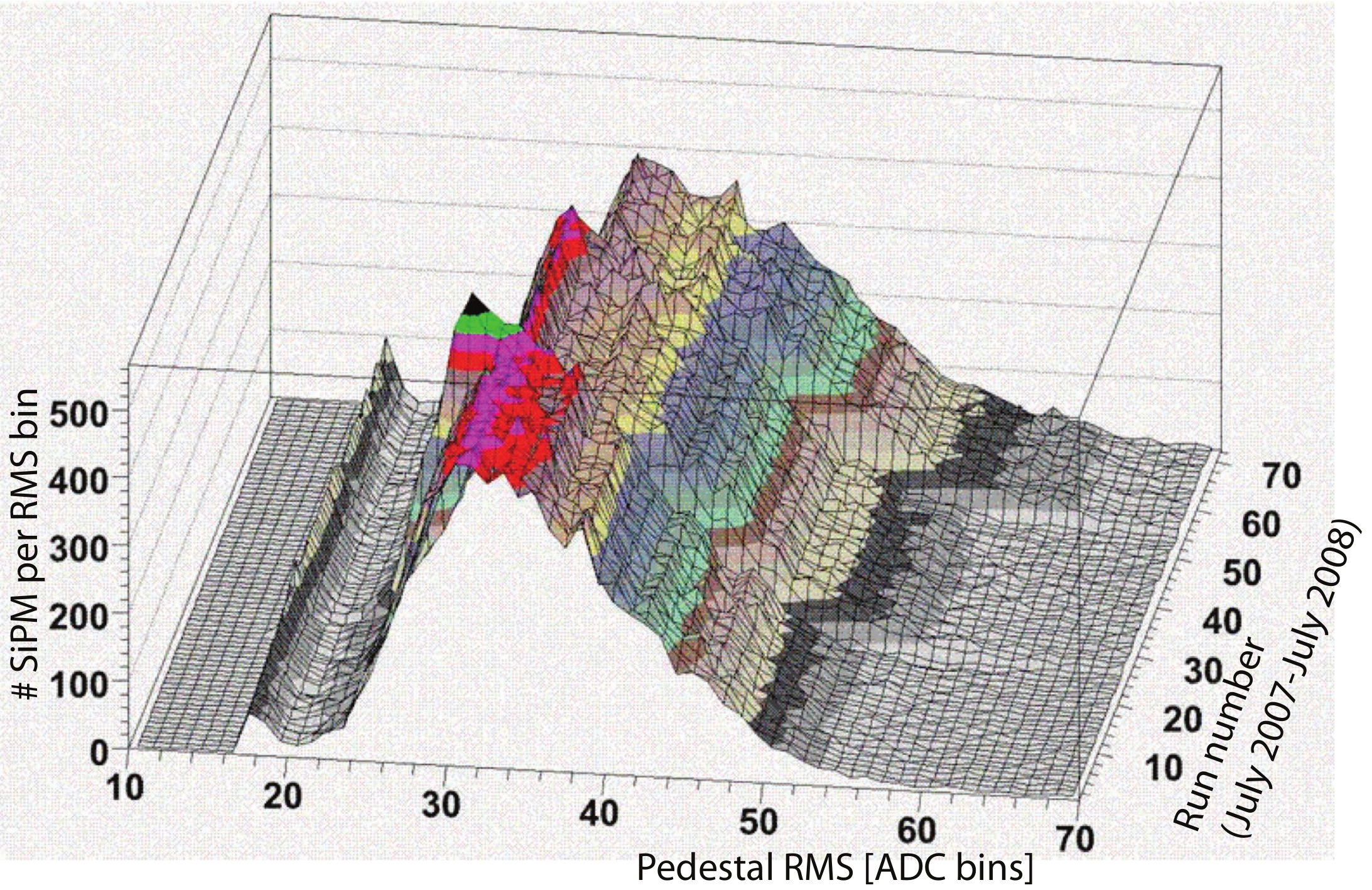}
\caption{SiPM noise during two years of operation in test beams. The pedestals are
  generally between 20 and 60 ADC bins and the number with a given  pedestal RMS is stable with time} 
  \label{2_dim_hist}
  \end{center}
\end{figure}

The performance of the 7608 SiPMs has been studied in three test beam periods, July 2007, May 2008 and July 2008. The pedestal distribution is an excellent indicator for working, non-functioning or noisy SiPMs. The pedestal RMS of all SiPMs for runs between July 2007 and July 2008 is shown in Figure ~\ref{2_dim_hist}.

The left peak corresponds to channels without electrical connection to SiPMs. For these channels the width of the pedestal distribution is about 18 ADC bins due to electronics noise. About $2\%$ of all SiPMs fall into this category because of initially bad soldering and subsequent broken connections caused by slight deformation of the PCB to which the SiPM leads are soldered. Such deformations occurred during detector movements. For example 43 SiPMs became disconnected  after transport from CERN to FNAL. By the end of July 2008, we found that 87 SiPMs had become disconnected permanently, and 115 SiPMs had intermittent disconnections. 

The distribution of the pedestal RMS does not change within periods of constant reverse bias voltage. Three periods with different reverse bias voltage are clearly seen in Figure ~\ref{2_dim_hist}. In addition, a spike during runs 25 and 26 is caused by the adjustment period of the reverse bias voltage. The average pedestal RMS values are 35, 37 and 40 ADC bins for the three periods that used different reverse bias voltages. These widths are much smaller than the typical distance between photoelectron peaks of 300--400 ADC bins and do not affect the SiPM gain calibration. In nine SiPMs the gain cannot be determined. This corresponds to $0.11\%$ of all channels. We also have a few SiPMs which show a sizable increase of the pedestal RMS with time but they are still usable. Thus, we can conclude that the SiPMs in the AHCAL demonstrate excellent performance and stability during four months of operation in 2007--2008. The number of unusable SiPMs is at the per-mil level. The soldering problem is a separate issue which will be fixed in the next prototype.

\subsection{Event Displays}
\label{event-display}

\begin{figure}[!t]
\includegraphics[width=75mm]{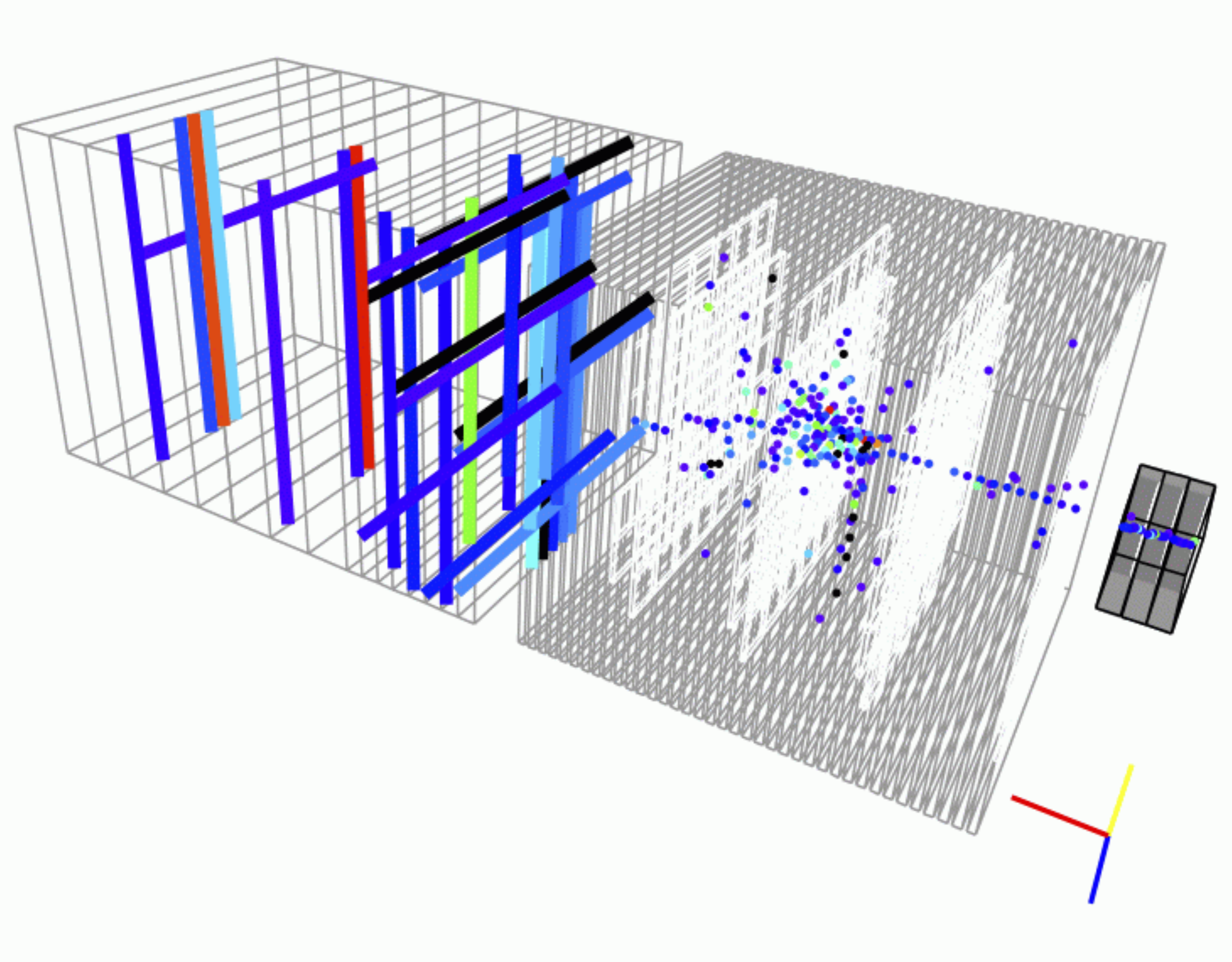}
\includegraphics[width=75mm]{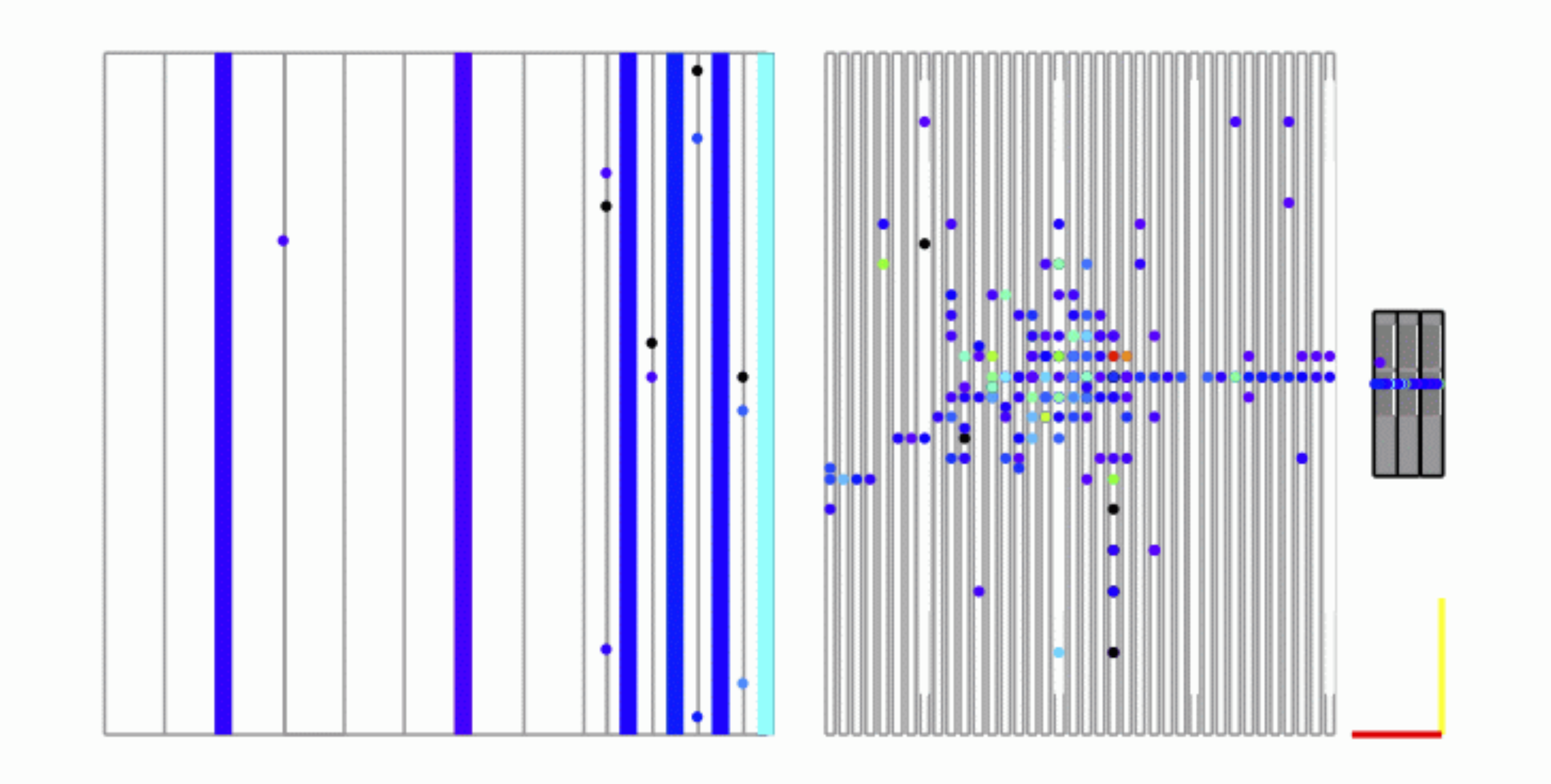}
  \caption{Event displays of a $20$~GeV/c pion from the online monitor, showing  a three dimensional view (left) and a side view (right). The beam is coming from the right-hand side. The shower starts in the middle of AHCAL. Hits in the TCMT are indicated by the colored bars.}
  \label{fig:eventdisplay}
\end{figure}

In order to demonstrate that the AHCAL prototype is capable of measuring the hadronic shower structure, Figure~\ref{fig:eventdisplay} shows a three-dimensional view (left) and a side view (right) of a pion shower that started in the AHCAL after leaving a track-like signature in the ECAL. The display illustrates the excellent imaging potential of the AHCAL and we are confident that this will allow us to test rigorously simulation models and reconstruction algorithms. 

\section{Conclusion} 
\label{conclusion}

We have constructed a 38-layer steel-scintillator tile analog hadron calorimeter prototype that has a total thickness of 5.3~$\lambda_n$ ($4.3\lambda_\pi$). This is the first detector that uses many thousands of SiPMs for readout. The non-linearity of the SiPMs can be reliably corrected. Commissioning and operation in the test beam at CERN have demonstrated that the calorimeter performs according to expectations. The two beam-test periods show the operational stability and robustness of the AHCAL. The high granularity gives us an excellent opportunity for studying hadronic shower shapes in great detail. We anticipate that the results of such studies, in combination with detailed simulations, will demonstrate the extent to which nearby particle showers can be separated from one another, and hence provide a direct test of the performance that may ultimately be achieved using the particle flow approach.


\section{Acknowledgments}    

We would like to thank the technicians and the engineers who
contributed to the design and construction of the prototypes, in particular P. Smirnov. 
We express our gratitude to the DESY, CERN and FNAL laboratories for hosting our test beam experiments, and to their staff for the efficient
accelerator operation and excellent support.
We would like to thank the HEP group of the University of
Tsukuba for the loan of drift chambers for the DESY test beam.
The authors would like to thank the RIMST (Zelenograd) group for their
help and sensors manufacturing.
This work was supported by the 
Bundesministerium f\"{u}r Bildung und Forschung, Germany;
by the DFG cluster of excellence 'Origin and Structure of the Universe' of Germany;
by the Helmholtz-Nachwuchsgruppen grant VH-NG-206;
by the BMBF, grant no. 05HS6VH1;
by the Alexander von Humboldt Foundation (Research Award IV, RUS1066839 GSA);
by joint Helmholtz Foundation and RFBR grant HRJRG-002, Russian Agency for Atomic Energy, ISTC grant 3090; 
by the Norwegian Research Council;
by joint Helmholtz Foundation and RFBR grant HRJRG-002,
SC Rosatom;
by Russian Grants  SS-1329.2008.2 and RFBR08-02-12100-OFI
and by the Russian Ministry of Education and Science contract
02.740.11.0239;
by CICYT,Spain;
by CRI(MST) of MOST/KOSEF in Korea;
by the US Department of Energy and the US National Science
Foundation;
by the Ministry of Education, Youth and Sports of the Czech Republic
under the projects AV0 Z3407391, AV0 Z10100502, LC527  and by the
Grant Agency of the Czech Republic under the project 202/05/0653;  
and by the Science and Technology Facilities Council, UK.





\end{document}